\documentclass[lettersize,journal]{IEEEtran}
\usepackage{amsmath,amsfonts}
\usepackage{algorithmic}
\usepackage{algorithm}
\usepackage{array}
\usepackage[caption=false,font=normalsize,labelfont=sf,textfont=sf]{subfig}
\usepackage{textcomp}
\usepackage{stfloats}
\usepackage{url}
\usepackage{booktabs}
\usepackage{verbatim}
\usepackage{graphicx}
\usepackage{amsfonts}
\usepackage{bbding}
\usepackage{pifont}
\usepackage{enumitem}
\usepackage{color}
\usepackage[table]{xcolor}
\usepackage{multirow}
\usepackage{makecell}
\usepackage{amssymb}
\usepackage{cite}
\usepackage{xspace}
\usepackage{hyperref} 
\hypersetup{colorlinks=true, linkcolor=red}
\newcommand{\eg}{\emph{e.g.,}\xspace}
\newcommand{\ie}{\emph{i.e.,}\xspace}

\begin{document}

\title{Personalized Recommendation Models in Federated Settings: A Survey}

\author{Chunxu Zhang, Guodong Long, Zijian Zhang, Zhiwei Li, Honglei Zhang, Qiang Yang, \textit{Fellow}, \textit{IEEE}, Bo Yang
        % <-this % stops a space
\thanks{Chunxu Zhang, Zijian Zhang and Bo Yang are with the College of Computer Science and Technology, Jilin University, Jilin, China (e-mail: zhangchunxu@jlu.edu.cn, zhangzijian@jlu.edu.cn, ybo@jlu.edu.cn).}% <-this % stops a space
\thanks{Guodong Long and Zhiwei li are with the Australian AI Institute, Faculty of Engineering and IT, University of Technology Sydney, Sydney, Australia (e-mail: guodong.long@uts.edu.au, zhiwei.li@student.uts.edu.au).}
\thanks{Honglei Zhang is with the School of Computer Science and
Technology, Beijing Jiaotong University, Beijing, China (e-mail: honglei.zhang@bjtu.edu.cn).}
\thanks{Qiang Yang is Professor Emeritus at the Department of Computer Science and Engineering, Hong Kong University of Science and Technology,
Hong Kong, and the Chief AI Officer of WeBank, Shenzhen, China (e-mail: qyang@cse.ust.hk).}
\thanks{\textit{Corresponding Author}: Bo Yang.}
}

% \markboth{Journal of \LaTeX\ Class Files,~Vol., No., 2025}%


\maketitle

\begin{abstract}
Federated recommender systems (FedRecSys) have emerged as a pivotal solution for privacy-aware recommendations, balancing growing demands for data security and personalized experiences. Current research efforts predominantly concentrate on adapting traditional recommendation architectures to federated environments, optimizing communication efficiency, and mitigating security vulnerabilities. However, user personalization modeling, which is essential for capturing heterogeneous preferences in this decentralized and non-IID data setting, remains underexplored. This survey addresses this gap by systematically exploring personalization in FedRecSys, charting its evolution from centralized paradigms to federated-specific innovations. We establish a foundational definition of personalization in a federated setting, emphasizing personalized models as a critical solution for capturing fine-grained user preferences. The work critically examines the technical hurdles of building personalized FedRecSys and synthesizes promising methodologies to meet these challenges. As the first consolidated study in this domain, this survey serves as both a technical reference and a catalyst for advancing personalized FedRecSys research.
\end{abstract}


\section{Introduction}

\subsection{Motivation}
Federated recommender systems (FedRecSys)~\cite{chai2020secure,yang2020federated,huang2021feddsr,wang2022fast,wu2022federated,zhang2023dual} have burgeoned as a remarkable paradigm to promote privacy-preserving recommendation services. By embodying recommender systems (RecSys)~\cite{bobadilla2013recommender,zangerle2022evaluating,zhang2019deep,wu2022graph,wu2024supporting} within the federated learning (FL) framework~\cite{mcmahan2017communication,zhang2021survey,kairouz2021advances,li2021survey,chai2024survey,liu2024vertical}, FedRecSys mitigates the risk of user privacy leakage with local data storage. Besides, the distributed optimization pattern enables service providers to effectively harness the vast computational resources of various devices. This balance between performance and privacy protection makes FedRecSys an attractive research avenue with significant potential for edge AI development.

\begin{figure}[!t]
\setlength{\abovecaptionskip}{-1mm}
\setlength{\belowcaptionskip}{3mm}
    \centering
    \includegraphics[width=1.0\linewidth]{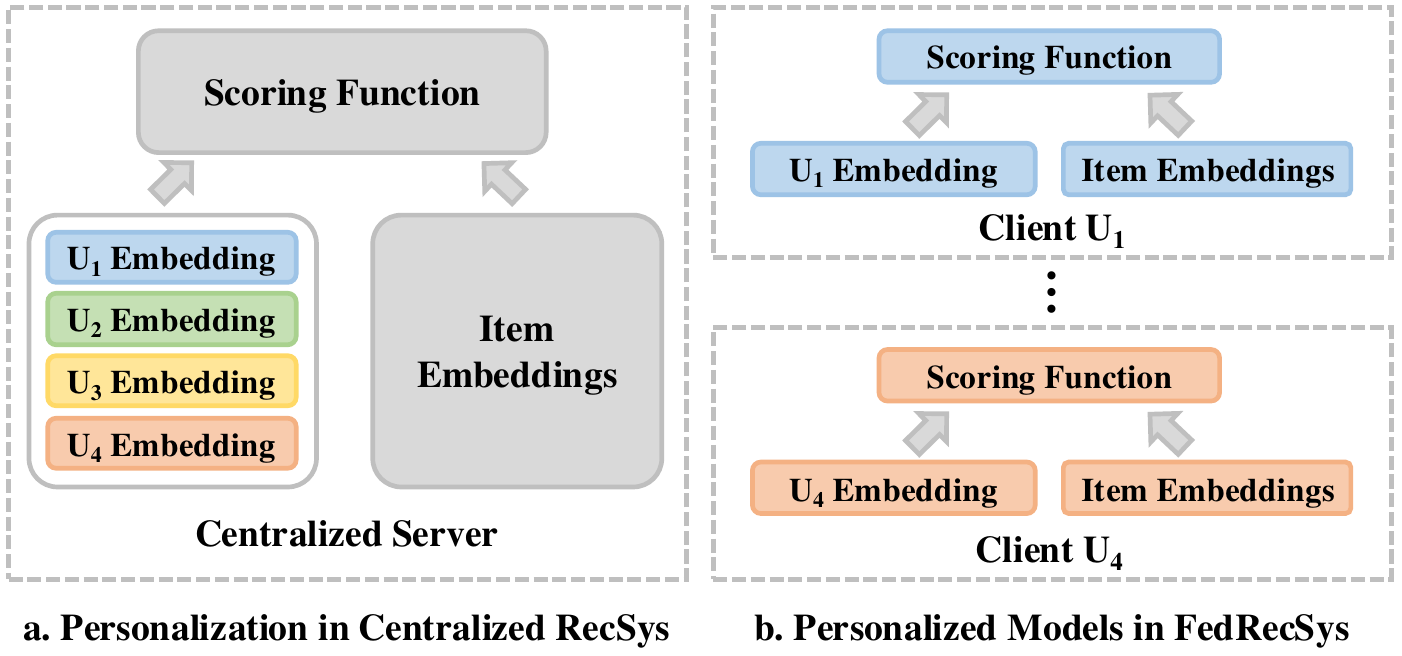}
    \caption{Personalization technique comparison in centralized and federated RecSys. The \textbf{colorful} module denotes the \textbf{user-specific} parameters and the \textbf{gray} module represents the \textbf{user-shared} parameters. FL's ability to collaboratively train multiple models across different devices naturally supports the development of personalized models, making it easier to tailor recommendations to individual user needs.}
    \label{fig:personalization}
\end{figure}

Current research in FedRecSys primarily derives from the perspectives of \textbf{RecSys} and \textbf{FL} views. 
It encompasses various \textit{{model architectures}}~\cite{ammad2019federated,perifanis2022federated} and \textit{{recommendation scenarios}}~\cite{meihan2022fedcdr,liu2022federated} within RecSys, as well as the inherent challenges of FL, such as \textit{{security}}~\cite{qu2023semi}, \textit{{robustness}}~\cite{zhang2022pipattack} and \textit{{efficiency}}~\cite{zhang2023lightfr}. Despite the significant progress in FedRecSys, we highlight an important yet often overlooked aspect, \ie \textbf{user personalization modeling}. Personalization lies at the heart of RecSys, enabling tailored services that adapt dynamically to user interests and requirements. This is especially crucial in FedRecSys, as the non-iid characteristic of data complicates the accurate capture of user preferences. Personalized models offer an effective solution by decoupling user-specific preferences, allowing for the introduction of user-specific parameters that capture unique interests that global models often miss due to statistical bias~\cite{jiang2022adaptive,zhang2024m3oe}. Moreover, they support continuous adaptation, allowing systems to update recommendations in response to evolving user preferences, which enhances both long-term user satisfaction and retention~\cite{qin2020multitask,guo2024multi}.

However, the potential for personalized modeling in FedRecSys has long been overlooked. The collaborative optimization process in FL, which trains multiple client models, naturally facilitates the development of personalized models. As shown in Figure \ref{fig:personalization}, traditional RecSys rely on a single and unified model for all users, only preserving user-specific embeddings to distinguish users. 
In contrast, FedRecSys can leverage the federated architecture to allow each client to tailor the item embeddings and scoring function to its local data, significantly enhancing user personalization modeling while maintaining privacy. This approach not only enhances the precision of user preference modeling but also mitigates the challenges posed by non-IID data, positioning it as especially effective for large-scale, decentralized systems.

% Furthermore, the significant non-iid nature of data in FedRecSys complicates user preferences capture. As a result, developing FedRecSys focusing on user personalized modeling is imperative to advance the studies to new frontiers.

In this paper, we provide a comprehensive examination of user personalization modeling in FedRecSys, especially from the perspective of \textbf{learning personalized models}. Specifically, we first build an extensive review of existing FedRecSys studies, offering insights into the status of the field and available code resources. Based on this foundation, we formulate a clear definition of personalization in FedRecSys and deeply explore its role in RecSys and FL, and highlight that learning personalized models has profound significance in FedRecSys. Furthermore, we dive into a comprehensive discussion about the challenges and solutions of learning personalized models in FedRecSys. Finally, we outline the future directions to accelerate the advancement of personalized FedRecSys.

\begin{figure*}[!t]
\setlength{\abovecaptionskip}{-1mm}
\setlength{\belowcaptionskip}{3mm}
    \centering
    \includegraphics[width=0.9\linewidth]{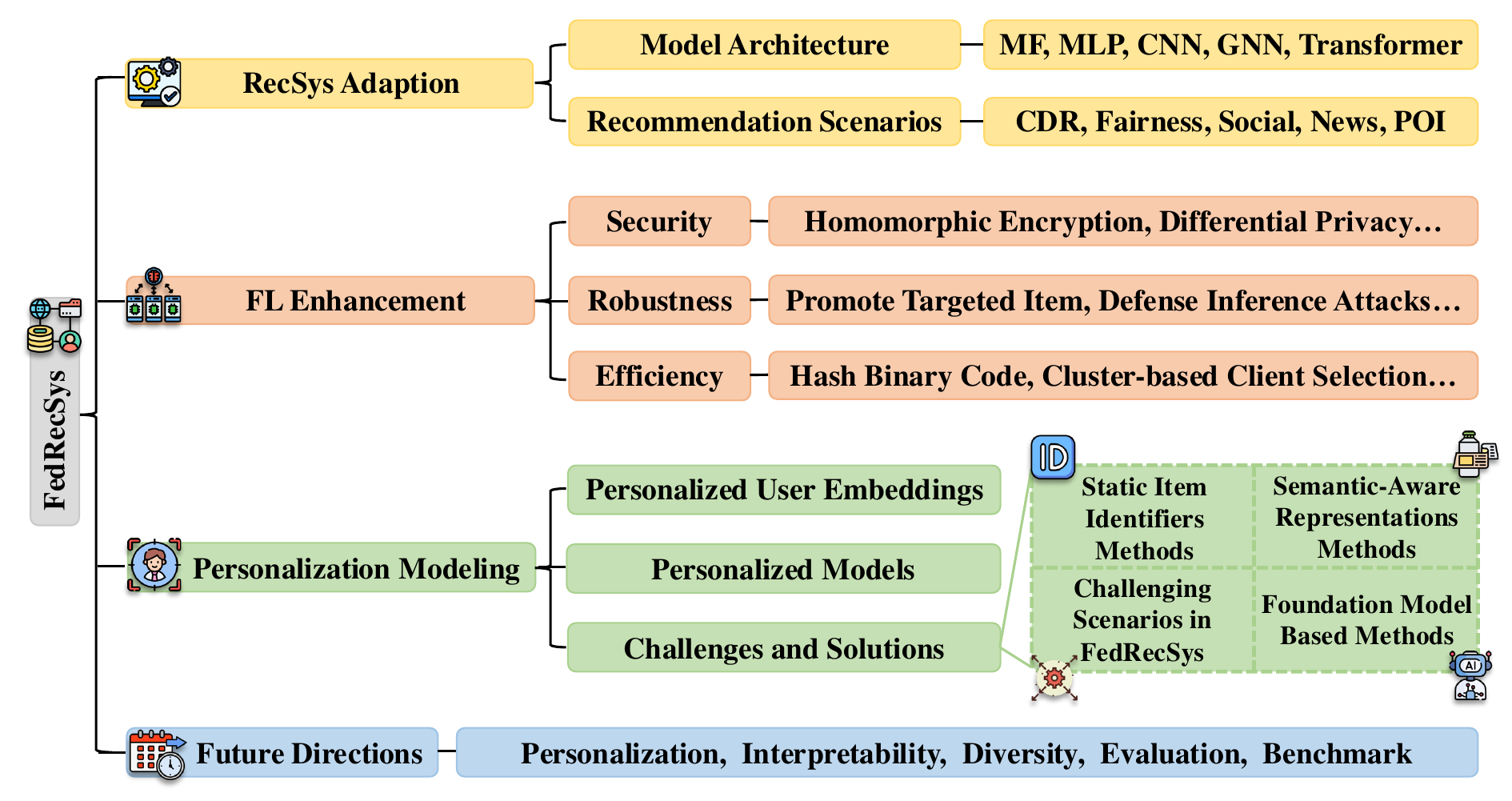}
    \caption{Overview of this paper. We summarize existing FedRecSys methods from two perspectives: \textbf{RecSys Adaptation} (focusing on model architectures and scenarios) and \textbf{FL Enhancement} (improving security, robustness, and efficiency). We then explore the role of \textbf{personalization modeling} in FedRecSys, emphasizing its potential for future development. Finally, we discuss challenges and solutions for personalized model-driven FedRecSys and outline promising \textbf{future directions} to advance research in this field.}
    \label{fig:organization}
\end{figure*}
\subsection{Related Surveys}
With the advancement of the field, several review papers have examined various facets of FedRecSys. 
% For instance, Yang et al.~\cite{yang2020federated} first systematically define FedRecSys and provide a thorough summary of related work from the FL perspective. 
For instance, Javeed et al.~\cite{javeed2023federated} and Harasic et al.~\cite{chronis2024survey} primarily focus on the challenges and solutions of FedRecSys from the standpoint of privacy and security. Works such as \cite{harasic2024recent,alamgir2022federated,sun2024survey,wang2024horizontal} provide valuable insights into the aspects of recommendation model architectures, FL paradigms, and common challenges encountered in FL. Li et al.~\cite{li2024navigating} delves into the emerging challenges that arise when integrating FedRecSys with cutting-edge foundation models. We compare our survey with existing reviews across key aspects of FedRecSys, using \textcolor{green}{\Checkmark} to denote covered topics and \textcolor{red}{\ding{55}} to indicate areas not addressed.

Existing review papers typically cover broad discussions of RecSys and FL, overlooking the critical aspect of user personalization modeling. Specifically, none explore the development of personalized models within the FL framework, neglecting the user-centric nature of personalization. Moreover, recent advancements in this area remain under-explored, and there is still a notable gap in providing consolidated code resources for practitioners. This paper seeks to address these gaps by offering an in-depth exploration of user personalization modeling in FedRecSys, emphasizing its significance, challenges, and the potential innovations that personalized models can bring to the field. By focusing on this crucial yet under-addressed area, we aim to make a timely and valuable contribution to the growing body of research on personalized FedRecSys.

\begin{table*}[ht]
\centering
\caption{Comparison of existing surveys about FedRecSys with This Survey Paper.}
\label{tab:comparison}
\begin{tabular}{c|c|c|c|c|c|c|c|c|c}
% \hline
\toprule
\multirow{2}{*}{\textbf{Year}} & \multirow{2}{*}{\textbf{References}} & \textbf{Model} & \textbf{Recommendation} & \multirow{2}{*}{\textbf{Security}} & \multirow{2}{*}{\textbf{Robustness}} & \multirow{2}{*}{\textbf{Efficiency}} & \textbf{Personalized} & \textbf{Objective} & \textbf{Code} \\
& & \textbf{Architecture} & \textbf{Scenario} & & & & \textbf{Model} & \textbf{Formulation} & \textbf{Resources} \\
% \hline
\midrule
% 2020 & Yang et al.~\cite{yang2020federated} & $\times$ & \textcolor{green}{\Checkmark} & $\times$ & $\times$ & $\times$ & $\times$ & \textcolor{green}{\Checkmark} & $\times$ \\
% \hline
2022 & Alamgir et al.~\cite{alamgir2022federated} & \textcolor{green}{\Checkmark} & \textcolor{red}{\ding{55}} & \textcolor{green}{\Checkmark} & \textcolor{green}{\Checkmark} & \textcolor{green}{\Checkmark} & \textcolor{red}{\ding{55}} & \textcolor{red}{\ding{55}} & \textcolor{red}{\ding{55}} \\
% \hline
\midrule
2023 & Javeed et al.~\cite{javeed2023federated} & \textcolor{red}{\ding{55}} & \textcolor{green}{\Checkmark} & \textcolor{green}{\Checkmark} & \textcolor{red}{\ding{55}} & \textcolor{red}{\ding{55}} & \textcolor{red}{\ding{55}} & \textcolor{red}{\ding{55}} & \textcolor{red}{\ding{55}} \\
% \hline
\midrule
\multirow{5}{*}{2024} & Chronis et al.~\cite{chronis2024survey} & \textcolor{red}{\ding{55}} & \textcolor{green}{\Checkmark} & \textcolor{green}{\Checkmark} & \textcolor{green}{\Checkmark} & \textcolor{red}{\ding{55}} & \textcolor{red}{\ding{55}} & \textcolor{red}{\ding{55}} & \textcolor{red}{\ding{55}} \\
& Harasic et al.~\cite{harasic2024recent} & \textcolor{green}{\Checkmark} & \textcolor{green}{\Checkmark} & \textcolor{red}{\ding{55}} & \textcolor{red}{\ding{55}} & \textcolor{red}{\ding{55}} & \textcolor{red}{\ding{55}} & \textcolor{red}{\ding{55}} & \textcolor{red}{\ding{55}} \\
& Sun et el.~\cite{sun2024survey} & \textcolor{green}{\Checkmark} & \textcolor{green}{\Checkmark} & \textcolor{green}{\Checkmark} & \textcolor{green}{\Checkmark} & \textcolor{green}{\Checkmark} & \textcolor{red}{\ding{55}} & \textcolor{red}{\ding{55}} & \textcolor{red}{\ding{55}} \\
& Wang et al.~\cite{wang2024horizontal} & \textcolor{green}{\Checkmark} & \textcolor{red}{\ding{55}} & \textcolor{green}{\Checkmark} & \textcolor{green}{\Checkmark} & \textcolor{green}{\Checkmark} & \textcolor{red}{\ding{55}} & \textcolor{red}{\ding{55}} & \textcolor{green}{\Checkmark} \\
& Li et al.~\cite{li2024navigating} & \textcolor{red}{\ding{55}} & \textcolor{red}{\ding{55}} & \textcolor{green}{\Checkmark} & \textcolor{red}{\ding{55}} & \textcolor{green}{\Checkmark} & \textcolor{red}{\ding{55}} & \textcolor{red}{\ding{55}} & \textcolor{red}{\ding{55}} \\
% \hline
% \bottomrule
\midrule
% \hline
\multicolumn{2}{c|}{\textbf{This Survey Paper}} & \textcolor{green}{\Checkmark} & \textcolor{green}{\Checkmark} & \textcolor{green}{\Checkmark} & \textcolor{green}{\Checkmark} & \textcolor{green}{\Checkmark} & \textcolor{green}{\Checkmark} & \textcolor{green}{\Checkmark} & \textcolor{green}{\Checkmark} \\
% \hline
\bottomrule
\end{tabular}
\end{table*}

\subsection{Contributions}
The \textbf{main contributions} of this paper are as follows:
\begin{itemize}
    % \item We extensively review foundational and recent advancements in FedRecSys, systematically analyzing the field through RecSys and FL with a focus on taxonomy construction and optimization objective formalization. Additionally, we curate an open-source code repository\footnote{\url{https://anonymous.4open.science/r/Personalized_FedRecSys}}, actively maintained to accelerate future research.
    % \item We propose the first formal definition of personalization in FedRecSys and establish a systematic objective optimization framework. This framework establishes a unified theoretical foundation for designing personalized FedRecSys and directly guides its technical roadmap.
    % \item We identify personalized models as the cornerstone of FedRecSys, offering a structured analysis of critical challenges and proposing potential solutions across three dimensions: embedding representation forms, common FedRecSys challenges, and emerging foundational models. These insights establish a robust foundation for advancing personalization in federated environments.
    % \item We outline pivotal pathways for advancing FedRecSys, focusing on the theoretical deepening of personalization, privacy-preserving enhancements and scalability optimization. Addressing these challenges will promote the practical deployment of FedRecSys with reliable and personalized services.

    \item We systematically review the advancements in FedRecSys from RecSys and FL, including taxonomy construction and optimization objective formalization. The FedRecSys paper repository with the open-source code\footnote{\url{https://anonymous.4open.science/r/Personalized_FedRecSys}} is made public for a clear overview.

    \item For the first time, we propose a formal definition of personalization in FedRecSys with a systematic optimization objective, which establishes a unified theoretical foundation for designing personalized FedRecSys.
    
    \item We identify personalized models as the cornerstone of FedRecSys, highlighting a structured analysis of critical challenges with potential solutions across three dimensions: embedding representation forms, common FedRecSys challenges, and emerging foundational models. These insights offer valuable practical guidance for implementing personalization in federated environments.
    % \item We outline pivotal pathways for advancing FedRecSys, focusing on the theoretical deepening of personalization, privacy-preserving enhancements and scalability optimization. Addressing these challenges will promote the practical deployment of FedRecSys with reliable and personalized services.
    
    % We outline key future research directions for FedRecSys, emphasizing critical areas for development, such as advancing personalization modeling, improving privacy-preserving techniques, and enhancing scalability and real-world applicability. By addressing these challenges, we aim to accelerate the advancement of FedRecSys, ensuring their effective personalization and successful deployment in practical scenarios.
\end{itemize}

% In this paper, we focus on discussing personalization modeling in the federated recommender system, especially from learning customized model parameters for individual users. Several previous works~\cite{zhang2023dual,zhang2024gpfedrec,li2023federated} have explored learning personalized item embedding for each client, whose key insight is that the users preserve different views of the same items. Along this line, this survey researches how personalized item embedding works in the federated recommender system and how to solve the challenges in various federated recommendation frameworks. Recently, there emerge several surveys discussing federated recommender system~\cite{yang2020federated,sun2024survey,wang2024horizontal}. However, they review existing federated recommendation models to provide a comprehensive categorization, and none of them pay attention to the personalization modeling discussion in the federated recommender system. This survey is the first to investigate the personalized federated recommender system in-depth and provide guidance on personalization implementation. Furthermore, this paper discerns the future directions worth exploring that shed light on the personalized federated recommender system deployment in practical applications.

\subsection{Organization}
The remainder of this paper is structured as follows. Section \ref{preliminary} presents the definition, optimization objective, and pipeline of FedRecSys for a comprehensive overview. In Section \ref{taxonomy}, existing FedRecSys are classified into two categories based on technical focuses: “RecSys Adaption” and “FL Enhancement”, with further detailed taxonomies for each. Section \ref{personalization} formally defines personalization in FedRecSys, and emphasizes personalized models as a crucial future direction. Section \ref{challenges_and_solutions} explores challenges and solutions in applying personalized models in FedRecSys across representative scenarios. Section \ref{future_directions} discusses promising future directions for personalized FedRecSys research. Finally, Section \ref{conlusion} concludes the paper. Figure \ref{fig:organization} summarizes the paper's overall structure.

\section{Preliminary}\label{preliminary}
In this section, we first provide the definition and universal optimization objective of FedRecSys, which can be instantiated with various federated recommendation models. Then, we introduce its optimization pipeline, offering a comprehensive overview by delineating the iterative workflow encompassing client training, server aggregation, and global synchronization.

\subsection{Definition and Optimization Objective}
% FedRecSys can be defined as distributed recommendation models trained collaboratively across multiple clients (\eg user devices) without sharing their local data. The key idea is to leverage the collective knowledge from all participating clients to improve the recommendation performance, while preserving the privacy and data ownership of individual clients.
\textsc{\textbf{Definition 1.}}
FedRecSys is a privacy-preserving machine learning paradigm that trains decentralized recommendation models through coordinated parameter aggregation across distributed clients (\eg user devices). By maintaining raw data localized on client nodes and exchanging encrypted model updates during collaborative training, the system achieves dual objectives: (a) enhancing recommendation accuracy through knowledge fusion from heterogeneous user behaviors, and (b) ensuring data sovereignty via cryptographic protocols that prevent private data exposure.

Let \(  \mathcal{U} \) and \( \mathcal{I} \) denote the user set and item set, respectively. Each client \( u \in U \) maintains private interaction records \( \mathcal{Y}_u \), and \( \mathcal{Y} = \bigcup_{u \in U} \mathcal{Y}_u \) is the complete set of user-item interactions. The FedRecSys aims to learn a global model by minimizing the following optimization objective:
\begin{equation}
\min_{\theta} \sum_{u \in U} \alpha_{u} \mathcal{L}_{u}(\theta; \mathcal{Y}_u)
\end{equation}
Here, \( \theta \) denotes the recommendation model parameters, \( \mathcal{L}_u \) is the local loss function (\eg MSE for explicit feedback~\cite{chai2020secure} or BCE for implicit feedback~\cite{perifanis2022federated}), and \( \alpha_u \) is the aggregation weight typically proportional to client data size \( \alpha_u = |\mathcal{Y}_u| / |\mathcal{Y}| \). Rigorous data locality means that \( \mathcal{Y}_u \) stays only on client 
\( u \)'s local device, thereby preserving user privacy through decentralized data governance. 

\begin{figure}[!t]
    \centering
    \includegraphics[width=1.\linewidth]{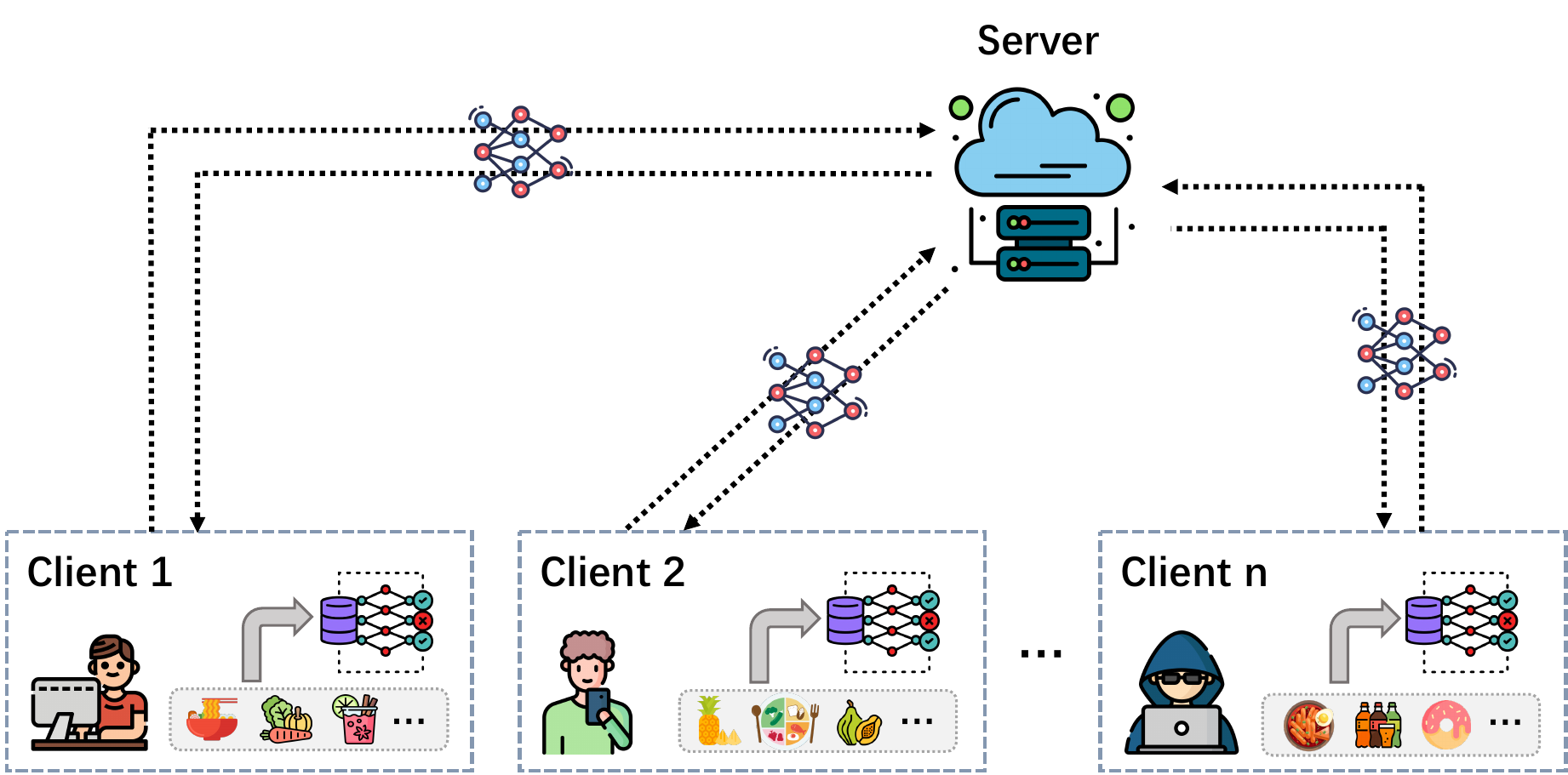}
    \caption{The framework of FedRecSys. The users (clients) store personal data and train the recommendation model locally. A cloud server orchestrates the global training by aggregating and distributing model parameters of all users iteratively. Once the training converges, each client device can predict the potentially interesting items for the user.}
    \label{fig:fedrec}
\end{figure}
\subsection{Optimization Pipeline}
To solve the optimization objective of FedRecSys, we can execute the below steps iteratively between client and server,
\begin{itemize}
    \item \textbf{Client-side model training}: Each client trains the recommendation model using its local data with standard model optimization techniques, such as SGD.
    \item \textbf{Server-side aggregation}: A centralized server aggregates the model updates from all clients, aiming to learn a global recommendation model that benefits the system.
    \item \textbf{Global synchronization}: The aggregated global model is then distributed back to all clients, allowing them to improve their local recommendation models.
\end{itemize}

The overall paradigm can be summarized in Figure~\ref{fig:fedrec}.

\begin{table*}[!t]
\centering
\caption{Summary of \textbf{matrix factorization} architecture-based FedRecSys. \textbf{Task} denotes the user-item interaction is formulated in either ``implicit" feedback (rating=1 for interacted items and rating=0 for un-interacted items) or ``explicit" feedback (the actual rating scores). We abbreviate Movielens as ML, Amazon as AMZ and Douban as DB.}
\begin{tabular}{p{60pt}<{\centering}|p{25pt}<{\centering}|p{120pt}
<{\centering}|p{200pt}
<{\centering}|p{55pt}
<{\centering}}
\toprule
% \hline 
\textbf{Publication} & \textbf{Task} & \textbf{Evaluation Metric} & \textbf{Dataset} &
\textbf{Code} \\
\midrule
FCF~\cite{ammad2019federated} & Implicit & Precision, Recall, F1, MAP, RMSE & Simulated Data, ML, In-house Production & Not Available \\
\midrule
FED-MVMF~\cite{flanagan2021federated} & Implicit & Precision, Recall, F1, MAP, NMR & ML-1M, BookCrossings, In-house Production & Not Available \\
\midrule
P-NSMF~\cite{hu2022federated} & Implicit & Precision, NDCG & ML-1M, Netflix5K5K, XING5K5K, AMZ-KindleStore & \href{https://github.com/PengQ94/P-NSMF}{Code Repository}\\
\midrule
FedRAP~\cite{li2023federated} & Implicit & HR, NDCG & ML-100K, ML-1M,  AMZ-Instant-Video, LastFM-2K, TaFeng Grocery, QB-article & \href{https://github.com/mtics/FedRAP}{Code Repository} \\
\midrule
FedMF~\cite{chai2020secure} & Explicit & Computation Time & ML & \href{https://github.com/Di-Chai/FedMF}{Code Repository} \\
% Available\footnote{\href{https://github.com/Di-Chai/FedMF}{FedMF Code: https://github.com/Di-Chai/FedMF}} \\
\midrule
FedRec++~\cite{liang2021fedrec++} & Explicit & MAE, RMSE & ML-100K, ML-1M, NF5K5K & Not Available \\
\midrule
FedRec~\cite{lin2020fedrec} & Explicit & MAE, RMSE & ML-100K, ML-1M & Not Available \\
\midrule
MetaMF~\cite{lin2020meta} & Explicit & MAE, MSE & DB, Hetrec-movielens, ML-1M, Ciao & Not Available \\
\midrule
Fedmf~\cite{du2021federated} & Explicit & RMSE, CDF &  Filmtrust, ML-100K & Not Available \\
\midrule
FCMF~\cite{yang2021fcmf} & Explicit & MAE, RMSE & ML-100K,  ML-1M, ML-10M, Netflix & Not Available \\
\midrule
F2MF~\cite{liu2022fairness} & Explicit & Recall, F1, NDCG &  ML-1M, AMZ-Movies & \href{https://github.com/CharlieMat/FedFairRec}{Code Repository} \\
% Available\footnote{\href{https://github.com/CharlieMat/FedFairRec}{F2MF Code: https://github.com/CharlieMat/FedFairRec}} \\
% Available\footnote{\href{https://github.com/PengQ94/P-NSMF}{P-NSMF Code: https://github.com/PengQ94/P-NSMF}} \\
\midrule
EIFedMF~\cite{chai2022efficient} & Explicit & RMSE & ML, NYC & Not Available \\
\midrule
LightFR~\cite{zhang2023lightfr} & Explicit & HR, NDCG & ML-1M, Filmtrust, DB-Movie, Ciao & Not Available \\
\midrule
FMFSS~\cite{zheng2023federated} & Explicit & RMSE, MAE & ML-100K, filmTrust, Epinions & Not Available \\
\midrule
FedRecon~\cite{singhal2021federated} & Explicit & RMSE, Accuracy & ML-1M & Not Available \\
% \hline
\bottomrule
\end{tabular}
\label{architecture_mf}
\end{table*}

\section{Taxonomy of FedRecSys Studies}\label{taxonomy}
Benefiting from its inherent privacy-preserving properties, FedRecSys have emerged as a robust paradigm for decentralized personalized services. Based on data distribution characteristics across recommendation scenarios, existing approaches can be categorized into three distinct types: horizontal FedRecSys, vertical FedRecSys, and transfer learning-based FedRecSys~\cite{li2024fedcore, wang2024horizontal}. While all three categories contribute to the advancement of privacy-aware recommendations, horizontal FL currently dominates research efforts due to its alignment with real-world cross-device collaboration scenarios. Our analysis therefore focuses primarily on this predominant paradigm.

The key insight of federated recommendation models is to encapsulate the RecSys within the FL framework so as to provide customized recommendation service while safeguarding user privacy. Based on the technical emphasis of existing FedRecSys studies, we categorize them into two primary research directions, each addressing distinct aspects of decentralized RecSys: (1) \textit{\textbf{RecSys Adaptation}}, which focuses on adapting recommendation model structures and scenario-specific mechanisms to decentralized settings, and (2) \textit{\textbf{FL Enhancement}}, which tackles intrinsic challenges of federated optimization including security, robustness, and efficiency. In the next subsections, we will conduct a comprehensive analysis of these research directions and provide detailed comparisons of technical approaches within each category.

% Based on the model designing emphasis of existing FedRecSys studies, we first divide them into two categories: \textit{\textbf{Enhance RecSys' Privacy Protection Ability}} and \textit{\textbf{Address Federated Optimization's Challenging Issues}}. Furthermore, we would summarize and compare the subcategories under each research branch in detail.

\subsection{RecSys Adaptation}
% The foundational paradigm for constructing federated recommendation systems involves reconfiguring centralized recommendation architectures within privacy-preserving federated learning frameworks. This adaptation enables distributed collaborative optimization while ensuring raw user data remains localized, thereby fundamentally addressing privacy leakage risks inherent in traditional recommendation paradigms. We systematically categorize existing works along two complementary dimensions: \textbf{\textit{Model Architecture Adaptation}}, focusing on redesigning neural network structures and learning protocols for federated deployment, and \textbf{\textit{Scenario-Specific Adaptation}}, tailoring recommendation mechanisms to heterogeneous application contexts such as cross-platform services and temporal interaction modeling. This taxonomy systematically addresses both algorithmic compatibility and practical deployment requirements in decentralized environments.

A simple approach to constructing FedRecSys is to adapt typical centralized recommendation models within the FL framework. This distributed optimization model enables users to store personal data locally, safeguarding privacy. Specifically, we categorize existing research from two perspectives: \textit{\textbf{model architecture}} and \textit{\textbf{recommendation scenario}}.

\subsubsection{From the Model Architecture Aspect}
In existing studies, matrix factorization-based architecture and neural network-based architecture are the two most prevalent embranchments.

\begin{table*}[!t]
\caption{Summary of \textbf{deep neural networ}k architecture-based FedRecSys. \textbf{Architecture} denotes the specific deep neural networks, including MLP (Multilayer Perceptron), CNN (Convolutional Neural Network), GNN (Graph Neural Network) and Transformer. We abbreviate Movielens as ML, Amazon as AMZ and Douban as DB.}
\centering
\begin{tabular}{p{55pt}<{\centering}|p{40pt}<{\centering}|p{25pt}<{\centering}|p{90pt}
<{\centering}|p{180pt}
<{\centering}|p{55pt}
<{\centering}}
% \hline 
\toprule
\textbf{Publication} & \textbf{Architecture} & \textbf{Task} & \textbf{Evaluation Metric} & \textbf{Dataset} &
\textbf{Code} \\
\midrule
PFedRec~\cite{zhang2023dual} & MLP & Implicit & HR, NDCG & ML-100K, ML-1M, Lastfm-2K, AMZ-Video & \href{https://github.com/Zhangcx19/IJCAI-23-PFedRec}{Code Repository} \\
\midrule
FedNCF~\cite{perifanis2022federated} & MLP & Implicit & HR, NDCG & ML-100K, ML-1M,  Lastfm-2K, Foursquare NY & Not Available \\
\midrule
FedFast~\cite{muhammad2020fedfast} & MLP & Implicit& HR, NDCG & ML-1M, ML-100K, TripAdvisor, Yelp & Not Available \\
\midrule
UC-FedRec~\cite{hu2024user} & MLP & Implicit & HR, NDCG & ML, DB & 
\href{https://github.com/HKUST-KnowComp/UC-FedRec}{Code Repository} \\
\midrule
IFedRec~\cite{zhang2023federated} & MLP & Implicit & Recall, Precision, NDCG & CiteULike,  XING & \href{https://github.com/Zhangcx19/IFedRec}{Code Repository} \\
\midrule
HPFL~\cite{wu2021hierarchical} & MLP & Explicit & AUC, ACC, MAE, RMSE, DOA, NDCG & ASSIST, ML & \href{https://github.com/bigdata-ustc/hierarchical-personalized-federated-learning}{Code Repository} \\
\midrule
FedPA~\cite{zhang2024federated} & MLP & Implicit & AUC, Precision & KuaiRand-Pure and small, KuaiSAR-S and R & \href{https://github.com/Zhangcx19/IJCAI-24-FedPA}{Code Repository} \\
\midrule
Dual-CPMF~\cite{duan2019jointrec} & CNN & Explicit & RMSE, Recall, Precision & ML & Not Available \\
\midrule
FedPOIRec~\cite{perifanis2023fedpoirec} & CNN & Implicit & Precision, Recall, MAP, F1 & Foursquare & Not Available \\
\midrule
FedPerGNN~\cite{wu2022federated} & GNN & Explicit & RMSE & ML-100K, ML-1M, ML-10M, Flixster, DB, Yahoo & \href{https://github.com/wuch15/FedPerGNN}{Code Repository} \\
\midrule
FedHGNN~\cite{yan2024federated} & GNN & Explicit & HR, NDCG & ACM, DBLP, Yelp, DB-Book & Not Available \\
\midrule
SemiDFEGL~\cite{qu2023semi} & GNN & Explicit & Recall, NDCG & ML-1M,  Yelp2018, Gowalla & Not Available \\
\midrule
P-GCN~\cite{hu2023privacy} & GNN & Implicit & Recall, NDCG &  Gowalla, Yelp2018, AMZ-Book & Not Available \\
\midrule
F$^2$PGNN~\cite{agrawal2024no} & GNN & Explicit & RMSE & ML-100K, ML-1M, AMZ-Movies & \href{https://github.com/nimeshagrawal/F2PGNN-AAAI24}{Code Repository} \\
\midrule
PPCDR~\cite{tian2024privacy} & GNN & Implicit & Recall, NDCG & Amazon, DB & Not Available \\
\midrule
DCI-PFGL\cite{xie2023dci} & GNN & Explicit & Accuracy & Ciao, Epinions & Not Available \\
\midrule
FedHGNN~\cite{sun2024federated} & GNN & Explicit & MAE, RMSE &  Filmtrust, Ciao, Epinionss & Not Available \\
\midrule
FeSoG~\cite{liu2022federated} & GNN & Explicit & MAE, RMSE & Ciao, Epinions, Filmtrust & \href{https://github.com/YangLiangwei/FeSoG}{Code Repository} \\ 
\midrule
FedGST~\cite{tangfedgst} & GNN & Explicit & NDCG, RMSE & FourSquare &  \href{https://github.com/yushuowiki/FedGST}{Code Repository} \\
\midrule
GPFedRec~\cite{zhang2024gpfedrec} & GNN & Implicit & HR, NDCG & ML-100K, ML-1M, Lastfm-2K, HetRec2011, DB & \href{https://github.com/Zhangcx19/GPFedRec}{Code Repository} \\
\midrule
KG-FedTrans4Rec~\cite{wei2023edge} & Transformer & Implicit & HR, NDCG & ML, Last FM, Book-Crossing & Not Available \\
\midrule
FLT-PR~\cite{belhadi2024federated} & Transformer & Implicit & Recall, NDCG & ML-1M, AMZ-book & Not Available \\
\midrule
RP$^3$FL~\cite{feng2024robust} & Transformer & Implicit & F1-score, Accuracy, AUC & ML-1M, Jester & Not Available \\
\midrule
MRFF~\cite{zhang2024multifaceted} & Transformer & Implicit & AUC, LogLoss & KuaiRand-Pure, KuaiSAR-R and S & \href{ https://github.com/Zhangcx19/AAAI-25-MRFF}{Code Repository} \\
% \hline
\bottomrule
\end{tabular}
\label{architecture_dnn}
\end{table*}

\textbf{Matrix factorization-based architecture.}
Matrix factorization (MF)~\cite{wang2012nonnegative} provides a principled framework for FedRecSys by decomposing user-item interactions into low-dimensional latent embeddings. In this architecture, the recommendation model comprises dual components: \textit{user embeddings} and \textit{item embeddings}. The predicted preference score for user $u$ on item $i$ is computed through their inner product:  
\begin{equation}
    \hat{y}_{ui} = \theta_u^\top \theta_i
\end{equation}
The federated optimization objective formalizes this process as follows:  
\begin{equation}
\min_{\theta} \sum_{u \in U} \alpha_u \left[ \sum_{(i,y_{ui}) \in \mathcal{Y}_u} L( y_{ui}, \hat{y}_{ui}) + \lambda \left( \|\theta_u\|_2^2 + \|\theta_i\|_2^2 \right) \right]
\end{equation}
where $\theta_i$ is aggregated across clients to share common knowledge and $\theta_u$ is retained privately on each device to maintain personalization. $\|\theta_u\|_2^2$ and $\|\theta_i\|_2^2$ represent the $L_2$ regularization, and the hyperparameter $\lambda > 0$ controls the trade-off between recommendation accuracy and model simplicity.

For instance, Muhammad et al.~\cite{ammad2019federated} pioneered the integration of collaborative filtering with FL through their federated matrix factorization framework. In this architecture, clients independently train local matrix factorization models utilizing their user-specific interaction data $\mathcal{Y}_u$. During each federated round, clients exclusively transmit item embedding parameters $\theta_i$ to the central server. The server aggregates these distributed item embeddings across all clients, thereby facilitating global knowledge integration. Subsequently, the updated global item embeddings are distributed back to clients for subsequent local training iterations. This FL cycle iterates until model convergence is achieved.

As the most prevalent architectural paradigm in FedRecSys research, matrix factorization serves as the foundational framework for numerous extensions. Subsequent innovations have extended this paradigm along two key dimensions: (1) \textit{privacy enhancement} through differential privacy mechanisms \cite{chai2020secure,liang2021fedrec++}, and (2) \textit{efficiency optimization} via communication-efficient protocols \cite{singhal2021federated,zhang2023lightfr}. We provide a comprehensive summary of these matrix factorization-based FedRecSys advancements in Table~\ref{architecture_mf}.

\begin{table*}[!t]
\caption{Summary of representative FedRecSys under \textbf{various recommendation scenarios}. We abbreviate Movielens as ML, Amazon as AMZ and Douban as DB.}
\centering
\begin{tabular}{p{75pt}<{\centering}|p{70pt}<{\centering}|p{90pt}
<{\centering}|p{160pt}
<{\centering}|p{55pt}
<{\centering}}
% \hline 
\toprule
\textbf{Publication} & \textbf{Scenario}& \textbf{Evaluation Metric} & \textbf{Dataset} &
\textbf{Code} \\
\midrule
PPCDR~\cite{tian2024privacy} & Cross-domain Rec & Recall, NDCG & AMZ, DB & Not Available \\
\midrule
FedCDR~\cite{meihan2022fedcdr} & Cross-domain Rec & MAE, RMSE & AMZ-review & Not Available \\
\midrule
P2FCDR~\cite{chen2023win} & Cross-domain Rec & HR, NDCG & AMZ & Not Available \\
\midrule
FPPDM~\cite{liu2023federated} & Cross-domain Rec & HR, NDCG & DB, AMZ & Not Available \\
\midrule
FedDCSR~\cite{zhang2024feddcsr} & Cross-domain Rec & HR, NDCG & AMZ & \href{https://github.com/orion-orion/FedDCSR}{Code Repository} \\
\midrule
PFCR~\cite{guo2024prompt} & Cross-domain Rec & Recall, NDCG & AMZ, OnlineRetail & \href{https://github.com/Sapphire-star/PFCR}{Code Repository} \\
\midrule
FedHCDR~\cite{zhang2024fedhcdr} & Cross-domain Rec & MRR, NDCG, HR & AMZ & \href{https://github.com/orion-orion/FedHCDR}{Code Repository} \\
\midrule
F2MF~\cite{liu2022fairness} & Rec Fairness & Recall, F1, NDCG &  ML-1M, AMZ-Movies & \href{https://github.com/CharlieMat/FedFairRec}{Code Repository} \\
\midrule
F$^2$PGNN~\cite{agrawal2024no} & Rec Fairness & RMSE & ML-100K, ML-1M, AMZ-Movies & \href{https://github.com/nimeshagrawal/F2PGNN-AAAI24}{Code Repository} \\
\midrule
RF$^2$~\cite{maeng2022towards} & Rec Fairness & AUC, MDAC & Taobao Ad Display, ML-20M & \href{https://github.com/facebookresearch/RF2}{Code Repository} \\
\midrule
Cali3F~\cite{zhu2022cali3f} & Rec Fairness & HR, NDCG & ML-1M, ML-100K,  Pinterest & Not Available \\
\midrule
CF-FedSR~\cite{luo2022towards} & Rec Fairness & HR, NDCG & AMZ, Wikipedia & Not Available \\
\midrule
FPFR~\cite{wang2024towards} & Rec Fairness & HR, NDCG & Filmtrust, AMZ-Electronic, Steam-200K, ML-100K, ML-1M & Not Available \\
\midrule
FedHGNN~\cite{sun2024federated} & Social Rec & MAE, RMSE &  Filmtrust, Ciao, Epinionss & Not Available \\
\midrule
FeSoG~\cite{liu2022federated} & Social Rec & MAE, RMSE & Ciao, Epinions, Filmtrust & \href{https://github.com/YangLiangwei/FeSoG}{Code Repository} \\ 
\midrule
T-PriDO~\cite{zhou2019privacy} & Social Rec & Average Reward, Average Regret & YFCC100M & Not Available \\
\midrule
DFSR~\cite{luo2022dual} & Social Rec & MAE, RMSE & Flixster, DB,  Filmtrust & Not Available \\
\midrule
FedNewsRec~\cite{qi2020privacy} & News Rec & AUC, MRR, NDCG & Adressa, Adressa & \href{https://github.com/taoqi98/FedNewsRec}{Code Repository} \\
\midrule
Efficient-FedRec~\cite{yi2021efficient} & News Rec & AUC, MRR, NDCG & MIND, Adressa & \href{https://github.com/yjw1029/Efficient-FedRec}{Code Repository} \\
\midrule
UA-FedRec~\cite{yi2023ua} & News Rec & AUC, MRR, NDCG & MIND, Feeds & \href{https://github.com/yjw1029/UA-FedRec}{Code Repository} \\
\midrule
PrivateRec~\cite{liu2023privaterec} & News Rec & AUC, MRR, NDCG & MIND, NewsFeeds & Not Available \\
\midrule
FINDING~\cite{yu2023federated} & News Rec & AUC, MRR, NDCG & Adressa, MIND & \href{https://github.com/yusanshi/FINDING}{Code Repository} \\
\midrule
RD-FedRec~\cite{huang2023randomization} & News Rec & AUC, MRR, NDCG & MIND, Adressa & Not Available \\
\midrule
FedPOIRec~\cite{perifanis2023fedpoirec} & POI Rec & Precision, Recall, MAP & Foursquare & Not Available \\
\midrule
FedGST~\cite{tangfedgst} & POI Rec & NDCG, RMSE & FourSquare &  \href{https://github.com/yushuowiki/FedGST}{Code Repository} \\
\midrule
PriRec~\cite{chen2020practical} & POI Rec & AUC & Foursquare, Koubei & Not Available \\
\midrule
RFPG~\cite{dong2022ranking} & POI Rec & Precision, Recall & Foursquare, Gowalla & Not Available \\
\midrule
PrefFedPOI~\cite{zhang2023fine} &  POI Rec & Accuracy, MRR & Foursquare, Weeplaces & \href{https://github.com/Leavesy/PrefFedPOI}{Code Repository} \\
\midrule
CPF-POI~\cite{ye2023adaptive} & POI Rec & Accuracy, MRR &  GeoLife, Gowalla & \href{https://github.com/Leavesy/CPF-POI}{Code Repository} \\
% \hline
\bottomrule
\end{tabular}
\label{recommendation_scenario}
\end{table*}
\textbf{Deep neural network-based architecture.}
Deep neural architectures enhance FedRecSys by learning hierarchical representations of user-item interactions~\cite{chen2017sampling,hao2023feature}. Compared to matrix factorization, the deep neural network-based architecture introduces additional \textit{neural network weights}, denoted as $W$. The prediction for user $u$ on item $i$ is formulated as:
\begin{equation}
    \hat{y}_{ui} = \sigma\left(W(\theta_u \oplus \theta_i)\right)
\end{equation}
where $\oplus$ denotes concatenation operation and $\sigma$ is the final activation function. The federated optimization objective is formulated as follows:
% \begin{equation}
% \begin{aligned}
% \min_{\theta_i, \mathbf{W}^G} \sum_{u \in U} \alpha_u \left[ \sum_{(i,y_{ui}) \in \mathcal{Y}_u} \mathcal{L}(y_{ui}, \hat{y}_{ui})
% + \lambda \left(||\theta_u||_2^2 + ||\theta_i||_2^2 + ||\mathbf{W}^G||_F^2 \right) \right]
% \end{aligned}
% \end{equation}

\begin{equation}
\begin{aligned}
\min_{\theta} \sum_{u \in U} \alpha_u \Biggl[ \sum_{(i,y_{ui}) \in \mathcal{Y}_u} \mathcal{L}(y_{ui}, \hat{y}_{ui})
+ \lambda \Biggl( & ||\theta_u||_2^2 + ||\theta_i||_2^2 \\
                  & + ||\mathbf{W}||_F^2 \Biggr) \Biggr]
\end{aligned}
\end{equation}

Perifanis et al.~\cite{perifanis2022federated} are the first to develop the federated neural collaborative filtering framework. In this method, they replace the inner product computation of user and item embeddings with nonlinear neural networks, aiming to enhance the representational power of the recommendation model. Perifanis et al.~\cite{perifanis2023fedpoirec} propose a federated recommendation model based on convolutional neural networks. By applying convolution operations on the embeddings of the products that users have interacted with in the short term, this method aims to uncover the sequential patterns in user behavior. Furthermore, Wu et al.~\cite{wu2022federated} present a federated recommendation model based on graph neural networks. They incorporate a third-party server to match the commonly interacted products among users, which allows them to effectively recover the connections between users. Feng et al.~\cite{feng2024robust} present a multimodal federated recommendation framework that fuses multiple modality data to promote recommendation accuracy.
These works, leveraging advanced deep learning techniques like CNNs, GNNs and Transformer, represent further advancements in the field, aiming to capture more sophisticated patterns in user-item interactions while maintaining privacy protection. We systematically compare these deep learning-based federated recommendation models in Table~\ref{architecture_dnn}.

\subsubsection{From the Recommendation Scenario Aspect}
The initial FedRecSys studies mainly focus on the fundamental recommendation scenario, such as the rating prediction~\cite{xie2022contrastive} and Top-K prediction tasks~\cite{chen2022learning}. With the development of the field, there are also works exploring how to extend the models to more complex recommendation scenarios, \eg cross-domain recommendation~\cite{meihan2022fedcdr,chen2023win}, fair recommendation~\cite{luo2022towards,zhu2022cali3f}, social recommendation~\cite{luo2022dual,zhou2019privacy}, news recommendation~\cite{yi2021efficient,qi2020privacy,liu2023privaterec}, POI prediction~\cite{perifanis2023fedpoirec,chen2020practical,zhang2023fine} and so on.

\begin{table*}[!t]
\caption{Summary of representative FedRecSys addressing federated optimization's \textbf{security} challenge. We abbreviate Movielens as ML, Amazon as AMZ and Douban as DB.}
\centering
\begin{tabular}{p{75pt}<{\centering}|p{150pt}
<{\centering}|p{180pt}
<{\centering}|p{55pt}
<{\centering}}
% \hline 
\toprule
\textbf{Publication} & \textbf{Technique} & \textbf{Dataset} &
\textbf{Code} \\
\midrule
FedMF~\cite{chai2020secure} & Homomorphic Encryption & ML & \href{https://github.com/Di-Chai/FedMF}{Code Repository} \\
\midrule
Fedmf~\cite{du2021federated} & Homomorphic Encryption &  Filmtrust, ML-100K & Not Available \\
\midrule
EIFedMF~\cite{chai2022efficient} & Homomorphic Encryption & ML, NYC & Not Available \\
\midrule
FedPOIRec~\cite{perifanis2023fedpoirec} & Homomorphic Encryption & Foursquare & Not Available \\
\midrule
FINDING~\cite{yu2023federated} & Homomorphic Encryption & Adressa, MIND & \href{https://github.com/yusanshi/FINDING}{Code Repository} \\
\midrule
FedGNN~\cite{wu2021fedgnn} & Homomorphic Encryption & Flixster, DB, Yahoo, ML-100K, ML-1M, ML-10M & Not Available \\
\midrule
PFedRec~\cite{zhang2023dual} & Differential Privacy & ML-100K, ML-1M, Lastfm-2K, AMZ-Video & \href{https://github.com/Zhangcx19/IJCAI-23-PFedRec}{Code Repository} \\
\midrule
FedRAP~\cite{li2023federated} & Differential Privacy & ML-100K, ML-1M,  AMZ-Instant-Video, LastFM-2K, TaFeng Grocery, QB-article & \href{https://github.com/mtics/FedRAP}{Code Repository} \\
\midrule
IFedRec~\cite{zhang2023federated} & Differential Privacy & CiteULike, XING & \href{https://github.com/Zhangcx19/IFedRec}{Code Repository} \\
\midrule
GPFedRec~\cite{zhang2024gpfedrec} & Differential Privacy & ML-100K, ML-1M, Lastfm-2K, HetRec2011, DB & \href{https://github.com/Zhangcx19/GPFedRec}{Code Repository} \\
\midrule
FL-MV-DSSM~\cite{huang2020federated} & Differential Privacy & ML-100K & Not Available \\
\midrule
FedPOIRec~\cite{perifanis2023fedpoirec} & Secret Sharing & Foursquare & Not Available \\
\midrule
Efficient-FedRec~\cite{yi2021efficient} & Secret Sharing & MIND, Adressa & \href{https://github.com/yjw1029/Efficient-FedRec}{Code Repository} \\
\midrule
Federated CF~\cite{wang2020federated} & Secret Sharing & ML-1M & Not Available \\
\midrule
FR-FMSS~\cite{lin2021fr} & Secret Sharing & -- & Not Available \\
\midrule
FedRec++~\cite{liang2021fedrec++} & Pseudo Item Generation & ML-100K, ML-1M, NF5K5K & Not Available \\
\midrule
FedRec~\cite{lin2020fedrec} & Pseudo Item Generation & ML-100K, ML-1M & Not Available \\
\midrule
SemiDFEGL~\cite{qu2023semi} & Pseudo Item Generation & ML-1M,  Yelp2018, Gowalla & Not Available \\
\midrule
FedMMF~\cite{yang2022practical} & Personalized Mask Generation & ML-100K,  ML-10M, LastFM & Not Available \\
\midrule
FedPerGNN~\cite{wu2022federated} & Differential Privacy, Pseudo Item Generation & ML-100K, ML-1M, ML-10M, Flixster, DB, Yahoo & \href{https://github.com/wuch15/FedPerGNN}{Code Repository} \\
\midrule
FMFSS~\cite{zheng2023federated} & Secret Sharing, Pseudo Item Generation & ML-100K, filmTrust, Epinions & Not Available \\
\midrule
FeSoG~\cite{liu2022federated} & Differential Privacy, Pseudo Item Generation & Ciao, Epinions, Filmtrust & \href{https://github.com/YangLiangwei/FeSoG}{Code Repository} \\
% \hline
\bottomrule
\end{tabular}
\label{security}
\end{table*}
For FedRecSys employed in various recommendation scenarios, the federated optimization objective can be expressed as the base recommendation loss combined with a specific scenario loss function,
\begin{gather}
    \min_{\theta} \left[ \sum_{u \in U} \alpha_u \underbrace{\mathcal{L}_u(\theta;\mathcal{Y}_u)}_{\text{Base loss}} + \underbrace{\mathcal{L}_{\text{scenario}}}_{\text{Scenario loss}} \right] \\
    s.t. \quad \mathcal{L}_{\text{scenario}} < \delta_{\text{scenario}} \notag
\end{gather}
where $\delta_{\text{scenario}}$ is a predefined threshold, and the scenario loss term must be within $\delta_{\text{scenario}}$. This constraint is crucial in federated settings, where clients may exhibit varying levels of tolerance for the same constraints, thereby requiring a global constraint to maintain consistency across the system.

For the cross-domain recommendation scenario~\cite{tang2012cross}, the scenario loss function can be formulated as follows,
\begin{equation}
  \mathcal{L}_{\text{cross\_domain}} = \|\mathbf{M}\theta_c^{(s)} - \theta_c^{(t)}\|_2^2  
\end{equation}
Here, $\mathbf{M}$ denotes the cross-domain transfer matrix, and $\theta_c^{(s)}$ and $\theta_c^{(t)}$ are the transferable model parameters of the source domain and target domain. For instance, Meihan et al.~\cite{meihan2022fedcdr} point out that FedRecSys cannot make recommendations for new users without any historical interactions. To this end, they propose a cross-domain federated recommender model that introduces beneficial information from the auxiliary domain to achieve new users' recommendations in the target domain. 

For the fair recommendation scenario~\cite{pitoura2022fairness}, the scenario loss function can be formulated as follows,
\begin{equation}
    \mathcal{L}_{\text{fair}} = \sum_{k=1}^K \Omega(\{\hat{y}_{ui}\}_{u \in \mathcal{G}_k})
\end{equation}
where $\mathcal{G}_k$ denotes the protected user groups $(k=1,...,K)$ and $\Omega(\cdot)$ is the fairness metric. For instance, Luo et al.~\cite{luo2022towards} propose a fairness-aware model aggregation algorithm, which adaptively captures client differences with a fairness coefficient during model aggregation so that the system can achieve fair recommendations. 

For the social recommendation scenario~\cite{cao2019social}, the scenario loss function can be formulated as follows,
\begin{equation}
    \mathcal{L}_{\text{social}} = \sum_{v \in \mathcal{S}_u} \|\theta^{(u)} - \theta^{(v)}\|_2^2
\end{equation}
where $\mathcal{S}_u$ denotes the social neighbor set of user $u$, and $\theta^{(u)}$ and $\theta^{(v)}$ are the model parameters of user $u$ and $v$, respectively.
For instance, Luo et al.~\cite{luo2022dual} focus on building FedRecSys enhanced with social network, which can strengthen user modeling by virtue of friend users with similar preferences. 

Moreover, the exploration of FedRecSys in diverse domains, such as news recommendation and POI prediction, showcases the growing applicability and potential impact of these advancements in real-world scenarios. We summarize the FedRecSys designed for various scenarios in Table~\ref{recommendation_scenario}. 

\subsection{FL Enhancement}
Building FedRecSys also entails grappling with the inherent challenges posed by the FL framework itself, encompassing issues like \textbf{\textit{security}}, \textbf{\textit{robustness}}, \textbf{\textit{efficiency}}. Next we will delve into the research goal and representative frameworks that address these specific facets in detail.

\begin{table*}[!t]
\caption{Summary of representative FedRecSys addressing federated optimization's \textbf{robustness} challenge. We abbreviate Movielens as ML, Amazon as AMZ and Douban as DB.}
\centering
\begin{tabular}{p{75pt}<{\centering}|p{25pt}<{\centering}|p{100pt}
<{\centering}|p{170pt}
<{\centering}|p{55pt}
<{\centering}}
% \hline 
\toprule
\textbf{Publication} & \textbf{Type} & \textbf{Target} & \textbf{Dataset} &
\textbf{Code} \\
\midrule
UA-FedRec~\cite{yi2023ua} & Attack & Degrade Model Performance & MIND, Feeds & \href{https://github.com/yjw1029/UA-FedRec}{Code Repository} \\
\midrule
PipAttack~\cite{zhang2022pipattack} & Attack & Promote Targeted Item & ML-1M, AMZ & Not Available \\
\midrule
FedAttack~\cite{wu2022fedattack} & Attack & Degrade Model Performance & ML-1M, Beauty & \href{https://github.com/wuch15/FedAttack}{Code Repository} \\
\midrule
FedRecAttack~\cite{rong2022fedrecattack} & Attack & Promote Targeted Item &  ML-100K, ML-1M, Steam-200K & \href{https://github.com/rdz98/FedRecAttack}{Code Repository} \\
\midrule
IMIA~\cite{yuan2023interaction} & Attack & Infer User-Item Interactions & ML-100K, Steam-200K, Amazon Cell Phone & Not Available \\
\midrule
ClusterAttack~\cite{yu2023untargeted} & Attack & Degrade Model Performance & ML-1M, Gowalla & \href{https://github.com/yflyl613/FedRec}{Code Repository} \\
\midrule
PIECK~\cite{zhang2024preventing} & Attack & Promote Targeted Item & ML-100K, ML-1M, Amazon Digital Music & Not Available \\
\midrule
A-ra \& A-hum~\cite{ijcai2022p306} & Attack & Generate Poisoned User Embedding & ML, AmazonDigitalMusic & \href{https://github.com/rdz98/PoisonFedDLRS}{Code Repository} \\
\midrule
PSMU~\cite{yuan2023manipulating} & Attack & Promote Targeted Item & ML-1M, AMZ Digital Music & Not Available \\
\midrule
PoisonFRS~\cite{yin2024poisoning} & Attack & Promote Targeted Item & Steam-200K, Yelp, ML-10M, ML-20M & Not Available \\
\midrule
HMTA~\cite{su2024revisit} & Attack & Promote Targeted Item & ML, AMZ, IJCAI & Not Available \\
\midrule
HidAttack~\cite{ali2024hidattack} & Attack & Promote Targeted Item & Amazon Appliances, ML-1M, YahooMusic & Not Available \\
\midrule
EIFedMF~\cite{chai2022efficient} & Defense & Defense Inference Attacks & ML, NYC & Not Available \\
\midrule
UC-FedRec~\cite{hu2024user} & Defense & Safeguard Users’ Attributes & ML, DB & 
\href{https://github.com/HKUST-KnowComp/UC-FedRec}{Code Repository} \\
\midrule
UNION~\cite{yu2023untargeted} & Defense & Safeguard Model Performance & ML-1M, Gowalla & \href{https://github.com/yflyl613/FedRec}{Code Repository} \\
\midrule
APM~\cite{zhang2023comprehensive} & Defense & Safeguard Users’ Attributes & ML-100K, ML-1M & Not Available \\
\midrule
CIRDP~\cite{liu2024defending} & Defense & Defense Inference Attacks & ML-1M, Lastfm-360K & Not Available \\
% \hline
\bottomrule
\end{tabular}
\label{robustness}
\end{table*}
\subsubsection{From the Security Aspect}
Although FL's training mechanism doesn't require clients to directly upload private data, inquisitive servers might infer sensitive information by monitoring changes in client model parameters. Thus, security has long been a key concern in FL research \cite{hao2021efficient,mothukuri2021survey}. Many FedRecSys studies focus on model design to enhance the system's privacy protection. Table \ref{security} summarizes representative FedRecSys that tackle the security challenge.

The security-enhanced FedRecSys extends the standard optimization framework with privacy-preserving mechanisms:
\begin{gather}
    \min_{\theta} \left[ \sum_{u \in U} \alpha_u  \underbrace{\mathcal{L}_u(\theta;\mathcal{Y}_u)}_{\text{Base loss}} + \underbrace{\mathcal{L}_{\text{security}}}_{\text{Privacy loss}} \right]\\
    s.t. \quad \mathcal{L}_{\text{security}} < \delta_{\text{security}} \notag
\end{gather}
where the privacy loss function $\mathcal{L}_{\text{security}}$ can be instantiated with a specific security enhancement technique, and it must remain within a predefined threshold $\delta_{\text{security}}$.

For example, the homomorphic encryption technique~\cite{park2022privacy} enables computations to be conducted on encrypted data without the need for decryption, thereby preserving data privacy. The optimization objective is to ensure the reversibility of encryption, which can be expressed as follows,
\begin{equation}
    \mathcal{L}_{\text{HE}} = ||\text{Decrypt}(\text{Encrypt}(\theta)) - \theta||_2^2
\end{equation}
where $\text{Encrypt}(\cdot)$ and $\text{Decrypt}(\cdot)$ denote the encrypt and decrypt operation. Chai et al.~\cite{chai2020secure} claim that uploading model gradient to the server makes it easy to leak users' data. To this end, they propose to integrate a homomorphic encryption technique into the federated matrix factorization framework to further enhance the system's privacy protection capability.

In a similar vein, Wu et al.~\cite{wu2022federated} suggest employing the local differential privacy technique~\cite{geyer2017differentially}. This involves introducing noise to the model parameters before transmission to the server, ensuring that the server receives a perturbed version, thereby alleviating privacy leakage. Generally, the optimization objective of local differential privacy is comprised of two components: a privacy protection term and a noise control term, which together balance the trade-off between ensuring privacy guarantees and minimizing the impact of noise on data utility. The objective can be formalized as follows,
\begin{equation}
    \mathcal{L}_{\text{LDP}} = \lambda_1 \cdot \text{PrivacyCost}(\theta; \epsilon) + \lambda_2 \cdot \text{NoisePenalty}(\theta; \epsilon)
\end{equation}
where $\epsilon$ denotes the privacy budget, which determines the noise intensity, typically drawn from a Laplace distribution.

% \[
% \mathcal{R}_{\text{mask}} = \|\mathbf{M}_u \mathcal{Y}_u - \mathcal{Y}_u\|_2^2 + \lambda \|\mathbf{M}_u - \mathbf{I}\|_F^2
% \]

% \[
% \mathcal{R}_{\text{pseudo}} = \|\nabla_i^{\text{pseudo}} - \nabla_i^{\text{real}}\|_2^2
% \]

Moreover, there are additional studies that develop specialized methods tailored to the recommendation task to enhance the security of the system. Yang et al.~\cite{yang2022practical} have developed a personalized mask mechanism to generate user-specific masks. This innovation allows the conversion of original user ratings into masked ratings, thereby enhancing the security of user rating information. Qu et al.~\cite{qu2023semi} propose to generate pseudo item gradients and send them along with the real item gradient to the server, which can effectively shield the real user interactions from exposure.

\begin{table*}[!t]
\caption{Summary of representative FedRecSys addressing federated optimization's \textbf{efficiency} challenge. We abbreviate Movielens as ML and Douban as DB.}
\centering
\begin{tabular}{p{75pt}<{\centering}|p{150pt}
<{\centering}|p{180pt}
<{\centering}|p{55pt}
<{\centering}}
% \hline 
\toprule
\textbf{Publication} & \textbf{Technique}& \textbf{Dataset} &
\textbf{Code} \\
\midrule
LightFR~\cite{zhang2023lightfr} & Hash Binary Code &  ML-1M, Filmtrust, DB-Movie, Ciao & Not Available \\
\midrule
FedFast~\cite{muhammad2020fedfast} & Cluster-based Client Selection & ML-1M, ML-100K, TripAdvisor,  Yelp & Not Available \\
\midrule
CF-FedSR~\cite{luo2022towards} & Cluster-based Client Selection & AMZ, Wikipedia & Not Available \\
\midrule
EIFedMF~\cite{chai2022efficient} & Reduce Transmission Parameters & ML, NYC & Not Available \\
\midrule
MOEFR~\cite{cui2022communication} & Reduce Transmission Parameters & ML-100K, Epinions & Not Available \\
\midrule
FCIS~\cite{zhang2024efvae} & Reduce Transmission Parameters & Citeulike-a, LastFM, Steam, ML-1M & \href{https://github.com/LukeZane118/EFVAE}{Code Repository} \\
\midrule
FNCF-MAB~\cite{ali2024communication} & Reduce Transmission Parameters & ML-1M, ML-100K, FilmTrust, YahooMusic & \href{https://github.com/waqar-uestc/fncf_mab}{Code Repository} \\
\midrule
FCF-BTS~\cite{khan2021payload} & Reduce Transmission Parameters & ML-1M,  Last-FM, MIND & Not Available \\
\midrule
FedGST~\cite{tangfedgst} & Contribution Oriented Client Selection & FourSquare &  \href{https://github.com/yushuowiki/FedGST}{Code Repository} \\
\midrule
Efficient-FedRec~\cite{yi2021efficient} & Decompose Model into Independent Modules & MIND, Adressa & \href{https://github.com/yjw1029/Efficient-FedRec}{Code Repository} \\
\midrule
FedMMR~\cite{li2024towards} & Decompose Model into Independent Modules & Baby, Sports and Clothing & Not Available \\
\midrule
FedKD~\cite{wu2022communication} & Knowledge Distillation & MIND, ADR & \href{https://github.com/wuch15/FedKD}{Code Repository} \\
\midrule
FedIS~\cite{ding2023efficient} & Fast-Convergent Aggregation & ML-1M, Lastfm-2K, Steam, Foursquare & \href{https://github.com/XuanangD/FedIS}{Code Repository} \\
\midrule
CoLR~\cite{nguyen2024towards} & Low Rank Decomposition & ML-1M, Pinterest & \href{https://github.com/NNHieu/CoLR-FedRec}{Code Repository} \\
\midrule
AeroRec~\cite{xia2024aerorec} & Self-Supervised Knowledge Distillation & ML-1M, ML-20M, Yelp & Not Available \\
\midrule
RFRecF~\cite{liu2024efficient} & Refined Optimization Algorithm & ML-100K, ML-1M, KuaiRec, Jester & \href{https://github.com/Applied-Machine-Learning-Lab/RFRec}{Code Repository} \\
% \hline 
\bottomrule
\end{tabular}
\label{efficiency}
\end{table*}
\subsubsection{From the Robustness Aspect}
% Robustness~\cite{luo2021feature,zhang2022fldetector} is a critical area of research within the realm of FedRecSys~\cite{yuan2023interaction}. Researchers typically explore the system's robustness from two key perspectives. One aspect involves crafting attack methods tailored for FedRecSys to assess performance under external threats like noise. Simultaneously, other researchers concentrate on devising defensive techniques to bolster the system's resilience against such adversarial attacks. The unified optimization objective of robustness-enhanced FedRecSys can be formulated as follows,
Robustness \cite{luo2021feature,zhang2022fldetector} is crucial in FedRecSys. Researchers explore robustness from two angles. Some create FedRecSys-specific attack methods to evaluate performance against external threats like noise. Others focus on defensive techniques to boost resilience. The unified optimization for a more robust FedRecSys is as follows,
\begin{equation}
    \min_{\theta} \left[ \sum_{u \in U} \alpha_u \underbrace{\mathcal{L}_u(\theta; \mathcal{Y}_u)}_{\text{Base loss}} + \underbrace{\mathbb{E}[\mathcal{A}(\theta)]}_{\text{Attack expectation}} + \underbrace{\mathcal{D}(\theta)}_{\text{Defense regularizer}} \right]
\end{equation}
The attack objective is to maximize model deviation from normal by perturbing targets via a disturbance function $f_{\text{attack}}(\cdot)$, while ensuring small perturbations to avoid detection with a stealth function $g_{\text{stealth}}(\cdot)$. We formulate it as follows,
\begin{equation}
    \mathcal{A}(\theta) = \sum_{t \in \mathcal{T}} f_{\text{attack}}(\theta_t) + \beta \cdot g_{\text{stealth}}(\nabla^{(u)})
\end{equation}
where $\mathcal{T}$ is the target set, $\theta_t$ is the target parameters (\eg item embeddings), and $\nabla^{(u)}$ is the model gradient of malicious $u$.

For example, Zhang et al.~\cite{zhang2022pipattack} introduce a backdoor attack technique to manipulate user preferences for specific items within FedRecSys. Their method involves training a classification model capable of tagging item popularity. To execute this attack, they first align the target item embeddings with those of popular items. Subsequently, a subset of malicious users uploads gradient information of the target items to the server during the optimization process. This strategic manipulation increases the visibility of the target item among users, influencing the FedRecSys to promote the target items.

On the other hand, the defense objective is to mitigate the malicious impact, such as by evaluating user trustworthiness with a trust evaluation function $\text{TrustScore}(\cdot)$, while enhancing model robustness by incorporating the stability constraints function $\text{Stability}(\cdot)$. We formulate it as follows,
\begin{equation}
    \mathcal{D}(\theta) = \sum_{u \in U} \text{TrustScore}(u) + \beta \cdot \text{Stability}(\theta)
\end{equation}

For instance, Yu et al.~\cite{yu2023untargeted} present a defense strategy to mitigate attacks on FedRecSys. They employ a contrastive learning task to steer the updating of item embeddings toward a uniform distribution. By assessing the uniformity of item embeddings, the server can efficiently screen out malicious gradients. This tactic can tackle challenges stemming from system attacks that often result in a decrease in recommendation performance. We summarize the representative FedRecSys addressing the robustness challenge in Table~\ref{robustness}.

\subsubsection{From the Efficiency Aspect}
In the framework of FL, the continuous exchange of model parameters between the server and clients poses communication efficiency as a primary bottleneck in federated optimization~\cite{zeng2023flbooster,yan2023criticalfl,zhang2024prototype}. Particularly in recommendation scenarios, the substantial number of clients further exacerbates this challenge. To address this issue, researchers have proposed enhanced federated optimization methods~\cite{khan2021payload} and model segmentation~\cite{yi2021efficient} techniques. These approaches effectively reduce the system's communication overhead by decreasing parameter transmission volume or optimizing model training strategies. 

The optimization objective for the efficiency-enhanced FedRecSys can be formally expressed as a multi-objective optimization problem, given by:
\begin{gather} 
\min_\theta \Biggl[ \sum_{u} \alpha_u \mathcal{L}_u(\theta; \mathcal{Y}_u) + \mathcal{L}_{\text{comm}}(\theta) + \mathcal{L}_{\text{mem}}(\theta) + \mathcal{L}_{\text{comp}}(\theta) \Biggr] \\ \notag
s.t. \quad \mathcal{L}_{\text{comm}}(\theta) + \mathcal{L}_{\text{mem}}(\theta) + \mathcal{L}_{\text{comp}}(\theta) > 0, \\ \notag
\mathcal{L}_{\text{comm}}(\theta) < \delta_{\text{comm}}, \>
\mathcal{L}_{\text{mem}}(\theta) < \delta_{\text{mem}}, \>
\mathcal{L}_{\text{comp}}(\theta) < \delta_{\text{comp}}
\end{gather}
Here, $\mathcal{L}_{\text{comm}}$, $\mathcal{L}_{\text{mem}}$, and $\mathcal{L}_{\text{comp}}$ denote the loss functions for communication, memory, and computation efficiency, respectively, while $\delta_{\text{comm}}$, $\delta_{\text{mem}}$, and $\delta_{\text{comp}}$ are predefined thresholds. Notably, the sum of these three efficiency-related losses is constrained to be greater than zero. This constrains that the model does not overly optimize for one objective at the expense of the others, maintaining a balance in multi-objective optimization, which aligns with the ``no free lunch" theorem~\cite{zhang2022no}.

For instance, Zhang et al.~\cite{zhang2023lightfr} propose utilizing hashing techniques to binarize continuous user/item embeddings into a discrete Hamming space, thereby reducing system computational complexity and communication overhead. In addressing the significant communication costs associated with directly transmitting large-scale models between terminals and servers, Wu et al.~\cite{wu2022communication} have designed a dynamic gradient approximation method based on singular value decomposition. This method decomposes the model into three smaller matrices, effectively compressing communication gradients in federated optimization and subsequently lowering system communication overhead. We summarize the representative FedRecSys addressing the efficiency challenge in Table~\ref{efficiency}.

\section{Personalization in FedRecSys}\label{personalization}
% In previous section, we systematically reviewed existing studies on FedRecSys, focusing on how to adapt RecSys to FL framework and address common challenges in this context. While these discussions are crucial for advancing the field, we believe that more attention should be given to the fundamental objective of RecSys—\textbf{user personalization modeling}. In this section, we begin by presenting a formal definition of personalization in FedRecSys from the view of learning personalized models. To provide a deeper understanding of this definition, we discuss the core components that shape personalization in the context of federated systems. We first examine the general concept of personalization in RecSys, followed by a review of common approaches to personalization modeling within FedRecSys. This leads to a discussion of personalized FL techniques, highlighting the unique opportunities they present for advancing personalized recommendations in federated settings. Ultimately, we propose that the future of FedRecSys lies in the development of adaptable and privacy-preserving personalized models that are inherently aligned with FL paradigm, improving both recommendation quality and user privacy.

In the previous section, we systematically reviewed existing FedRecSys research, mainly on adapting RecSys to the FL framework and solving common challenges. However, we think more focus should be on RecSys' fundamental goal—\textbf{user personalization modeling}. In this section, we first formally define personalization in FedRecSys from the perspective of learning personalized models. To better understand this definition, we discuss the key elements of personalization in federated systems. We start with the general concept of personalization in RecSys, then review common personalization modeling methods in FedRecSys. This leads to a discussion of personalized FL techniques and the unique advantages they offer for personalized recommendations in federated settings. Finally, we suggest that the future of FedRecSys lies in developing adaptable, privacy-preserving personalized models that fit the FL paradigm, thus enhancing both recommendation quality and user privacy.

\subsection{Definition of Personalization in FedRecSys}
\textsc{\textbf{Definition 2.}}
Personalization in FedRecSys refers to the capability of learning user-specific model components while collaboratively training a global recommendation model under federated constraints. Specifically, each client $u \in \mathcal{U}$ maintains a personalized model $\mathcal{F}_u=\{\theta, \phi_u\}$, where $\theta$ is the global parameters shared across all clients and $\phi_u$ is the personalized parameters unique to client $u$. This dual-parameter architecture enables: (1) \textit{Knowledge Sharing}: Global parameters $\theta$ capture cross-user patterns through federated aggregation, and (2) \textit{Local Adaptation}: Personalized parameters $\phi_u$ encodes client-specific preferences inferred from private interaction data $\mathcal{Y}_u$.

The unified optimization objective of personalized FedRecSys is formulated as a bi-level optimization problem,
\begin{gather}\label{ob_of_pfedrecsys}
\quad \min_{\theta} \sum_{u \in U} \alpha_u \mathcal{L}_u(\theta, \phi_u^*; \mathcal{Y}_u) \\ \notag
where \quad \phi_u^* = \arg\min_{\phi_u} \mathcal{L}_u(\theta, \phi_u; \mathcal{Y}_u)) 
\end{gather}
% This framework implements privacy-preserving personalized recommendations in a systematic manner under federated constraints, utilizing a dual-layer optimization and parameter isolation mechanism. It preserves the collaborative benefits of FL while effectively harnessing the user adaptation capabilities of personalized models.
This framework achieves privacy-preserving personalized recommendations within federated constraints. It uses a dual-layer optimization and parameter isolation mechanism, maintaining FL's collaborative advantages and effectively harnessing the user adaptation capabilities of personalized models.

% Personalization in Federated Recommender Systems (FedRecSys) is the process of tailoring recommendation models to individual users by dynamically incorporating their unique preferences, historical interactions, and contextual information. This adaptation occurs within the constraints of FL, where user data remains decentralized and private, ensuring that personalization is achieved without compromising data security or privacy.

\subsection{Personalization in RecSys}
Personalization lies at the heart of modern RecSys, enabling the transformation of generic content delivery into tailored experiences that align with individual user preferences~\cite{ko2022survey}. By dynamically adapting to users’ unique behavioral patterns and contextual needs, personalized systems enhance relevance and foster long-term satisfaction, which are essential for success in data-driven environments. Effective personalization hinges on two foundational tasks: (1) \textit{Granular User Representation}, which learns low-dimensional embeddings that encode stable preferences and transient interests, and (2) \textit{Multi-Relational Interaction Modeling}, which decodes complex user-item, item-item, and user-context relationships.

% Personalization is central to RecSys, enabling the delivery of tailored recommendations that align with individual preferences~\cite{ko2022survey}. Rather than providing generic suggestions, personalized systems capture the intricate and dynamic nature of user interests, effectively filtering content to prioritize the most relevant items. This ability to offer context-aware and adaptive recommendations underscores the significance of personalized RecSys in today’s data-driven landscape.

% Personalization in RecSys focuses on adapting recommendations to the unique preferences and behaviors of users. This is achieved through the analysis of various data signals, such as historical interactions and contextual information, which help reveal the underlying patterns of user interests. User preferences may be explicitly stated, such as through ratings or purchase history, or implicitly inferred from behavioral patterns. The key challenge lies not in the volume of data, but in effectively integrating and modeling the complex relationships between users, items, and contextual factors to deliver accurate, contextually relevant recommendations. Effective personalization necessitates advanced algorithms capable of capturing these dynamic interactions and providing recommendations that resonate with users’ evolving needs.

To implement personalized experience, RecSys achieve this through a variety of technical paradigms. These include content-based models~\cite{javed2021review,perez2021content}, which rely on item features to match users with similar content, collaborative filtering models~\cite{martins2020deep,papadakis2022collaborative}, which identify patterns in user-item interactions, and hybrid models~\cite{thorat2015survey,walek2020hybrid} that combine multiple approaches for more robust personalization. Furthermore, deep learning-based models~\cite{da2020recommendation,wu2020sse} and graph-based models~\cite{he2020lightgcn,shuai2022review} are increasingly adopted for their ability to capture complex, non-linear relationships between users and items. Each method represents user preferences differently, ensuring personalized recommendations are relevant in meeting individual needs, thereby enhancing user satisfaction and engagement. This transformative approach has become ubiquitous across a wide range of application domains, including e-commerce~\cite{hussien2021recommendation,ji2021you}, content platforms~\cite{deldjoo2020recommender,wang2020fine}, and social networks~\cite{fan2019graph,yu2021self}.

A unifying thread across these methods is their use of user embeddings to parameterize individual preferences. However, in centralized frameworks, storing all embeddings on servers creates a tension between effective personalization and privacy risks. This shows the need for better paradigms that balance personalized modeling with decentralized requirements, which we'll discuss in the upcoming FedRecSys sections.

\subsection{Common Personalization Modeling Strategy in FedRecSys}

Traditional FL frameworks mandate clients to transmit entire local model parameters for global aggregation \cite{mcmahan2017communication}. In FedRecSys, this brings high privacy risks as user-item interaction patterns are in model parameters, especially via user ID embeddings. To address this, FedRecSys often uses a parameter decoupling strategy. They keep user embeddings private on the clients and selectively share item embeddings and neural network weights for global aggregation \cite{ammad2019federated,chai2020secure,wu2022federated}, which is similar to centralized recommendation architectures \cite{he2017neural,cailightgcl} in maintaining personalized user representations. As a result, the federated framework accomplishes two objectives: (1) safeguarding user privacy through the localized management of personalized features and (2) facilitating global knowledge distillation by aggregating common parameters. This balance validates FedRecSys as a practical privacy-preserving collaborative learning framework for recommendation scenarios.

\subsection{Personalized FL}
% Personalized federated learning (PFL) emerges as a pivotal evolution of conventional FL, addressing the fundamental challenge of statistical heterogeneity in decentralized environments where clients exhibit divergent data distributions \cite{zhang2021parameterized,hu2020personalized,shamsian2021personalized,pillutla2022federated}. While standard FL aggregates local model updates to construct a universal global model under the implicit assumption of client data homogeneity, real-world applications—such as RecSys and healthcare—often encounter significant cross-client distribution shifts. These shifts render a unified global model suboptimal, prone to either underfitting client-specific patterns or overfitting to dominant data characteristics while marginalizing minority clients.

% The essence of PFL lies in its capacity to facilitate client-specific model adaptation while preserving federated privacy guarantees. Diverging from the ``one-model-fits-all" paradigm, PFL enables each client to derive a tailored model optimized for its unique data distribution through a distributed optimization framework. This is achieved by orchestrating collaborative learning across clients without centralizing raw data, thereby harmonizing personalized performance with privacy preservation. Current methodologies predominantly adopt two strategic paradigms: \textit{\textbf{global model personalization}} and \textit{\textbf{personalized model learning}}.

Personalized federated learning (PFL) represents a crucial advancement over conventional FL, specifically designed to tackle the core issue of statistical heterogeneity in decentralized settings~\cite{zhang2021parameterized,hu2020personalized,shamsian2021personalized,pillutla2022federated}. Standard FL, which aggregates local model updates to build a universal global model under the implicit assumption of client data homogeneity, fails to account for cross-client distribution shifts. PFL, in contrast, enables client-specific model adaptation while maintaining federated privacy. Departing from the ``one-model-fits-all" approach, it empowers each client to create a model optimized for its own data characteristics, which effectively balances performance and privacy protection. Current PFL methodologies mainly follow two strategic paradigms: \textit{\textbf{global model personalization}} and \textit{\textbf{personalized model learning}}.

% Personalized FL is a key advancement in FL, designed to address challenges arising from heterogeneous data distributions across clients in decentralized environments~\cite{zhang2021parameterized,hu2020personalized,shamsian2021personalized,pillutla2022federated}. In traditional FL, a global model is created by aggregating updates from local models trained on clients' data, assuming data distributions are similar across clients. However, in many real-world scenarios, such as RecSys or healthcare, clients' data distributions can vary significantly. As a result, a global model may either underfit the unique data distributions of clients or overfit to the majority client data, neglecting outliers.

% The core of personalized FL lies in enabling each client to learn a model tailored to its own data distribution. Instead of enforcing a shared global model, personalized FL allows each client to optimize a unique set of parameters best suited to its local data. This is made possible by FL’s distributed optimization framework, which supports the simultaneous training of multiple models across clients while maintaining data privacy. By leveraging this decentralized structure—where data remains on the client device—personalized FL enables personalized model learning without compromising privacy, allowing each client to independently optimize a model that reflects its distinct data characteristics. Two common strategies for designing a personalized FL paradigm are \textit{\textbf{global model personalization}} and \textit{\textbf{learning personalized models}}. 

\textbf{Global model personalization.} 
This approach first trains a global model via standard FL protocols, then fine-tunes it locally for client-specific adaptation~\cite{mansour2020three}. Furthermore, there are two categories of methods, which are designed from the data and model perspectives. The data-based methods~\cite{wu2020fedhome,wang2020optimizing,li2021fedsae} usually focus on reducing the data statistic heterogeneity among clients. Model-based methods~\cite{li2021model,fallah2020personalized,yang2020fedsteg} aim to learn a capable global model for better adaption with clients.

\textbf{Learning personalized models.} 
This paradigm re-engineers the FL architecture to inherently support client-specific models~\cite{arivazhagan2019federated}. Specifically, the methods can be further classified into two branches, including architecture-based methods and similarity-based methods. In general, the architecture-based methods either decouple the models with partial layers of personalization or deploy customized models on each client~\cite{liang2020think,zhu2021data}. The similarity-based methods~\cite{smith2017federated,sattler2020clustered} discover the relationships among clients and utilize similar clients to promote personalization modeling.

\begin{figure*}[!t]
\setlength{\abovecaptionskip}{1mm}
\setlength{\belowcaptionskip}{3mm}
    \centering
    \includegraphics[width=1.\linewidth]{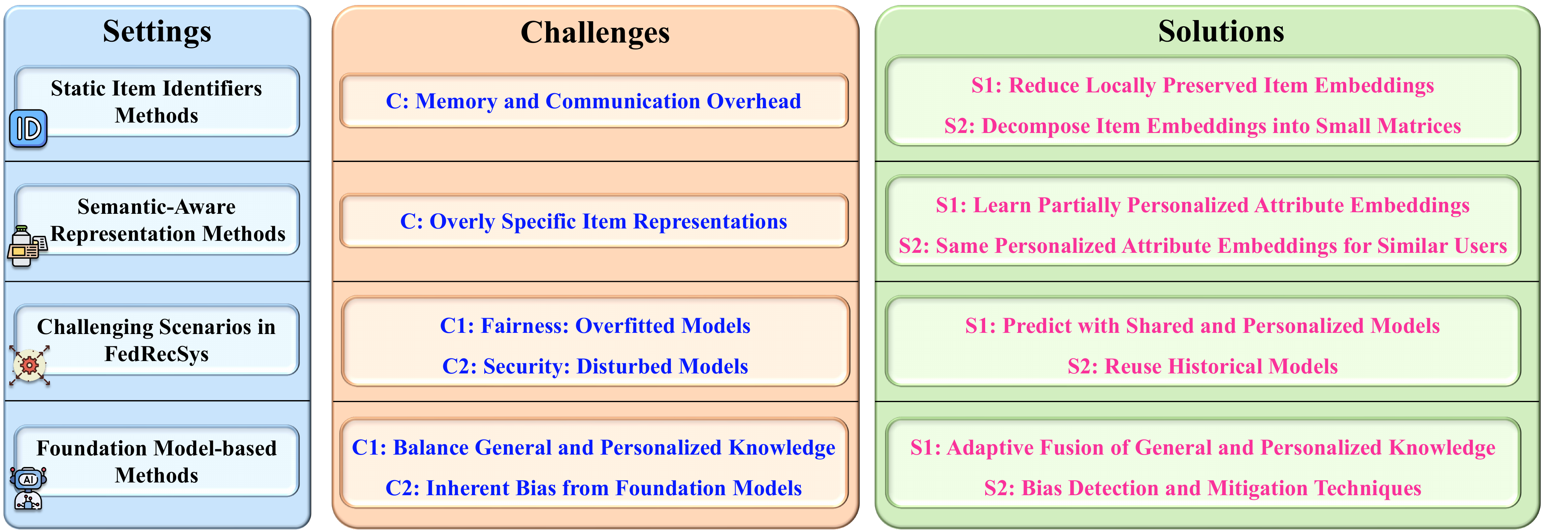}
    \caption{Challenges (\textcolor{blue}{\textbf{C}}) and solutions (\textcolor{magenta}{\textbf{S}}) summary for developing personalized models-driven FedRecSys.}
    \label{fig:challenge_solution}
\end{figure*}
\subsection{New Perspective for Personalized FedRecSys}
The prevailing user embedding-centric paradigm in FedRecSys exhibits a critical methodological gap: it offers an insufficient framework for modeling personalized user-item interactions. Although localized user embeddings capture some individual preferences, they operate under the limiting assumption that item semantics and interaction dynamics can be modeled uniformly across all clients. This approach fundamentally disregards two empirically validated phenomena: (1) users interpret identical items through personalized cognitive lenses, and (2) cross-client heterogeneity manifests not only in user preferences but also in how interactions reveal those preferences. These limitations necessitate a paradigm shift toward \textbf{personalized models}, where both representational spaces (users/items) and interaction mechanisms (scoring functions, attention layers) adapt to localized contexts.

Recent advances substantiate this perspective. The PFedRec framework \cite{zhang2023dual} pioneers personalized model components by enabling clients to reinterpret items through privatized representations and adapt scoring functions to localized rating patterns. This dual personalization resolves semantic mismatches between global assumptions and user cognition. Subsequent innovations extend this principle: Dual-view architectures \cite{li2023federated} synergize global and personalized item embeddings to preserve common knowledge while capturing perception biases, while graph-enhanced methods \cite{zhang2024gpfedrec} inject social contextualization into personalized representations and refine user-specific scoring functions through federated relational learning. Collectively, these advancements establish personalized model adaptation as a critical pathway for FedRecSys, achieving effective privacy preservation while fundamentally redefining the capacity to model heterogeneous user-item interactions at scale.

For better understanding, Figure~\ref{fig:personalization} contrasts the personalization techniques in centralized and federated RecSys. In FedRecSys, the process of learning personalized models is in line with the federated optimization framework. This framework enables the simultaneous learning of distinct model parameters for each client. From the perspective of recommendation tasks, personalized models allow for a more detailed portrayal of how individual users perceive and interact with items through adaptive parameterization. This can potentially result in more accurate user preference modeling. Moreover, this approach helps deal with the data heterogeneity typical in federated settings. Each client can develop model components customized to its local user population and behavioral patterns. By enabling personalization across multiple model components (such as representations and interaction functions), learning personalized models increases the flexibility of FedRecSys. This makes them more capable of adapting to diverse user preferences across distributed data sources.

\section{Challenges and Solutions for Personalized FedRecSys}\label{challenges_and_solutions}
This section systematically analyzes the challenges and potential solutions in deploying personalized models for FedRecSys, through a structured examination of four critical dimensions. First, we analyze the fundamental components of personalized architectures, distinguishing between \textbf{static item identifiers} (\eg item ID embeddings) and \textbf{semantic-aware representations} (\eg attribute-based embeddings), which collectively establish the basis for client-specific adaptation. Subsequently, we investigate how \textbf{challenging FedRecSys scenarios} (\eg fairness and security) are exacerbated by model heterogeneity when transitioning from conventional architectures to personalized models. We then address the frontier challenge of \textbf{foundation model-based FedRecSys}, where the fusion of large pre-trained models and personalized architectures creates tension between preserving universal knowledge and accommodating localized adaptations. 

We synthesize these perspectives through the framework in Figure~\ref{fig:challenge_solution}, mapping core challenges to potential solutions. This structured analysis shows that personalized models, when combined with multi-granular adaptation mechanisms, can effectively address these challenges, improving recommendation performance while maintaining the privacy-preserving characteristics inherent in FL architectures.

% This section discusses the challenges and solutions involved in personalized FedRecSys. We explore this topic from four key perspectives. First, we examine two primary forms of item embeddings used in RecSys: \textbf{item ID embeddings} and \textbf{item attribute embeddings}, which represent the foundational approaches to item representation in federated recommendation models. Next, we delve into \textbf{challenges faced by FedRecSys}, such as fairness, security and robustness, and how these are amplified when personalized models are introduced. The final perspective focuses on \textbf{foundation model-based FedRecSys}, where the integration of large pre-trained models adds further complexity when combined with personalized models. These perspectives together provide a comprehensive view of the current landscape, ranging from fundamental challenges to the most advanced methodologies. This structured exploration offers a thorough understanding of how personalized models can be effectively integrated into FedRecSys, outlining both practical and cutting-edge solutions. A summary of the specific challenges and solutions for each perspective is provided in Figure~\ref{fig:challenge_solution}, offering an overarching view. By addressing these challenges across the identified settings, personalized models strategies hold the potential to significantly enhance the performance and capabilities of FedRecSys.

\subsection{Static Item Identifiers Methods}
\begin{figure}[!t]
\setlength{\abovecaptionskip}{1mm}
\setlength{\belowcaptionskip}{-3mm}
    \centering
    \includegraphics[width=0.6\linewidth]{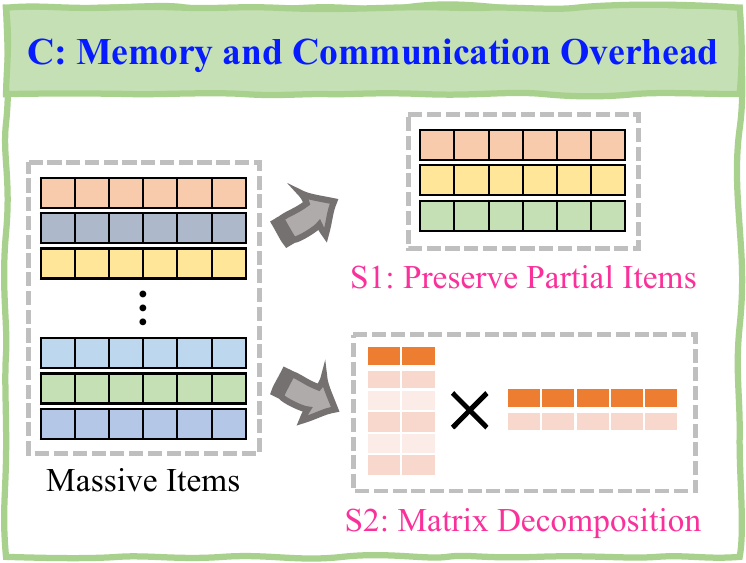}
    \caption{Solution schematic diagram to \textbf{memory and computation overhead challenge} for static item identifiers methods.}
    \label{fig:c1_solution}
\end{figure}

% In recommendation models, learning representations for both users and items is crucial for making personalized item recommendations that align with individual user preferences. One effective approach to item representation learning involves leveraging item ID features, which assigns a unique embedding vector to each item ID, enabling the system to accurately distinguish between diverse items in the catalog. This ID-based approach excels at capturing the distinct identity of each item, allowing the system to uncover subtle relationships between items and facilitate more precise recommendations. By doing so, item ID embedding learning has become a fundamental component of many modern RecSys architectures~\cite{wu2021self,xia2022hypergraph,ren2023disentangled}. We summarize the challenges and solutions associated with static item identifiers methods in Figure~\ref{fig:c1_solution}, and the following sections provide a detailed discussion.
In recommendation models, learning user and item representations is vital for personalized recommendations. An effective way to learn item representations is by using item ID features. Each item ID is assigned a unique embedding vector, enabling the system to clearly differentiate among various items. This ID-based method is excellent at capturing an item's distinct identity, which has been a key part of many modern RecSys architectures \cite{wu2021self,xia2022hypergraph,ren2023disentangled}. Figure \ref{fig:c1_solution} summarizes the challenges and solutions of static item identifier methods, with detailed discussion in the following sections.

\textbf{Challenges.} RecSys often deal with an enormous number of items \cite{roy2022systematic}. For example, an e-commerce platform like Amazon may have millions of items across various categories, from electronics and clothing to books and home goods. Similarly, a streaming service like Netflix could have tens of thousands of movies, and other video content available for users to enjoy. 
% Even a music streaming platform like Spotify can have millions of songs, albums, and artist profiles in its catalog. 
In FedRecSys, these vast item catalogs present \textit{\textbf{challenges in memory and communication}} (\textcolor{blue}{\textbf{Challenge}}). Client devices, with their limited memory and processing power, cannot store the entire item embedding matrix locally. Moreover, the federated optimization process, which repeatedly transfers the full set of model parameters between the server and clients, generates substantial communication overhead, hampering the efficiency of the FL workflow \cite{xia2021survey}.

\textit{Challenge Formulation.} We formulate the challenge as an optimization problem, which will guide the design of potential solutions. Specifically, we refine the optimization objective in Equation~\ref{ob_of_pfedrecsys} by introducing the memory loss $\mathcal{L}_{\text{mem}}$ and communication loss $\mathcal{L}_{\text{comm}}$ for personalized item embeddings $\phi_u^I$, finally formulated as a multi-objective optimization problem,
\begin{gather}\label{ob_c1}
    \min_{\theta} \sum_{u \in U} \alpha_u  \mathcal{L}_u(\theta, \phi_u^*; \mathcal{Y}_u) \\ \notag
% \begin{aligned}
    where \> \phi_u^* = \arg\min_{\phi_u} \Biggl[ \mathcal{L}_u(\theta, \phi_u; \mathcal{Y}_u) + \mathcal{L}_{\text{mem}}(\phi_u^I)
    + \mathcal{L}_{\text{comm}}(\phi_u^I) \Biggr] \\ \notag
% \end{aligned} \\ \notag
s.t. \quad \mathcal{L}_{\text{mem}}(\phi_u^I) + \mathcal{L}_{\text{comm}}(\phi_u^I) > 0, \\ \notag
\mathcal{L}_{\text{mem}}(\phi_u^I) < \delta_{\text{mem}}, \>
\mathcal{L}_{\text{comm}}(\phi_u^I) < \delta_{\text{comm}} \> (\forall u \in U)
\end{gather}
Notably, both losses must adhere to the multi-objective balance constraint and remain within the predefined thresholds.

\textbf{Solutions.} 
To tackle memory and communication issues in FedRecSys due to large item catalogs, a viable strategy is \textit{\textbf{reducing the size of item embeddings stored on clients}} (\textcolor{magenta}{\textbf{Solution 1}}). Instead of keeping the entire item embedding set, clients can store only embeddings of items they've interacted with. This substantially cuts local memory needs, as the client-side item embedding set is much smaller than the full catalog.

Moreover, \textit{\textbf{decomposing the item embedding matrix into smaller sub-matrices}} (\textcolor{magenta}{\textbf{Solution 2}}) on client devices is another effective approach \cite{hulora,nguyen2024towards}. This not only conserves local memory but also allows for the transfer of these decomposed sub-matrices during federated optimization, thus reducing communication overhead. By using partial item retention and matrix decomposition, FedRecSys can efficiently handle extensive item inventories. It overcomes memory and communication bandwidth limitations on individual devices, enabling the system to scale and provide personalized recommendations despite the resource constraints of the distributed architecture.

% To address memory and communication challenges in FedRecSys with massive item catalogs, a promising approach is to \textit{\textbf{reduce the scale of item embeddings stored on client}} (\textcolor{magenta}{\textbf{Solution 1}}). Instead of maintaining the complete item embedding, each client can store only the embeddings of the items they have interacted with. This significantly reduces the local memory requirements, as the item embedding set on each client is much smaller than the full item catalog. 

% Furthermore, \textit{\textbf{the item embedding matrix can be decomposed into smaller submatrices}} (\textcolor{magenta}{\textbf{Solution 2}}) on the client devices~\cite{hulora,nguyen2024towards}. This conserves local memory resources and allows the transmission of decomposed submatrices during federated optimization, which effectively lowers the communication overhead. By adopting these techniques of partial item retention and matrix decomposition, FedRecSys can efficiently manage vast item inventories while overcoming the limitations of memory capacity and communication bandwidth on individual devices. This enables the system to scale and deliver personalized recommendations without being constrained by the resource constraints of the distributed architecture.

\subsection{Semantic-Aware Representations Methods}
\begin{figure}[!t]
\setlength{\abovecaptionskip}{-1mm}
\setlength{\belowcaptionskip}{3mm}
    \centering
    \includegraphics[width=0.6\linewidth]{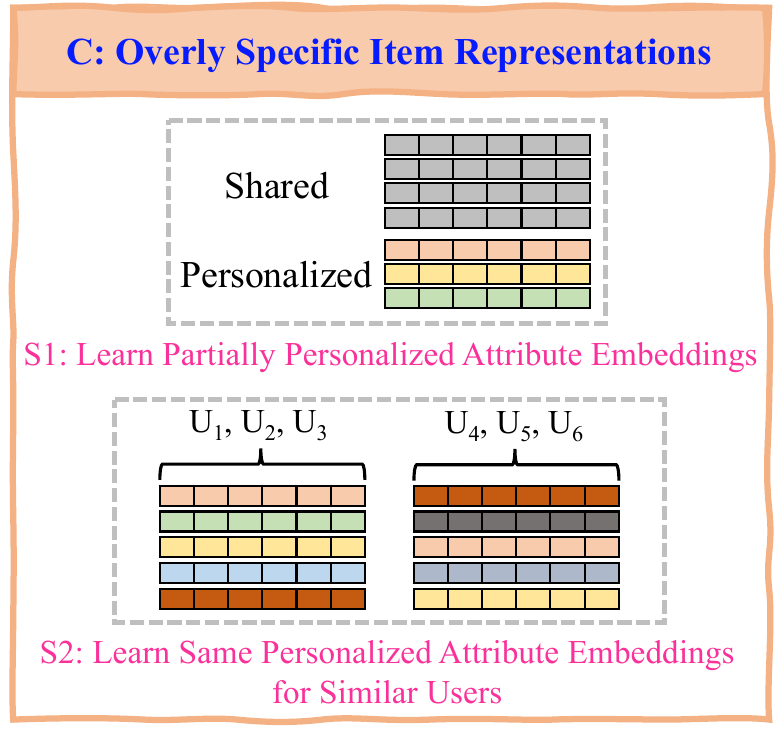}
    \caption{Solution schematic diagram to \textbf{overly specific item representations challenge} for item attribute embedding-based methods.}
    \label{fig:c2_solution}
\end{figure}

Item attributes are crucial in RecSys. Unlike relying only on item IDs, attribute information offers detailed item descriptions. This helps the system better grasp item traits and relationships, leading to more accurate recommendations. When dealing with cold-start users or items, using item attributes can overcome the lack of interaction data in cold-start scenarios \cite{lu2020meta,wei2021contrastive}. Also, item attributes can explain recommendation results. Showing users that recommended items match their preference traits boosts user understanding and trust in the recommendations \cite{afchar2022explainability,lyu2022knowledge}. These advantages highlight the importance of using item attributes in FedRecSys modeling \cite{lei2020interactive,chen2023bias}. Figure \ref{fig:c2_solution} summarizes the challenges and solutions of semantic-aware representation methods, with detailed discussion in the following sections.

% Item attributes play a pivotal role in RecSys construction. In contrast to relying solely on item identifiers, attribute information encompasses more detailed item descriptions, enabling system to better comprehend item characteristics and the associations between items. This enhanced understanding facilitates the system in making more accurate recommendations. Furthermore, when the system faces the challenge of cold-start users or items, leveraging item attribute information can address the insufficient interaction data inherent in cold-start recommendations, thereby overcoming this critical problem~\cite{lu2020meta,wei2021contrastive}. Additionally, item attributes can aid in explaining recommendation results by informing users that the recommended item matches certain characteristics of their preferences, enhancing user understanding and trust in the recommendations~\cite{afchar2022explainability,lyu2022knowledge}. These benefits underscore the significance of utilizing item attributes as an important direction in FedRecSys modeling~\cite{lei2020interactive,chen2023bias}. We summarize the challenges and solutions associated with semantic-aware representation methods in Figure~\ref{fig:c2_solution}, and the following sections provide a detailed discussion.

\textbf{Challenges.} 
Combining item attributes with multiple item embedding vectors allows for a detailed breakdown of item characteristics, offering a comprehensive item description \cite{sun2020multi,liu2024multimodal}. In short video recommendations, for instance, each video has rich attribute information. This includes discrete features like video type and category, along with numerical features such as view and download counts. These diverse attributes provide a multifaceted view of video details and can aid in knowledge transfer among users. However, \textit{\textbf{learning personalized attribute embeddings for each user might lead to overly specific item representations}} (\textcolor{blue}{\textbf{Challenge}}). This could impede the system's ability to collaboratively model user preferences, thus harming recommendation performance

% Item attributes information, when combined with multiple item embedding vectors, enables a fine-grained decomposition of item characteristics, providing a comprehensive description of the item~\cite{sun2020multi,liu2024multimodal}. Taking the short-video recommendation scenario as an example, each video possesses a wealth of attribute information, such as discrete features like video type, soundtrack type, and video category, as well as numerical features like view count, comment count, and download count. These diverse attributes offer a multifaceted introduction to the video details. The abundant attribute information can facilitate knowledge transfer across users. However, \textit{\textbf{learning personalized attribute embeddings for each user may result in overly specific item representations}} (\textcolor{blue}{\textbf{Challenge}}), which would hinder the system's ability to collaboratively model user preferences and negatively impact recommendation performance.

\textit{Challenge Formulation.} We formulate the challenge as an optimization problem, which will guide the design of potential solutions. Following the formulation in Equation~\ref{ob_c1}, we define the optimization objective by integrating a generality loss about personalized item embeddings $\phi_u^I$, given as,
\begin{gather}\label{ob_c2}
\min_{\theta} \sum_{u \in U} \alpha_u  \mathcal{L}_u(\theta, \phi_u^*; \mathcal{Y}_u) \\ \notag
where \quad \phi_u^* = \arg\min_{\phi_u} \left[ \mathcal{L}_u(\theta, \phi_u; \mathcal{Y}_u) + \mathcal{L}_{\text{gene}}(\phi_u^I) \right] \\ \notag
s.t. \quad \mathcal{L}_{\text{gene}}(\phi_u^I) < \delta_{\text{gene}} \> (\forall u \in U)
\end{gather}
where $\delta_{\text{gene}}$ is the predefined threshold, and the optimization objective ensures that FedRecSys improves the generalization of attribute embeddings while minimizing recommendation loss, thus avoiding overly specific item representations.

\begin{figure}[!t]
\setlength{\abovecaptionskip}{-1mm}
\setlength{\belowcaptionskip}{3mm}
    \centering
    \includegraphics[width=1\linewidth]{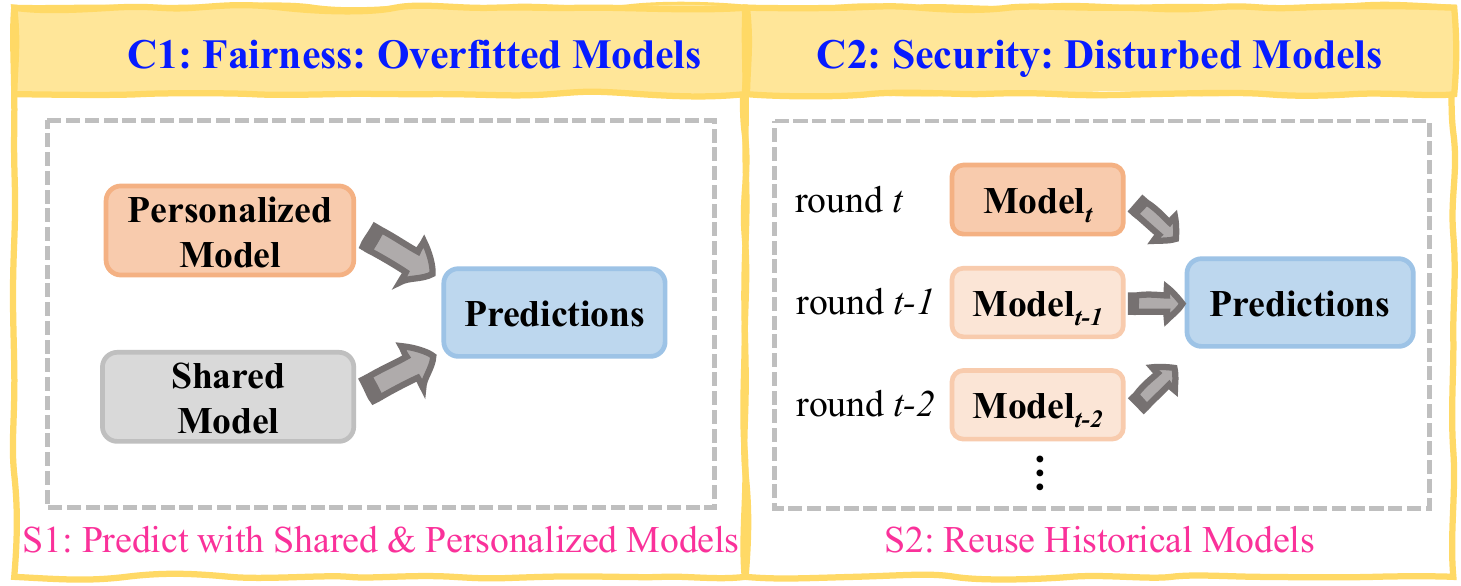}
    \caption{Solution schematic diagram to \textbf{fairness: overfitted models} and \textbf{security: disturbed models} for challenging scenarios in FedRecSys.}
    \label{fig:c3_solution}
\end{figure}
\textbf{Solutions.} 
Among item attributes, those significantly affecting user preferences often vary by user. Drawing from PFL concepts of learning partially personalized parameters \cite{liang2020think,singhal2021federated}, partitioning personalized attributes during federated optimization is key. To prevent issues from learning fully personalized attribute embeddings for each user, \textit{\textbf{users can learn only a subset of personalized attribute embeddings}} (\textcolor{magenta}{\textbf{Solution 1}}). This way, personalized embeddings capture user-specific preferences, while shared embeddings utilize general attribute information for collaborative preference learning.

Moreover, users can be grouped by similarity, enabling \textit{\textbf{users in the same group to learn identical personalized attribute embeddings}} (\textcolor{magenta}{\textbf{Solution 2}}), which strengthens the role of similar users in mining user interests \cite{he2024co}. This approach, selectively learning personalized and shared embeddings, balances capturing user-specific preferences with using general attribute info, thereby enhancing recommendation performance.

% Among item attributes, those that can have a critical influence on user preferences are often user-specific. Therefore, we can borrow the idea in PFL works that learn partially personalized parameters and take others as common parameters~\cite{liang2020think,singhal2021federated}. Similarly, it is crucial to partition the personalized attributes during the federated optimization process. To avoid the potential detrimental impact of learning full personalized attribute embeddings for each user, \textit{\textbf{each user can learn only a subset of personalized attribute embeddings}} (\textcolor{magenta}{\textbf{Solution 1}}). On one hand, learning personalized item attribute embeddings can enhance the modeling of user-specific preferences. On the other hand, learning shared item attribute embeddings across users can effectively leverage the general information embedded in the attributes to facilitate collaborative learning of user preferences. 

% Furthermore, users can be grouped based on similarity, allowing \textit{\textbf{users within the same group to learn the same personalized attribute embeddings}} (\textcolor{magenta}{\textbf{Solution 2}}), further strengthening the role of similar users in user interest mining~\cite{he2024co}. This selective and collaborative approach to learning personalized and shared item attribute embeddings can strike a balance between capturing user-specific preferences and leveraging general attribute information, ultimately improving the recommendation performance.

\subsection{Challenging Scenarios in FedRecSys}
% In FedRecSys research, substantial efforts have been directed toward resolving challenges arising from the interplay of recommendation dynamics and federated optimization frameworks, particularly in fairness-aware optimization \cite{luo2022towards,zhu2022cali3f} and secure federated architectures \cite{ribero2022federating,qu2023semi}.
% , and robust learning paradigms \cite{yuan2023interaction,yu2023untargeted}. 
% These challenges constitute critical research pillars in FedRecSys, demanding systematic solutions across algorithmic, architectural, and protocol design dimensions. \textbf{Fairness} preservation ensures equitable treatment of diverse user groups and mitigates systemic biases in recommendation outcomes, a prerequisite for maintaining user trust and engagement. \textbf{Security} emerges as an equally critical imperative due to the decentralized aggregation of sensitive user interaction data, necessitating rigorous privacy-preserving techniques. By systematically addressing these challenges, FedRecSys can deliver recommendations with provable reliability and transparency, and foster user confidence through privacy-compliant personalization.

In FedRecSys research, significant efforts have been dedicated to overcoming challenges at the intersection of recommendation dynamics and federated optimization frameworks, especially in fairness-aware optimization \cite{luo2022towards,zhu2022cali3f} and secure federated architectures \cite{ribero2022federating,qu2023semi}. These challenges are key research areas in FedRecSys, requiring comprehensive solutions in algorithm, architecture, and protocol design. \textbf{Fairness} is crucial as it ensures equal treatment of different user groups and reduces biases in recommendations, which is essential for user trust. \textbf{Security} is equally vital because sensitive user interaction data is aggregated in a decentralized manner, calling for strict privacy-preserving techniques. By addressing these challenges systematically, FedRecSys can provide reliable and transparent recommendations, and build user confidence through privacy-compliant personalization. Figure \ref{fig:c3_solution} summarizes the challenges and solutions of integrating personalized models into foundation model-based methods, with detailed discussion in the following sections.

% In federated recommendation research, a significant body of work focuses on addressing the challenges brought by recommendation scenarios and federated optimization frameworks, such as fairness~\cite{luo2022towards,zhu2022cali3f}, security~\cite{ribero2022federating,qu2023semi}, and robustness~\cite{yuan2023interaction,yu2023untargeted}. These challenges hold paramount importance in the FedRecSys. Fair recommendation ensures equitable treatment of users and mitigates biases, which is crucial for preserving user trust and satisfaction. Security is a primary concern, as FL aggregates potentially sensitive user data across multiple devices, necessitating the deployment of privacy-preserving techniques. Furthermore, robustness is essential, as federated systems must be able to withstand diverse adversarial attacks and effectively handle noisy or unreliable data originating from participating clients. By addressing these challenges, FedRecSys can deliver reliable recommendations that users can trust, ultimately enhancing the overall quality and real-world impact of the recommendation services. We summarize the challenges and solutions associated with integrating personalized models into challenging scenarios in FedRecSys in Figure~\ref{fig:c3_solution}, and the following sections provide a detailed discussion.

\textbf{Challenges.} 
To tackle challenging scenarios in FedRecSys, specific strategies are needed to boost federated optimization frameworks. However, implementing these strategies might conflict with personalized model learning.

For example, unfairness in federated recommendation occurs when the server gives preference to ``high-quality” clients during global aggregation, sidelining ``low-quality” clients. To counter this, some studies \cite{MLSYS2020_1f5fe839} suggest adjusting local iteration counts according to client capabilities, increasing low-capability clients' participation in global aggregation. But this can lead to \textit{\textbf{overfitting in personalized models}} (\textcolor{blue}{\textbf{Challenge 1}}) of high-capability clients. Their more frequent local updates may cause personalized parameters to over-converge, reducing the model's overall predictive power.

In privacy-enhanced FedRecSys, client privacy leakage risk is often reduced by adding noise to shared parameters \cite{wu2022federated}. While this safeguards client privacy, the introduced noise creates uncertainties that can \textit{\textbf{diminish the quality of personalized models}} (\textcolor{blue}{\textbf{Challenge 2}}). Thus, devising solutions that can address common scenario issues while maintaining personalized model effectiveness is vital for FedRecSys' progress.

\textit{Challenge Formulation.} We formulate the challenge as an optimization problem, which will guide the design of potential solutions. Building on the formulation in Equation~\ref{ob_c2}, we define the optimization objective by incorporating a versatility loss on the personalized parameters $\phi_u$, aiming to enhance the stability of personalized models when integrating techniques for diverse challenging scenarios,
\begin{gather}\label{ob_c3}
\min_{\theta} \sum_{u \in U} \alpha_u 
\mathcal{L}_u(\theta, \phi_u^*; \mathcal{Y}_u) \\ \notag
where \quad \phi_u^* = \arg\min_{\phi_u} \left[ \mathcal{L}_u(\theta, \phi_u; \mathcal{Y}_u) + \mathcal{L}_{\text{vers}}(\phi_u) \right] \\ \notag
s.t. \quad \mathcal{L}_{\text{vers}}(\phi_u) < \delta_{\text{vers}} \> (\forall u \in U)
\end{gather}
where $\delta_{\text{vers}}$ is the predefined threshold.

% In the context of FedRecSys, a common attack scenario involves malicious users uploading perturbed model parameters to disrupt the federated optimization process. To mitigate this, some works~\cite{yu2023untargeted} have proposed optimizing item embeddings towards a uniform distribution during FL, enabling the server to filter out malicious gradients by detecting embedding uniformity. However, this objective conflicts with the goal of learning personalized models, as the resulting \textit{\textbf{homogeneous models may fail to capture the unique preferences of individual clients}} (\textcolor{blue}{\textbf{Challenge 3}}), compromising the effectiveness of personalization. 

\textbf{Solutions.} 
The core of learning personalized models in challenging FedRecSys scenarios is effectively balancing personalization and scenario-specific strategies. In this subsection, we'll explore solutions for two key scenarios in FedRecSys: fairness and security.

In fair FedRecSys, to address personalized model overfitting, clients can \textit{\textbf{use global shared models in tandem with their personalized models}} (\textcolor{magenta}{\textbf{Solution to Challenge 1}}) for recommendation prediction \cite{deng2020adaptive,li2023federated}. Global shared models contain general information. Augmenting overly specific local models with them balances the use of common and personalized data, reducing the negative impact of overfitted local models on recommendation performance.

For privacy-enhanced FedRecSys, to counter noise interference on personalized models, clients can \textit{\textbf{collect their unperturbed local personalized models from previous iterations}} (\textcolor{magenta}{\textbf{Solution to Challenge 2}}) and include them in the final recommendation \cite{luo2023gradma}. This approach uses clean historical models to counter noise while maintaining privacy protection.

\subsection{Foundation Model-based Methods}

\begin{figure}[!t]
\setlength{\abovecaptionskip}{-1mm}
\setlength{\belowcaptionskip}{3mm}
    \centering
    \includegraphics[width=0.65\linewidth]{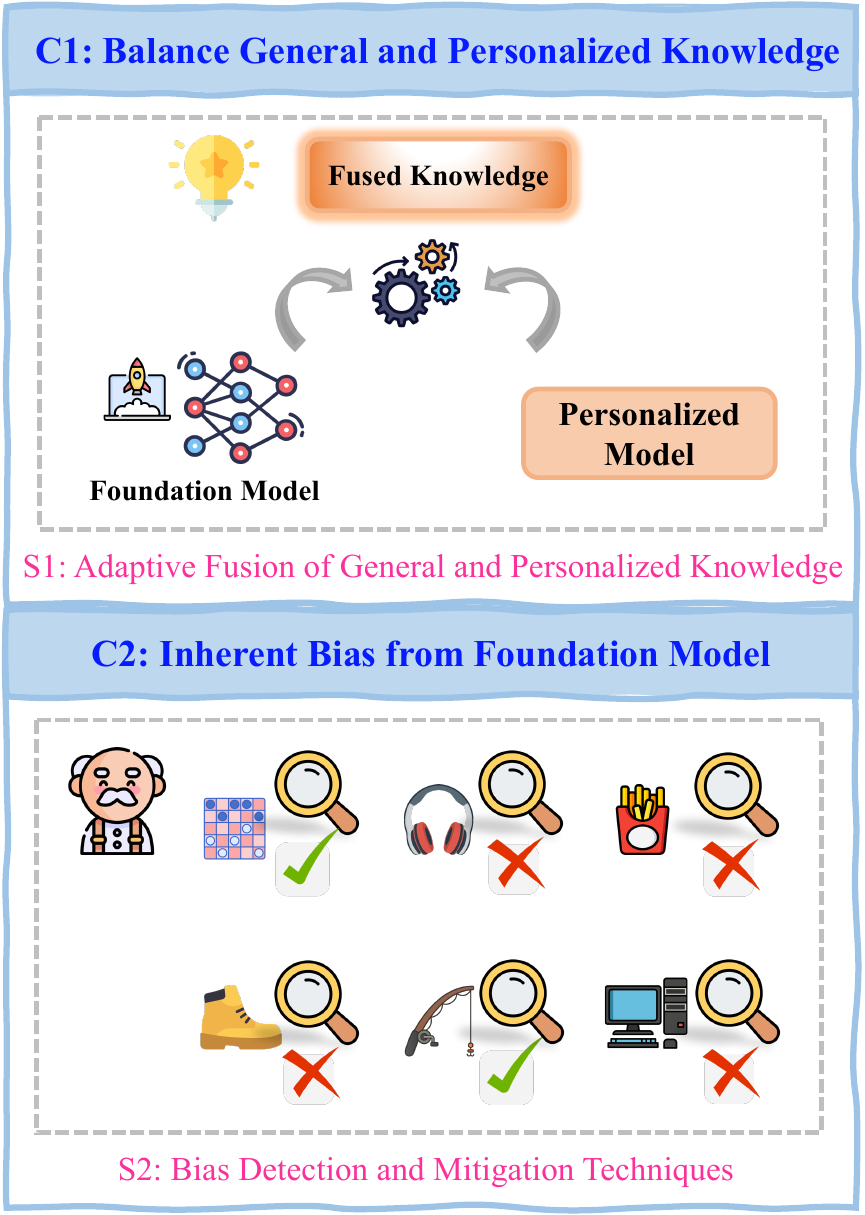}
    \caption{Solution schematic diagram to \textbf{balance general and personalized knowledge}, and \textbf{inherent bias from foundation model challenges} for foundation model-based methods.}
    \label{fig:c4_solution}
\end{figure}

Foundation models \cite{radford2019language,bommasani2021opportunities,achiam2023gpt,10787102,liang2024foundation}, like large language models, are powerful tools adaptable to various tasks via fine-tuning or prompting. They've shown remarkable capabilities in natural language processing \cite{alayrac2022flamingo}, generation \cite{yang2023uniaudio}, and reasoning \cite{kojima2022large}, capturing rich semantic and contextual data information. Recently, research on foundation model-based FedRecSys \cite{zhang2024federated,li2024navigating,zhang2024multifaceted,zhao2024llm} has revealed significant advantages. By fine-tuning these models on federated data, clients can boost personalized recommendations, leveraging the foundation models' broad knowledge. Moreover, it can enhance cold-start performance, and transfer learning, facilitating effective knowledge transfer across different recommendation tasks and domains. In summary, integrating foundation models with FL could revolutionize personalized RecSys. It can lead to more accurate and diverse recommendations tailored to individual users. Figure \ref{fig:c4_solution} summarizes the challenges and solutions of integrating personalized models into foundation model-based methods, with detailed discussion in the following sections.

\textbf{Challenges.} 
% While foundation models trained on rich datasets often contain a wealth of general knowledge, which can benefit the optimization of FedRecSys, learning personalized models in foundation model-based FedRecSys poses significant challenges~\cite{zhuang2023foundation,ren2024advances}. On one hand, \textit{\textbf{balancing the general knowledge inherent in the foundation model and the personalized models trained on user data}} (\textcolor{blue}{\textbf{Challenge 1}}) is a daunting challenge. Additionally, \textit{\textbf{the foundation models may inherently contain biases, which can have a detrimental impact on the learning of personalized models}} (\textcolor{blue}{\textbf{Challenge 2}}). Addressing these challenges is crucial for developing effective foundation model-based FedRecSys that can provide accurate and relevant recommendations to users.
Although foundation models trained on extensive datasets possess abundant general knowledge beneficial for FedRecSys, learning personalized models within foundation model-based FedRecSys is fraught with challenges \cite{zhuang2023foundation,ren2024advances}. Firstly, \textit{\textbf{striking a balance between the general knowledge in the foundation model and the personalized models derived from user data}} (\textcolor{blue}{\textbf{Challenge 1}}) is a formidable task. Secondly, \textit{\textbf{foundation models may harbor inherent biases, which can adversely affect the learning of personalized models}} (\textcolor{blue}{\textbf{Challenge 2}}). Solving these challenges is essential for creating effective foundation model-based FedRecSys.

\textit{Challenge Formulation.} We formulate the challenge as an optimization problem, which will guide the design of potential solutions. Specifically, we refine the optimization objective in Equation~\ref{ob_of_pfedrecsys} by introducing the balance loss $\mathcal{L}_{\text{bal}}$ and bias detection loss $\mathcal{L}_{\text{det}}$ for personalized parameters $\phi_u$, finally formulated as a multi-objective optimization problem,
\begin{gather}\label{ob_c4}
\begin{aligned}
    \min_{\theta} \sum_{u \in U} \alpha_u \mathcal{L}_u(\theta, \phi_u^*; \mathcal{Y}_u)
\end{aligned} \\ \notag
% \begin{aligned}
    where \> \phi_u^* = \arg\min_{\phi_u} \Biggl[ \mathcal{L}_u(\theta, \phi_u; \mathcal{Y}_u) + \mathcal{L}_{\text{bal}}(\phi_u)
    + \mathcal{L}_{\text{det}}(\phi_u) \Biggr] \\
% \end{aligned} \\ \notag
s.t. \quad \mathcal{L}_{\text{bal}}(\phi_u) + \mathcal{L}_{\text{det}}(\phi_u;) > 0 \\ \notag
\mathcal{L}_{\text{bal}}(\phi_u) < \delta_{\text{bal}}, \>
\mathcal{L}_{\text{det}}(\phi_u) < \delta_{\text{det}} \> (\forall u \in U)
\end{gather}
where $\delta_{\text{bias}}$ and $\delta_{\text{det}}$ are the predefined thresholds and FMs denote the foundation models.

\textbf{Solutions.} To balance the utilization of general knowledge from the foundation model and personalized models learned from user data, a hybrid architecture can be crafted. This approach would \textit{\textbf{combine the general knowledge with the personalized models in an adaptive fusion manner}} (\textcolor{magenta}{\textbf{Solution to Challenge 1}}), seamlessly integrating both types of information~\cite{zhang2024federated}. To mitigate the inherent biases in the foundation model, \textit{\textbf{bias detection and mitigation techniques}} (\textcolor{magenta}{\textbf{Solution to Challenge 2}}) can be incorporated. This may involve adversarial debiasing, calibrated data augmentation, or bias-aware loss functions~\cite{hong2021federated,xu2023bias}. These methods reduce bias impact, ensuring fair and unbiased personalized model learning. By employing hybrid architecture and bias mitigation techniques, FedRecSys can effectively blend general knowledge with unique user preferences.

% \begin{enumerate}
%     \item For id embedding-based fedrec methods. \underline{\textit{Challenge}}: storage and communication efficiency of id embeddings due to large item amount. \underline{\textit{Solution}}: each client preserves partial items; factorize item embedding into small matrixes.
%     \item For attribute embedding-based fedrec methods. \underline{\textit{Challenge}}: item attributes are generally abundant and if the clients personalize all attributes on the device, the learned attribute embeddings may be too specific and lead to a negative effect on model performance. \underline{\textit{Solution}}: learn partially personalized attribute embeddings or the user group share the same personalized embeddings.
%     \item For specific issues in fedrec methods, such as fairness and robustness. \underline{\textit{Challenge}}: the optimization objective of specific issue may be orthometric with the personalized item embeddigns. \underline{\textit{Solution}}: devise new optimization objective by integrating both specific issue and personalization modeling.
%     \item For foundation model-based fedrec methods. \underline{\textit{Challenge}}: balance between common knowledge learned from pre-training and personalized item embeddings. \underline{\textit{Solution}}: develop effective fusion mechanism.
% \end{enumerate}
\vspace{-2mm}
\section{Promising Future Directions}\label{future_directions}
% Personalization modeling lies at the heart of RecSys, and in the context of federated recommendation settings, strengthening user-centric personalization is vital to align with the fundamental goal of the recommendation task. Importantly, this can also amplify the inherent benefits of FL's distributed optimization, rendering it an essential component for developing sophisticated and practical FedRecSys. In this section, we explore promising future research directions focused on personalized FedRecSys.
Personalization modeling is central to RecSys. In federated recommendation, enhancing user-centric personalization is crucial to meet the core objective of recommendation tasks. Significantly, it also maximizes the advantages of FL's distributed optimization, making it an essential element for advanced and practical FedRecSys. Here, we explore prospective research directions for personalized FedRecSys.

\vspace{-2mm}
\subsection{New Personalized FedRecSys Modeling Methods}
% Existing personalized federated recommendation models typically learn user-specific models for each client. However, in some scenarios, learning highly personalized user-level models may be overly specific and not optimal for improving recommendation performance. In future research, we can explore alternative approaches to modeling personalized models. For example, we could group users into clusters and learn cluster-level personalized models, allowing similar users within the same group to share the same models. This can enhance the collaborative modeling among like-minded users in the system. Additionally, we can design models at different granularities, and leverage a hierarchical composition of these multi-granularity personalized models. This can help the system depict more comprehensive and accurate user preferences. By moving beyond user-specific s and exploring group-level or multi-granular personalization, we can develop federated recommendation models that better balance the trade-offs between personalization and generalization, ultimately leading to more robust and effective recommendations.
Existing personalized FedRecSys typically generate user-specific models for each client. However, highly personalized user-level models may over-specialize in certain scenarios, hampering recommendation performance. Future research can explore alternative personalized model-building approaches. For instance, user clustering for cluster-level models enables similar users to share models, enhancing collaborative modeling. Designing models at different granularities and using hierarchical compositions can better represent user preferences. By moving beyond user-specific models to explore group-level or multi-granular personalization, we can develop FedRecSys that balance personalization and generalization more effectively, leading to better recommendations.

% \subsection{Combination with Other Personalized Techniques}
% Personalized item embeddings, while meaningful for user personalization modeling, may not fully capture the nuances and complexity of individual user preferences. Future work can explore integrating personalized item embeddings with complementary personalization approaches, such as personalized feature extraction or personalized model architectures~\cite{ye2023heterogeneous}. For instance, the clients can combine personalized item embeddings with their personalized neural network layers, allowing the model to capture both the unique item preferences of each user as well as their personalized feature transformations. By seamlessly integrating personalized item embeddings with other personalization techniques in the federated learning setting, the recommendations can be highly tailored to the individual user's tastes and behaviors, leading to significant improvements in user satisfaction and engagement.

\subsection{Personalization Interpretability}
% Explainability has become a crucial consideration in RecSys research~\cite{zhang2020explainable,chen2024large}, as the need for transparent, accountable, and user-centric AI systems continues to grow across various domains. This is also particularly important in FedRecSys built on personalized models, as it can enhance user trust and enable better understanding and control of personalization process. Personalized models can be complex and opaque, making it difficult to understand why the model is making certain recommendations. Providing interpretability into how the personalized models capture user preferences allows users to understand why certain items are recommended based on their personal preferences reflected in the models. This in turn enables the system to provide explanations for recommendations, further enhancing user trust, as well as enabling more meaningful user control and customization of their personalized experience. Additionally, interpretably personalized models allow for better model debugging and refinement by developers. Overall, interpretably personalized models are crucial for building FedRecSys that are controllable and aligned with user needs.
Explainability has become a pivotal aspect in RecSys research \cite{zhang2020explainable,chen2024large}, especially in the context of growing demand for transparent and user-centric AI across domains. This is particularly crucial in FedRecSys relying on personalized models. Personalized models can be intricate and opaque, making it challenging to discern the basis of recommendations. By providing interpretability, users can understand how their preferences are translated into recommended items. This enhances user trust, enables better user-controlled personalization, and aids developers in debugging. Overall, interpretable personalized models are essential for building FedRecSys that aligns with user needs.

\subsection{Recommendation Diversity}
% Ensuring recommendation diversity has emerged as a vital focus in RecSys research. It helps mitigate filter bubbles, enhancing user satisfaction through serendipitous discoveries, while also supporting broader business objectives like increased engagement and sales~\cite{kunaver2017diversity,castells2021novelty}. Moreover, diverse recommendations are essential for promoting fairness and ethical AI, ensuring equal exposure and reducing biases in the recommendation process. Personalized models in FedRecSys tend to create user-specific representations that inadvertently reinforce filter bubbles and limit user exposure to diverse content. Incorporating recommendation diversity in FedRecSys built on personalized models enables users to discover novel items and experiences beyond their immediate preferences. This supports exploratory user behavior and enhances overall satisfaction by preventing the boredom and fatigue that can arise from overly homogeneous suggestions. Furthermore, as user preferences evolve over time, diverse recommendations can better cater to these dynamic interests, maintaining long-term engagement with the RecSys. By balancing personalized models with recommendation diversity, FedRecSys can provide a well-rounded experience that encourages exploration and adapts to changing user needs. 
Ensuring recommendation diversity is a key focus in RecSys research. It mitigates filter bubbles, boosts user satisfaction via serendipitous finds, and serves business goals like increased engagement and sales \cite{kunaver2017diversity,castells2021novelty}. Additionally, it promotes fairness and ethical AI by ensuring equal exposure and reducing biases. In FedRecSys, personalized models may create user-specific representations that reinforce filter bubbles and limit content diversity. Incorporating diversity allows users to discover items beyond their preferences, preventing boredom from homogeneous suggestions. It also caters to evolving user interests, maintaining long-term engagement. By balancing personalized models with recommendation diversity, FedRecSys can offer a comprehensive experience that encourages exploration and adapts to changing user needs.

\subsection{Practical Scenarios Evaluation}
% Current FedRecSys research primarily relies on public datasets, often lacking validation in real-world online environments. This disconnect can result in research findings that are difficult to apply in practical, large-scale deployments. On one hand, public datasets may not fully capture user behaviors, leading to biased findings that are difficult to apply to live systems. On the other hand, these datasets also cannot replicate the complex dynamics of real-world scenarios, such as diverse user profiles and the need for real-time responsiveness. To address these limitations, there is a pressing need to validate FedRecSys within real-world industrial settings. This can help identify and tackle unique challenges faced by these systems in large-scale, live deployments, including data heterogeneity, privacy constraints, and scalability. By fostering stronger collaborations between academia and industry, the transfer of cutting-edge federated recommendation techniques into practical solutions can be facilitated. This synergistic approach bridges the gap between theoretical advancements and tangible deployments, ensuring the personalization technologies are aligned with real-world business needs and user preferences.
Current FedRecSys research mainly uses public datasets, lacking validation in real-world online settings. This gap makes it hard to apply research findings in large-scale practical deployments. Public datasets may not fully represent user behaviors, leading to biased results, and cannot replicate real-world complexities like diverse user profiles and real-time requirements.
There is an urgent need to validate FedRecSys in industrial settings. This helps address challenges in large-scale live deployments, such as data heterogeneity, privacy issues, and scalability.
Collaboration between academia and industry can facilitate the transfer of advanced federated recommendation techniques into practical solutions. This approach bridges the gap between theory and practice, ensuring personalization technologies meet real-world business and user needs.

\subsection{Benchmark Construction}
% Despite growing research interest in FedRecSys, open-source code and standardized experimental frameworks remain limited. This lack of shared resources poses challenges for the research community. Without a comprehensive benchmark, it is difficult to perform fair comparisons of different federated recommendation algorithms. Researchers may implement their own versions, leading to inconsistencies and hindering replication. This fragmentation impedes progress and slows down the development of FedRecSys. Furthermore, the absence of a common benchmark dataset and evaluation protocols makes it challenging to assess real-world applicability. Researchers may rely on synthetic or limited-scope datasets, failing to capture practical deployment complexities. Developing a well-designed federated recommender benchmark can address these issues and drive the field forward. By providing standardized datasets, metrics, and protocols, it would enable researchers to compare algorithms on a level playing field, fostering healthy competition and innovation. In conclusion, a federated recommender benchmark is crucial to unlock the full potential of this technology and create meaningful impact for end-users.
Despite rising interest in FedRecSys, open-source code and standardized experimental frameworks are scarce. This lack of shared resources challenges the research community. Without a comprehensive benchmark, it is difficult to perform fair comparisons of different FedRecSys. Researchers may implement their own versions, leading to inconsistencies and hindering replication. This fragmentation impedes progress and slows down the development of FedRecSys. Developing a well-designed federated recommendation benchmark can solve these problems. Standardized datasets, metrics, and protocols allow fair algorithm comparisons, spurring competition and innovation. In summary, a benchmark is crucial for realizing FedRecSys' potential and benefiting end-users.

% \textbf{Privacy-Preserving of the Shared Item Embeddings.}
% \begin{itemize}
%     \item Problem: sharing the item embeddings may lead to private user interaction exposure.
%     \item Method: add privacy protection technique; factorize item embedding into small matrixes.
% \end{itemize}

% \textbf{Other Personalized Item Embeddings Modeling Methods.}
% \begin{itemize}
%     \item Problem: in addition to user-specific item embeddings, are there other methods?
%     \item Method: user-group-level personalized item embeddings; multi-grained personalized item embeddings.
% \end{itemize}

% \textbf{Combination with Other Personalization Modeling Techniques.}
% \begin{itemize}
%     \item Problem: 
%     \item Method:
% \end{itemize}

% \textbf{Personalization Interpretability}

% \textbf{Benchmark}

% \textbf{Practical Dataset}

% \textbf{Recommendation Diversity}

\section{Conclusion}\label{conlusion}
% In this survey, we present a comprehensive review of personalization within the context of FedRecSys, marking the first in-depth discussion of personalized FedRecSys and offering a valuable reference for the academic community. We begin by synthesizing the latest comprehensive reviews of the field, offering a clear understanding of the current landscape and available resources in FedRecSys. Building upon this foundation, we offer a first definition of personalization in FedRecSys, highlighting its crucial role in improving the relevance and effectiveness of recommendations. We then delve deeper into the importance of incorporating personalization within federated environments, demonstrating its critical influence on improving recommendation quality. Moreover, we highlight personalized models as a promising direction for future research, providing an in-depth exploration of the associated challenges and suggesting practical solutions to address them. The insights and roadmap presented in this survey are poised to drive meaningful progress, fostering stronger synergies between academic research and real-world applications of personalized recommendation technologies. Overall, this survey serves as an authoritative reference for researchers and practitioners seeking to understand the state-of-the-art in personalized FedRecSys. By consolidating the latest developments and charting the path forward, this work lays the foundation for accelerating innovation and unlocking the transformative potential of personalized recommendations in FL environments.

This survey provides the first systematic examination of personalization in FedRecSys. We commence by integrating the latest comprehensive reviews of the field, providing a lucid understanding of the current FedRecSys landscape and available resources. On this basis, we define personalization in FedRecSys for the first time, underlining its vital role in enhancing recommendation relevance and effectiveness. Additionally, we identify personalized models as a promising future research avenue, deeply exploring related challenges and proposing practical solutions. This work offers both a conceptual framework for researchers and practical insights for implementing privacy-aware RecSys, advancing the development of personalized FedRecSys.

{
    \bibliographystyle{IEEEtran}
    \bibliography{ref}

\begin{thebibliography}{100}
\providecommand{\url}[1]{#1}
\csname url@samestyle\endcsname
\providecommand{\newblock}{\relax}
\providecommand{\bibinfo}[2]{#2}
\providecommand{\BIBentrySTDinterwordspacing}{\spaceskip=0pt\relax}
\providecommand{\BIBentryALTinterwordstretchfactor}{4}
\providecommand{\BIBentryALTinterwordspacing}{\spaceskip=\fontdimen2\font plus
\BIBentryALTinterwordstretchfactor\fontdimen3\font minus \fontdimen4\font\relax}
\providecommand{\BIBforeignlanguage}[2]{{%
\expandafter\ifx\csname l@#1\endcsname\relax
\typeout{** WARNING: IEEEtran.bst: No hyphenation pattern has been}%
\typeout{** loaded for the language `#1'. Using the pattern for}%
\typeout{** the default language instead.}%
\else
\language=\csname l@#1\endcsname
\fi
#2}}
\providecommand{\BIBdecl}{\relax}
\BIBdecl

\bibitem{chai2020secure}
D.~Chai, L.~Wang, K.~Chen, and Q.~Yang, ``Secure federated matrix factorization,'' \emph{IEEE Intelligent Systems}, vol.~36, no.~5, pp. 11--20, 2020.

\bibitem{yang2020federated}
L.~Yang, B.~Tan, V.~W. Zheng, K.~Chen, and Q.~Yang, ``Federated recommendation systems,'' \emph{Federated Learning: Privacy and Incentive}, pp. 225--239, 2020.

\bibitem{huang2021feddsr}
W.~Huang, J.~Liu, T.~Li, T.~Huang, S.~Ji, and J.~Wan, ``Feddsr: Daily schedule recommendation in a federated deep reinforcement learning framework,'' \emph{IEEE Transactions on Knowledge and Data Engineering}, vol.~35, no.~4, pp. 3912--3924, 2021.

\bibitem{wang2022fast}
Q.~Wang, H.~Yin, T.~Chen, J.~Yu, A.~Zhou, and X.~Zhang, ``Fast-adapting and privacy-preserving federated recommender system,'' \emph{The VLDB Journal}, vol.~31, no.~5, pp. 877--896, 2022.

\bibitem{wu2022federated}
C.~Wu, F.~Wu, L.~Lyu, T.~Qi, Y.~Huang, and X.~Xie, ``A federated graph neural network framework for privacy-preserving personalization,'' \emph{Nature Communications}, vol.~13, no.~1, p. 3091, 2022.

\bibitem{zhang2023dual}
C.~Zhang, G.~Long, T.~Zhou, P.~Yan, Z.~Zhang, C.~Zhang, and B.~Yang, ``Dual personalization on federated recommendation,'' in \emph{Proceedings of the Thirty-Second International Joint Conference on Artificial Intelligence}, 2023, pp. 4558--4566.

\bibitem{bobadilla2013recommender}
J.~Bobadilla, F.~Ortega, A.~Hernando, and A.~Guti{\'e}rrez, ``Recommender systems survey,'' \emph{Knowledge-based systems}, vol.~46, pp. 109--132, 2013.

\bibitem{zangerle2022evaluating}
E.~Zangerle and C.~Bauer, ``Evaluating recommender systems: survey and framework,'' \emph{ACM Computing Surveys}, vol.~55, no.~8, pp. 1--38, 2022.

\bibitem{zhang2019deep}
S.~Zhang, L.~Yao, A.~Sun, and Y.~Tay, ``Deep learning based recommender system: A survey and new perspectives,'' \emph{ACM computing surveys (CSUR)}, vol.~52, no.~1, pp. 1--38, 2019.

\bibitem{wu2022graph}
S.~Wu, F.~Sun, W.~Zhang, X.~Xie, and B.~Cui, ``Graph neural networks in recommender systems: a survey,'' \emph{ACM Computing Surveys}, vol.~55, no.~5, pp. 1--37, 2022.

\bibitem{wu2024supporting}
L.~Wu, Z.~Li, H.~Zhao, Z.~Huang, Y.~Han, J.~Jiang, and E.~Chen, ``Supporting your idea reasonably: A knowledge-aware topic reasoning strategy for citation recommendation,'' \emph{IEEE Transactions on Knowledge and Data Engineering}, 2024.

\bibitem{mcmahan2017communication}
B.~McMahan, E.~Moore, D.~Ramage, S.~Hampson, and B.~A. y~Arcas, ``Communication-efficient learning of deep networks from decentralized data,'' in \emph{Artificial Intelligence and Statistics}.\hskip 1em plus 0.5em minus 0.4em\relax PMLR, 2017, pp. 1273--1282.

\bibitem{zhang2021survey}
C.~Zhang, Y.~Xie, H.~Bai, B.~Yu, W.~Li, and Y.~Gao, ``A survey on federated learning,'' \emph{Knowledge-Based Systems}, vol. 216, p. 106775, 2021.

\bibitem{kairouz2021advances}
P.~Kairouz, H.~B. McMahan, B.~Avent, A.~Bellet, M.~Bennis, A.~N. Bhagoji, K.~Bonawitz, Z.~Charles, G.~Cormode, R.~Cummings \emph{et~al.}, ``Advances and open problems in federated learning,'' \emph{Foundations and trends{\textregistered} in machine learning}, vol.~14, no. 1--2, pp. 1--210, 2021.

\bibitem{li2021survey}
Q.~Li, Z.~Wen, Z.~Wu, S.~Hu, N.~Wang, Y.~Li, X.~Liu, and B.~He, ``A survey on federated learning systems: Vision, hype and reality for data privacy and protection,'' \emph{IEEE Transactions on Knowledge and Data Engineering}, vol.~35, no.~4, pp. 3347--3366, 2021.

\bibitem{chai2024survey}
D.~Chai, L.~Wang, L.~Yang, J.~Zhang, K.~Chen, and Q.~Yang, ``A survey for federated learning evaluations: Goals and measures,'' \emph{IEEE Transactions on Knowledge and Data Engineering}, 2024.

\bibitem{liu2024vertical}
Y.~Liu, Y.~Kang, T.~Zou, Y.~Pu, Y.~He, X.~Ye, Y.~Ouyang, Y.-Q. Zhang, and Q.~Yang, ``Vertical federated learning: Concepts, advances, and challenges,'' \emph{IEEE Transactions on Knowledge and Data Engineering}, 2024.

\bibitem{ammad2019federated}
M.~Ammad-Ud-Din, E.~Ivannikova, S.~A. Khan, W.~Oyomno, Q.~Fu, K.~E. Tan, and A.~Flanagan, ``Federated collaborative filtering for privacy-preserving personalized recommendation system,'' \emph{arXiv preprint arXiv:1901.09888}, 2019.

\bibitem{perifanis2022federated}
V.~Perifanis and P.~S. Efraimidis, ``Federated neural collaborative filtering,'' \emph{Knowledge-Based Systems}, vol. 242, p. 108441, 2022.

\bibitem{meihan2022fedcdr}
W.~Meihan, L.~Li, C.~Tao, E.~Rigall, W.~Xiaodong, and X.~Cheng-Zhong, ``Fedcdr: federated cross-domain recommendation for privacy-preserving rating prediction,'' in \emph{Proceedings of the 31st ACM International Conference on Information \& Knowledge Management}, 2022, pp. 2179--2188.

\bibitem{liu2022federated}
Z.~Liu, L.~Yang, Z.~Fan, H.~Peng, and P.~S. Yu, ``Federated social recommendation with graph neural network,'' \emph{ACM Transactions on Intelligent Systems and Technology (TIST)}, vol.~13, no.~4, pp. 1--24, 2022.

\bibitem{qu2023semi}
L.~Qu, N.~Tang, R.~Zheng, Q.~V.~H. Nguyen, Z.~Huang, Y.~Shi, and H.~Yin, ``Semi-decentralized federated ego graph learning for recommendation,'' in \emph{Proceedings of the ACM Web Conference 2023}, 2023, pp. 339--348.

\bibitem{zhang2022pipattack}
S.~Zhang, H.~Yin, T.~Chen, Z.~Huang, Q.~V.~H. Nguyen, and L.~Cui, ``Pipattack: Poisoning federated recommender systems for manipulating item promotion,'' in \emph{Proceedings of the Fifteenth ACM International Conference on Web Search and Data Mining}, 2022, pp. 1415--1423.

\bibitem{zhang2023lightfr}
H.~Zhang, F.~Luo, J.~Wu, X.~He, and Y.~Li, ``Lightfr: Lightweight federated recommendation with privacy-preserving matrix factorization,'' \emph{ACM Transactions on Information Systems}, vol.~41, no.~4, pp. 1--28, 2023.

\bibitem{jiang2022adaptive}
Y.~Jiang, Q.~Li, H.~Zhu, J.~Yu, J.~Li, Z.~Xu, H.~Dong, and B.~Zheng, ``Adaptive domain interest network for multi-domain recommendation,'' in \emph{Proceedings of the 31st ACM International Conference on Information \& Knowledge Management}, 2022, pp. 3212--3221.

\bibitem{zhang2024m3oe}
Z.~Zhang, S.~Liu, J.~Yu, Q.~Cai, X.~Zhao, C.~Zhang, Z.~Liu, Q.~Liu, H.~Zhao, L.~Hu \emph{et~al.}, ``M3oe: Multi-domain multi-task mixture-of experts recommendation framework,'' in \emph{Proceedings of the 47th International ACM SIGIR Conference on Research and Development in Information Retrieval}, 2024, pp. 893--902.

\bibitem{qin2020multitask}
Z.~Qin, Y.~Cheng, Z.~Zhao, Z.~Chen, D.~Metzler, and J.~Qin, ``Multitask mixture of sequential experts for user activity streams,'' in \emph{Proceedings of the 26th ACM SIGKDD International Conference on Knowledge Discovery \& Data Mining}, 2020, pp. 3083--3091.

\bibitem{guo2024multi}
X.~Guo, M.~Ha, X.~Tao, S.~Li, Y.~Li, Z.~Zhu, Z.~Shen, and L.~Ma, ``Multi-task learning with sequential dependence toward industrial applications: A systematic formulation,'' \emph{ACM Transactions on Knowledge Discovery from Data}, vol.~18, no.~5, pp. 1--29, 2024.

\bibitem{javeed2023federated}
D.~Javeed, M.~S. Saeed, P.~Kumar, A.~Jolfaei, S.~Islam, and A.~N. Islam, ``Federated learning-based personalized recommendation systems: An overview on security and privacy challenges,'' \emph{IEEE Transactions on Consumer Electronics}, 2023.

\bibitem{chronis2024survey}
C.~Chronis, I.~Varlamis, Y.~Himeur, A.~N. Sayed, T.~M. Al-Hasan, A.~Nhlabatsi, F.~Bensaali, and G.~Dimitrakopoulos, ``A survey on the use of federated learning in privacy-preserving recommender systems,'' \emph{IEEE Open Journal of the Computer Society}, 2024.

\bibitem{harasic2024recent}
M.~Harasic, F.-S. Keese, D.~Mattern, and A.~Paschke, ``Recent advances and future challenges in federated recommender systems,'' \emph{International Journal of Data Science and Analytics}, vol.~17, no.~4, pp. 337--357, 2024.

\bibitem{alamgir2022federated}
Z.~Alamgir, F.~K. Khan, and S.~Karim, ``Federated recommenders: methods, challenges and future,'' \emph{Cluster Computing}, vol.~25, no.~6, pp. 4075--4096, 2022.

\bibitem{sun2024survey}
Z.~Sun, Y.~Xu, Y.~Liu, W.~He, L.~Kong, F.~Wu, Y.~Jiang, and L.~Cui, ``A survey on federated recommendation systems,'' \emph{IEEE Transactions on Neural Networks and Learning Systems}, 2024.

\bibitem{wang2024horizontal}
L.~Wang, H.~Zhou, Y.~Bao, X.~Yan, G.~Shen, and X.~Kong, ``Horizontal federated recommender system: A survey,'' \emph{ACM Computing Surveys}, vol.~56, no.~9, pp. 1--42, 2024.

\bibitem{li2024navigating}
Z.~Li and G.~Long, ``Navigating the future of federated recommendation systems with foundation models,'' \emph{arXiv preprint arXiv:2406.00004}, 2024.

\bibitem{flanagan2021federated}
A.~Flanagan, W.~Oyomno, A.~Grigorievskiy, K.~E. Tan, S.~A. Khan, and M.~Ammad-Ud-Din, ``Federated multi-view matrix factorization for personalized recommendations,'' in \emph{Machine learning and knowledge discovery in databases: European conference, ECML PKDD 2020, Ghent, Belgium, September 14--18, 2020, Proceedings, Part II}.\hskip 1em plus 0.5em minus 0.4em\relax Springer, 2021, pp. 324--347.

\bibitem{hu2022federated}
P.~Hu, E.~Yang, W.~Pan, X.~Peng, and Z.~Ming, ``Federated one-class collaborative filtering via privacy-aware non-sampling matrix factorization,'' \emph{Knowledge-Based Systems}, vol. 253, p. 109441, 2022.

\bibitem{li2023federated}
Z.~Li, G.~Long, and T.~Zhou, ``Federated recommendation with additive personalization,'' in \emph{The Twelfth International Conference on Learning Representations}.

\bibitem{liang2021fedrec++}
F.~Liang, W.~Pan, and Z.~Ming, ``Fedrec++: Lossless federated recommendation with explicit feedback,'' in \emph{Proceedings of the AAAI conference on artificial intelligence}, vol.~35, no.~5, 2021, pp. 4224--4231.

\bibitem{lin2020fedrec}
G.~Lin, F.~Liang, W.~Pan, and Z.~Ming, ``Fedrec: Federated recommendation with explicit feedback,'' \emph{IEEE Intelligent Systems}, vol.~36, no.~5, pp. 21--30, 2020.

\bibitem{lin2020meta}
Y.~Lin, P.~Ren, Z.~Chen, Z.~Ren, D.~Yu, J.~Ma, M.~d. Rijke, and X.~Cheng, ``Meta matrix factorization for federated rating predictions,'' in \emph{Proceedings of the 43rd International ACM SIGIR Conference on Research and Development in Information Retrieval}, 2020, pp. 981--990.

\bibitem{du2021federated}
Y.~Du, D.~Zhou, Y.~Xie, J.~Shi, and M.~Gong, ``Federated matrix factorization for privacy-preserving recommender systems,'' \emph{Applied soft computing}, vol. 111, p. 107700, 2021.

\bibitem{yang2021fcmf}
E.~Yang, Y.~Huang, F.~Liang, W.~Pan, and Z.~Ming, ``Fcmf: Federated collective matrix factorization for heterogeneous collaborative filtering,'' \emph{Knowledge-Based Systems}, vol. 220, p. 106946, 2021.

\bibitem{liu2022fairness}
S.~Liu, Y.~Ge, S.~Xu, Y.~Zhang, and A.~Marian, ``Fairness-aware federated matrix factorization,'' in \emph{Proceedings of the 16th ACM conference on recommender systems}, 2022, pp. 168--178.

\bibitem{chai2022efficient}
D.~Chai, L.~Wang, K.~Chen, and Q.~Yang, ``Efficient federated matrix factorization against inference attacks,'' \emph{ACM Transactions on Intelligent Systems and Technology (TIST)}, vol.~13, no.~4, pp. 1--20, 2022.

\bibitem{zheng2023federated}
X.~Zheng, M.~Guan, X.~Jia, L.~Sun, and Y.~Luo, ``Federated matrix factorization recommendation based on secret sharing for privacy preserving,'' \emph{IEEE Transactions on Computational Social Systems}, 2023.

\bibitem{singhal2021federated}
K.~Singhal, H.~Sidahmed, Z.~Garrett, S.~Wu, J.~Rush, and S.~Prakash, ``Federated reconstruction: Partially local federated learning,'' \emph{Advances in Neural Information Processing Systems}, vol.~34, pp. 11\,220--11\,232, 2021.

\bibitem{li2024fedcore}
Z.~Li, X.~Wu, W.~Pan, Y.~Ding, Z.~Wu, S.~Tan, Q.~Xu, Q.~Yang, and Z.~Ming, ``Fedcore: Federated learning for cross-organization recommendation ecosystem,'' \emph{IEEE Transactions on Knowledge and Data Engineering}, 2024.

\bibitem{muhammad2020fedfast}
K.~Muhammad, Q.~Wang, D.~O'Reilly-Morgan, E.~Tragos, B.~Smyth, N.~Hurley, J.~Geraci, and A.~Lawlor, ``Fedfast: Going beyond average for faster training of federated recommender systems,'' in \emph{Proceedings of the 26th ACM SIGKDD international conference on knowledge discovery \& data mining}, 2020, pp. 1234--1242.

\bibitem{hu2024user}
Q.~Hu and Y.~Song, ``User consented federated recommender system against personalized attribute inference attack,'' in \emph{Proceedings of the 17th ACM International Conference on Web Search and Data Mining}, 2024, pp. 276--285.

\bibitem{zhang2023federated}
C.~Zhang, G.~Long, T.~Zhou, Z.~Zhang, P.~Yan, and B.~Yang, ``When federated recommendation meets cold-start problem: Separating item attributes and user interactions,'' in \emph{Proceedings of the ACM on Web Conference 2024}, 2024, pp. 3632--3642.

\bibitem{wu2021hierarchical}
J.~Wu, Q.~Liu, Z.~Huang, Y.~Ning, H.~Wang, E.~Chen, J.~Yi, and B.~Zhou, ``Hierarchical personalized federated learning for user modeling,'' in \emph{Proceedings of the Web Conference 2021}, 2021, pp. 957--968.

\bibitem{zhang2024federated}
C.~Zhang, G.~Long, H.~Guo, X.~Fang, Y.~Song, Z.~Liu, G.~Zhou, Z.~Zhang, Y.~Liu, and B.~Yang, ``Federated adaptation for foundation model-based recommendations,'' in \emph{Proceedings of the Thirty-Third International Joint Conference on Artificial Intelligence, {IJCAI-24}}, 2024, pp. 5453--5461.

\bibitem{duan2019jointrec}
S.~Duan, D.~Zhang, Y.~Wang, L.~Li, and Y.~Zhang, ``Jointrec: A deep-learning-based joint cloud video recommendation framework for mobile iot,'' \emph{IEEE Internet of Things Journal}, vol.~7, no.~3, pp. 1655--1666, 2019.

\bibitem{perifanis2023fedpoirec}
V.~Perifanis, G.~Drosatos, G.~Stamatelatos, and P.~S. Efraimidis, ``Fedpoirec: Privacy-preserving federated poi recommendation with social influence,'' \emph{Information Sciences}, vol. 623, pp. 767--790, 2023.

\bibitem{yan2024federated}
B.~Yan, Y.~Cao, H.~Wang, W.~Yang, J.~Du, and C.~Shi, ``Federated heterogeneous graph neural network for privacy-preserving recommendation,'' in \emph{Proceedings of the ACM on Web Conference 2024}, 2024, pp. 3919--3929.

\bibitem{hu2023privacy}
P.~Hu, Z.~Lin, W.~Pan, Q.~Yang, X.~Peng, and Z.~Ming, ``Privacy-preserving graph convolution network for federated item recommendation,'' \emph{Artificial Intelligence}, vol. 324, p. 103996, 2023.

\bibitem{agrawal2024no}
N.~Agrawal, A.~K. Sirohi, S.~Kumar \emph{et~al.}, ``No prejudice! fair federated graph neural networks for personalized recommendation,'' in \emph{Proceedings of the AAAI Conference on Artificial Intelligence}, vol.~38, no.~10, 2024, pp. 10\,775--10\,783.

\bibitem{tian2024privacy}
C.~Tian, Y.~Xie, X.~Chen, Y.~Li, and X.~Zhao, ``Privacy-preserving cross-domain recommendation with federated graph learning,'' \emph{ACM Transactions on Information Systems}, vol.~42, no.~5, pp. 1--29, 2024.

\bibitem{xie2023dci}
B.~Xie, C.~Hu, H.~Huang, J.~Yu, and H.~Xia, ``Dci-pfgl: Decentralized cross-institutional personalized federated graph learning for iot service recommendation,'' \emph{IEEE Internet of Things Journal}, 2023.

\bibitem{sun2024federated}
H.~Sun, Z.~Tu, D.~Sui, B.~Zhang, and X.~Xu, ``A federated social recommendation approach with enhanced hypergraph neural network,'' \emph{ACM Transactions on Intelligent Systems and Technology}, 2024.

\bibitem{tangfedgst}
T.~Tang, M.~Hou, S.~Yu, Z.~Cai, Z.~Han, G.~Oatley, and V.~Saikrishna, ``Fedgst: An efficient federated graph neural network for spatio-temporal poi recommendation,'' \emph{ACM Transactions on Sensor Networks}.

\bibitem{zhang2024gpfedrec}
C.~Zhang, G.~Long, T.~Zhou, Z.~Zhang, P.~Yan, and B.~Yang, ``Gpfedrec: Graph-guided personalization for federated recommendation,'' in \emph{Proceedings of the 30th ACM SIGKDD Conference on Knowledge Discovery and Data Mining}, 2024, pp. 4131--4142.

\bibitem{wei2023edge}
S.~Wei, S.~Meng, Q.~Li, X.~Zhou, L.~Qi, and X.~Xu, ``Edge-enabled federated sequential recommendation with knowledge-aware transformer,'' \emph{Future Generation Computer Systems}, vol. 148, pp. 610--622, 2023.

\bibitem{belhadi2024federated}
A.~Belhadi, Y.~Djenouri, F.~A. de~Alcantara~Andrade, and G.~Srivastava, ``Federated constrastive learning and visual transformers for personal recommendation,'' \emph{Cognitive Computation}, vol.~16, no.~5, pp. 2551--2565, 2024.

\bibitem{feng2024robust}
C.~Feng, D.~Feng, G.~Huang, Z.~Liu, Z.~Wang, and X.-G. Xia, ``Robust privacy-preserving recommendation systems driven by multimodal federated learning,'' \emph{IEEE Transactions on Neural Networks and Learning Systems}, 2024.

\bibitem{zhang2024multifaceted}
C.~Zhang, G.~Long, H.~Guo, Z.~Liu, G.~Zhou, Z.~Zhang, Y.~Liu, and B.~Yang, ``Multifaceted user modeling in recommendation: A federated foundation models approach,'' in \emph{Proceedings of the AAAI Conference on Artificial Intelligence}, 2025.

\bibitem{wang2012nonnegative}
Y.-X. Wang and Y.-J. Zhang, ``Nonnegative matrix factorization: A comprehensive review,'' \emph{IEEE Transactions on knowledge and data engineering}, vol.~25, no.~6, pp. 1336--1353, 2012.

\bibitem{chen2023win}
G.~Chen, X.~Zhang, Y.~Su, Y.~Lai, J.~Xiang, J.~Zhang, and Y.~Zheng, ``Win-win: a privacy-preserving federated framework for dual-target cross-domain recommendation,'' in \emph{Proceedings of the AAAI Conference on Artificial Intelligence}, vol.~37, no.~4, 2023, pp. 4149--4156.

\bibitem{liu2023federated}
W.~Liu, C.~Chen, X.~Liao, M.~Hu, J.~Yin, Y.~Tan, and L.~Zheng, ``Federated probabilistic preference distribution modelling with compactness co-clustering for privacy-preserving multi-domain recommendation.'' in \emph{IJCAI}, 2023, pp. 2206--2214.

\bibitem{zhang2024feddcsr}
H.~Zhang, D.~Zheng, X.~Yang, J.~Feng, and Q.~Liao, ``Feddcsr: Federated cross-domain sequential recommendation via disentangled representation learning,'' in \emph{Proceedings of the 2024 SIAM International Conference on Data Mining (SDM)}.\hskip 1em plus 0.5em minus 0.4em\relax SIAM, 2024, pp. 535--543.

\bibitem{guo2024prompt}
L.~Guo, Z.~Lu, J.~Yu, Q.~V.~H. Nguyen, and H.~Yin, ``Prompt-enhanced federated content representation learning for cross-domain recommendation,'' in \emph{Proceedings of the ACM on Web Conference 2024}, 2024, pp. 3139--3149.

\bibitem{zhang2024fedhcdr}
H.~Zhang, D.~Zheng, L.~Zhong, X.~Yang, J.~Feng, Y.~Feng, and Q.~Liao, ``Fedhcdr: Federated cross-domain recommendation with hypergraph signal decoupling,'' in \emph{Joint European Conference on Machine Learning and Knowledge Discovery in Databases}.\hskip 1em plus 0.5em minus 0.4em\relax Springer, 2024, pp. 350--366.

\bibitem{maeng2022towards}
K.~Maeng, H.~Lu, L.~Melis, J.~Nguyen, M.~Rabbat, and C.-J. Wu, ``Towards fair federated recommendation learning: Characterizing the inter-dependence of system and data heterogeneity,'' in \emph{Proceedings of the 16th ACM Conference on Recommender Systems}, 2022, pp. 156--167.

\bibitem{zhu2022cali3f}
Z.~Zhu, S.~Si, J.~Wang, and J.~Xiao, ``Cali3f: Calibrated fast fair federated recommendation system,'' in \emph{2022 International Joint Conference on Neural Networks (IJCNN)}.\hskip 1em plus 0.5em minus 0.4em\relax IEEE, 2022, pp. 1--8.

\bibitem{luo2022towards}
S.~Luo, Y.~Xiao, Y.~Liu, C.~Li, and L.~Song, ``Towards communication efficient and fair federated personalized sequential recommendation,'' in \emph{2022 5th International Conference on Information Communication and Signal Processing (ICICSP)}.\hskip 1em plus 0.5em minus 0.4em\relax IEEE, 2022, pp. 1--6.

\bibitem{wang2024towards}
S.~Wang, H.~Tao, J.~Li, X.~Ji, Y.~Gao, and M.~Gong, ``Towards fair and personalized federated recommendation,'' \emph{Pattern Recognition}, vol. 149, p. 110234, 2024.

\bibitem{zhou2019privacy}
P.~Zhou, K.~Wang, L.~Guo, S.~Gong, and B.~Zheng, ``A privacy-preserving distributed contextual federated online learning framework with big data support in social recommender systems,'' \emph{IEEE Transactions on Knowledge and Data Engineering}, vol.~33, no.~3, pp. 824--838, 2019.

\bibitem{luo2022dual}
L.~Luo and B.~Liu, ``Dual-contrastive for federated social recommendation,'' in \emph{2022 International Joint Conference on Neural Networks (IJCNN)}.\hskip 1em plus 0.5em minus 0.4em\relax IEEE, 2022, pp. 1--8.

\bibitem{qi2020privacy}
T.~Qi, F.~Wu, C.~Wu, Y.~Huang, and X.~Xie, ``Privacy-preserving news recommendation model learning,'' in \emph{Findings of the Association for Computational Linguistics: EMNLP 2020}, 2020, pp. 1423--1432.

\bibitem{yi2021efficient}
J.~Yi, F.~Wu, C.~Wu, R.~Liu, G.~Sun, and X.~Xie, ``Efficient-fedrec: Efficient federated learning framework for privacy-preserving news recommendation,'' in \emph{Proceedings of the 2021 Conference on Empirical Methods in Natural Language Processing}, 2021, pp. 2814--2824.

\bibitem{yi2023ua}
J.~Yi, F.~Wu, B.~Zhu, J.~Yao, Z.~Tao, G.~Sun, and X.~Xie, ``Ua-fedrec: untargeted attack on federated news recommendation,'' in \emph{Proceedings of the 29th ACM SIGKDD Conference on Knowledge Discovery and Data Mining}, 2023, pp. 5428--5438.

\bibitem{liu2023privaterec}
R.~Liu, Y.~Cao, Y.~Wang, L.~Lyu, Y.~Chen, and H.~Chen, ``Privaterec: Differentially private model training and online serving for federated news recommendation,'' in \emph{Proceedings of the 29th ACM SIGKDD Conference on Knowledge Discovery and Data Mining}, 2023, pp. 4539--4548.

\bibitem{yu2023federated}
S.~L. Yu, Q.~Liu, F.~Wang, Y.~Yu, and E.~Chen, ``Federated news recommendation with fine-grained interpolation and dynamic clustering,'' in \emph{Proceedings of the 32nd ACM International Conference on Information and Knowledge Management}, 2023, pp. 3073--3082.

\bibitem{huang2023randomization}
X.~Huang, Y.~Luo, L.~Liu, W.~Zhao, and S.~Fu, ``Randomization is all you need: A privacy-preserving federated learning framework for news recommendation,'' \emph{Information Sciences}, vol. 637, p. 118943, 2023.

\bibitem{chen2020practical}
C.~Chen, J.~Zhou, B.~Wu, W.~Fang, L.~Wang, Y.~Qi, and X.~Zheng, ``Practical privacy preserving poi recommendation,'' \emph{ACM Transactions on Intelligent Systems and Technology (TIST)}, vol.~11, no.~5, pp. 1--20, 2020.

\bibitem{dong2022ranking}
Q.~Dong, B.~Liu, X.~Zhang, J.~Qin, B.~Wang, and J.~Qian, ``Ranking-based federated poi recommendation with geographic effect,'' in \emph{2022 international joint conference on neural networks (IJCNN)}.\hskip 1em plus 0.5em minus 0.4em\relax IEEE, 2022, pp. 1--8.

\bibitem{zhang2023fine}
X.~Zhang, Z.~Ye, J.~Lu, F.~Zhuang, Y.~Zheng, and D.~Yu, ``Fine-grained preference-aware personalized federated poi recommendation with data sparsity,'' in \emph{Proceedings of the 46th International ACM SIGIR Conference on Research and Development in Information Retrieval}, 2023, pp. 413--422.

\bibitem{ye2023adaptive}
Z.~Ye, X.~Zhang, X.~Chen, H.~Xiong, and D.~Yu, ``Adaptive clustering based personalized federated learning framework for next poi recommendation with location noise,'' \emph{IEEE Transactions on Knowledge and Data Engineering}, 2023.

\bibitem{chen2017sampling}
T.~Chen, Y.~Sun, Y.~Shi, and L.~Hong, ``On sampling strategies for neural network-based collaborative filtering,'' in \emph{Proceedings of the 23rd ACM SIGKDD International Conference on Knowledge Discovery and Data Mining}, 2017, pp. 767--776.

\bibitem{hao2023feature}
Y.~Hao, T.~Zhang, P.~Zhao, Y.~Liu, V.~S. Sheng, J.~Xu, G.~Liu, and X.~Zhou, ``Feature-level deeper self-attention network with contrastive learning for sequential recommendation,'' \emph{IEEE transactions on knowledge and data engineering}, vol.~35, no.~10, pp. 10\,112--10\,124, 2023.

\bibitem{xie2022contrastive}
X.~Xie, F.~Sun, Z.~Liu, S.~Wu, J.~Gao, J.~Zhang, B.~Ding, and B.~Cui, ``Contrastive learning for sequential recommendation,'' in \emph{2022 IEEE 38th international conference on data engineering (ICDE)}.\hskip 1em plus 0.5em minus 0.4em\relax IEEE, 2022, pp. 1259--1273.

\bibitem{chen2022learning}
Y.~Chen, H.~Guo, Y.~Zhang, C.~Ma, R.~Tang, J.~Li, and I.~King, ``Learning binarized graph representations with multi-faceted quantization reinforcement for top-k recommendation,'' in \emph{Proceedings of the 28th ACM SIGKDD Conference on Knowledge Discovery and Data Mining}, 2022, pp. 168--178.

\bibitem{wu2021fedgnn}
C.~Wu, F.~Wu, Y.~Cao, Y.~Huang, and X.~Xie, ``Fedgnn: Federated graph neural network for privacy-preserving recommendation,'' \emph{ICML Workshop}, 2021.

\bibitem{huang2020federated}
M.~Huang, H.~Li, B.~Bai, C.~Wang, K.~Bai, and F.~Wang, ``A federated multi-view deep learning framework for privacy-preserving recommendations,'' \emph{arXiv e-prints}, pp. arXiv--2008, 2020.

\bibitem{wang2020federated}
L.~Wang, Z.~Huang, Q.~Pei, and S.~Wang, ``Federated cf: Privacy-preserving collaborative filtering cross multiple datasets,'' in \emph{ICC 2020-2020 IEEE International Conference on Communications (ICC)}.\hskip 1em plus 0.5em minus 0.4em\relax IEEE, 2020, pp. 1--6.

\bibitem{lin2021fr}
Z.~Lin, W.~Pan, and Z.~Ming, ``Fr-fmss: Federated recommendation via fake marks and secret sharing,'' in \emph{Proceedings of the 15th ACM Conference on Recommender Systems}, 2021, pp. 668--673.

\bibitem{yang2022practical}
L.~Yang, J.~Zhang, D.~Chai, L.~Wang, K.~Guo, K.~Chen, and Q.~Yang, ``Practical and secure federated recommendation with personalized mask,'' in \emph{International Workshop on Trustworthy Federated Learning}.\hskip 1em plus 0.5em minus 0.4em\relax Springer, 2022, pp. 33--45.

\bibitem{tang2012cross}
J.~Tang, S.~Wu, J.~Sun, and H.~Su, ``Cross-domain collaboration recommendation,'' in \emph{Proceedings of the 18th ACM SIGKDD international conference on Knowledge discovery and data mining}, 2012, pp. 1285--1293.

\bibitem{pitoura2022fairness}
E.~Pitoura, K.~Stefanidis, and G.~Koutrika, ``Fairness in rankings and recommendations: an overview,'' \emph{The VLDB Journal}, pp. 1--28, 2022.

\bibitem{cao2019social}
D.~Cao, X.~He, L.~Miao, G.~Xiao, H.~Chen, and J.~Xu, ``Social-enhanced attentive group recommendation,'' \emph{IEEE Transactions on Knowledge and Data Engineering}, vol.~33, no.~3, pp. 1195--1209, 2019.

\bibitem{wu2022fedattack}
C.~Wu, F.~Wu, T.~Qi, Y.~Huang, and X.~Xie, ``Fedattack: Effective and covert poisoning attack on federated recommendation via hard sampling,'' in \emph{Proceedings of the 28th ACM SIGKDD Conference on Knowledge Discovery and Data Mining}, 2022, pp. 4164--4172.

\bibitem{rong2022fedrecattack}
D.~Rong, S.~Ye, R.~Zhao, H.~N. Yuen, J.~Chen, and Q.~He, ``Fedrecattack: Model poisoning attack to federated recommendation,'' in \emph{2022 IEEE 38th International Conference on Data Engineering (ICDE)}.\hskip 1em plus 0.5em minus 0.4em\relax IEEE, 2022, pp. 2643--2655.

\bibitem{yuan2023interaction}
W.~Yuan, C.~Yang, Q.~V.~H. Nguyen, L.~Cui, T.~He, and H.~Yin, ``Interaction-level membership inference attack against federated recommender systems,'' in \emph{Proceedings of the ACM Web Conference 2023}, 2023, pp. 1053--1062.

\bibitem{yu2023untargeted}
Y.~Yu, Q.~Liu, L.~Wu, R.~Yu, S.~L. Yu, and Z.~Zhang, ``Untargeted attack against federated recommendation systems via poisonous item embeddings and the defense,'' in \emph{Proceedings of the AAAI Conference on Artificial Intelligence}, vol.~37, no.~4, 2023, pp. 4854--4863.

\bibitem{zhang2024preventing}
J.~Zhang, H.~Li, D.~Rong, Y.~Zhao, K.~Chen, and L.~Shou, ``Preventing the popular item embedding based attack in federated recommendations,'' in \emph{2024 IEEE 40th International Conference on Data Engineering (ICDE)}.\hskip 1em plus 0.5em minus 0.4em\relax IEEE, 2024, pp. 2179--2191.

\bibitem{ijcai2022p306}
D.~Rong, Q.~He, and J.~Chen, ``Poisoning deep learning based recommender model in federated learning scenarios,'' in \emph{Proceedings of the Thirty-First International Joint Conference on Artificial Intelligence, {IJCAI-22}}, L.~D. Raedt, Ed.\hskip 1em plus 0.5em minus 0.4em\relax International Joint Conferences on Artificial Intelligence Organization, 2022, pp. 2204--2210.

\bibitem{yuan2023manipulating}
W.~Yuan, Q.~V.~H. Nguyen, T.~He, L.~Chen, and H.~Yin, ``Manipulating federated recommender systems: Poisoning with synthetic users and its countermeasures,'' in \emph{Proceedings of the 46th International ACM SIGIR Conference on Research and Development in Information Retrieval}, 2023, pp. 1690--1699.

\bibitem{yin2024poisoning}
M.~Yin, Y.~Xu, M.~Fang, and N.~Z. Gong, ``Poisoning federated recommender systems with fake users,'' in \emph{Proceedings of the ACM on Web Conference 2024}, 2024, pp. 3555--3565.

\bibitem{su2024revisit}
J.~Su, C.~Chen, W.~Liu, Z.~Lin, S.~Shen, W.~Wang, and X.~Zheng, ``Revisit targeted model poisoning on federated recommendation: Optimize via multi-objective transport,'' in \emph{Proceedings of the 47th international acm sigir conference on research and development in information retrieval}, 2024, pp. 1722--1732.

\bibitem{ali2024hidattack}
W.~Ali, K.~Umer, X.~Zhou, and J.~Shao, ``Hidattack: An effective and undetectable model poisoning attack to federated recommenders,'' \emph{IEEE Transactions on Knowledge and Data Engineering}, 2024.

\bibitem{zhang2023comprehensive}
S.~Zhang, W.~Yuan, and H.~Yin, ``Comprehensive privacy analysis on federated recommender system against attribute inference attacks,'' \emph{IEEE Transactions on Knowledge and Data Engineering}, 2023.

\bibitem{liu2024defending}
X.~Liu, Y.~Chen, and S.~Pang, ``Defending against membership inference attack for counterfactual federated recommendation with differentially private representation learning,'' \emph{IEEE Transactions on Information Forensics and Security}, 2024.

\bibitem{hao2021efficient}
M.~Hao, H.~Li, G.~Xu, H.~Chen, and T.~Zhang, ``Efficient, private and robust federated learning,'' in \emph{Proceedings of the 37th Annual Computer Security Applications Conference}, 2021, pp. 45--60.

\bibitem{mothukuri2021survey}
V.~Mothukuri, R.~M. Parizi, S.~Pouriyeh, Y.~Huang, A.~Dehghantanha, and G.~Srivastava, ``A survey on security and privacy of federated learning,'' \emph{Future Generation Computer Systems}, vol. 115, pp. 619--640, 2021.

\bibitem{park2022privacy}
J.~Park and H.~Lim, ``Privacy-preserving federated learning using homomorphic encryption,'' \emph{Applied Sciences}, vol.~12, no.~2, p. 734, 2022.

\bibitem{geyer2017differentially}
R.~C. Geyer, T.~Klein, and M.~Nabi, ``Differentially private federated learning: A client level perspective,'' \emph{arXiv preprint arXiv:1712.07557}, 2017.

\bibitem{cui2022communication}
Z.~Cui, J.~Wen, Y.~Lan, Z.~Zhang, and J.~Cai, ``Communication-efficient federated recommendation model based on many-objective evolutionary algorithm,'' \emph{Expert Systems with Applications}, vol. 201, p. 116963, 2022.

\bibitem{zhang2024efvae}
L.~Zhang, Q.~Rong, X.~Ding, G.~Li, and L.~Yuan, ``Efvae: Efficient federated variational autoencoder for collaborative filtering,'' in \emph{Proceedings of the 33rd ACM International Conference on Information and Knowledge Management}, 2024, pp. 3176--3185.

\bibitem{ali2024communication}
W.~Ali, M.~Ammad-ud din, X.~Zhou, Y.~Zhang, and J.~Shao, ``Communication-efficient federated neural collaborative filtering with multi-armed bandits,'' \emph{ACM Transactions on Recommender Systems}, 2024.

\bibitem{khan2021payload}
F.~K. Khan, A.~Flanagan, K.~E. Tan, Z.~Alamgir, and M.~Ammad-Ud-Din, ``A payload optimization method for federated recommender systems,'' in \emph{Proceedings of the 15th ACM Conference on Recommender Systems}, 2021, pp. 432--442.

\bibitem{li2024towards}
G.~Li, X.~Ding, L.~Yuan, L.~Zhang, and Q.~Rong, ``Towards resource-efficient and secure federated multimedia recommendation,'' in \emph{ICASSP 2024-2024 IEEE International Conference on Acoustics, Speech and Signal Processing (ICASSP)}.\hskip 1em plus 0.5em minus 0.4em\relax IEEE, 2024, pp. 5515--5519.

\bibitem{wu2022communication}
C.~Wu, F.~Wu, L.~Lyu, Y.~Huang, and X.~Xie, ``Communication-efficient federated learning via knowledge distillation,'' \emph{Nature communications}, vol.~13, no.~1, p. 2032, 2022.

\bibitem{ding2023efficient}
X.~Ding, G.~Li, L.~Yuan, L.~Zhang, and Q.~Rong, ``Efficient federated item similarity model for privacy-preserving recommendation,'' \emph{Information Processing \& Management}, vol.~60, no.~5, p. 103470, 2023.

\bibitem{nguyen2024towards}
N.-H. Nguyen, T.-A. Nguyen, T.~Nguyen, V.~T. Hoang, D.~D. Le, and K.-S. Wong, ``Towards efficient communication federated recommendation system via low-rank training,'' \emph{arXiv preprint arXiv:2401.03748}, 2024.

\bibitem{xia2024aerorec}
T.~Xia, J.~Ren, W.~Rao, Q.~Zu, W.~Wang, S.~Chen, and Y.~Zhang, ``Aerorec: an efficient on-device recommendation framework using federated self-supervised knowledge distillation,'' in \emph{IEEE INFOCOM 2024-IEEE Conference on Computer Communications}.\hskip 1em plus 0.5em minus 0.4em\relax IEEE, 2024, pp. 121--130.

\bibitem{liu2024efficient}
L.~Liu, W.~Wang, X.~Zhao, Z.~Zhang, C.~Zhang, S.~Lin, Y.~Wang, L.~Zou, Z.~Liu, X.~Wei \emph{et~al.}, ``Efficient and robust regularized federated recommendation,'' in \emph{Proceedings of the 33rd ACM International Conference on Information and Knowledge Management}, 2024, pp. 1452--1461.

\bibitem{luo2021feature}
X.~Luo, Y.~Wu, X.~Xiao, and B.~C. Ooi, ``Feature inference attack on model predictions in vertical federated learning,'' in \emph{2021 IEEE 37th International Conference on Data Engineering (ICDE)}.\hskip 1em plus 0.5em minus 0.4em\relax IEEE, 2021, pp. 181--192.

\bibitem{zhang2022fldetector}
Z.~Zhang, X.~Cao, J.~Jia, and N.~Z. Gong, ``Fldetector: Defending federated learning against model poisoning attacks via detecting malicious clients,'' in \emph{Proceedings of the 28th ACM SIGKDD Conference on Knowledge Discovery and Data Mining}, 2022, pp. 2545--2555.

\bibitem{zeng2023flbooster}
Z.~Zeng, Y.~Du, Z.~Fang, L.~Chen, S.~Pu, G.~Chen, H.~Wang, and Y.~Gao, ``Flbooster: A unified and efficient platform for federated learning acceleration,'' in \emph{2023 IEEE 39th International Conference on Data Engineering (ICDE)}.\hskip 1em plus 0.5em minus 0.4em\relax IEEE, 2023, pp. 3140--3153.

\bibitem{yan2023criticalfl}
G.~Yan, H.~Wang, X.~Yuan, and J.~Li, ``Criticalfl: A critical learning periods augmented client selection framework for efficient federated learning,'' in \emph{Proceedings of the 29th ACM SIGKDD Conference on Knowledge Discovery and Data Mining}, 2023, pp. 2898--2907.

\bibitem{zhang2024prototype}
C.~Zhang, Y.~Xie, T.~Chen, W.~Mao, and B.~Yu, ``Prototype similarity distillation for communication-efficient federated unsupervised representation learning,'' \emph{IEEE Transactions on Knowledge and Data Engineering}, 2024.

\bibitem{zhang2022no}
X.~Zhang, H.~Gu, L.~Fan, K.~Chen, and Q.~Yang, ``No free lunch theorem for security and utility in federated learning,'' \emph{ACM Transactions on Intelligent Systems and Technology}, vol.~14, no.~1, pp. 1--35, 2022.

\bibitem{ko2022survey}
H.~Ko, S.~Lee, Y.~Park, and A.~Choi, ``A survey of recommendation systems: recommendation models, techniques, and application fields,'' \emph{Electronics}, vol.~11, no.~1, p. 141, 2022.

\bibitem{javed2021review}
U.~Javed, K.~Shaukat, I.~A. Hameed, F.~Iqbal, T.~M. Alam, and S.~Luo, ``A review of content-based and context-based recommendation systems,'' \emph{International Journal of Emerging Technologies in Learning (iJET)}, vol.~16, no.~3, pp. 274--306, 2021.

\bibitem{perez2021content}
Y.~P{\'e}rez-Almaguer, R.~Yera, A.~A. Alzahrani, and L.~Mart{\'\i}nez, ``Content-based group recommender systems: A general taxonomy and further improvements,'' \emph{Expert Systems with Applications}, vol. 184, p. 115444, 2021.

\bibitem{martins2020deep}
G.~B. Martins, J.~P. Papa, and H.~Adeli, ``Deep learning techniques for recommender systems based on collaborative filtering,'' \emph{Expert Systems}, vol.~37, no.~6, p. e12647, 2020.

\bibitem{papadakis2022collaborative}
H.~Papadakis, A.~Papagrigoriou, C.~Panagiotakis, E.~Kosmas, and P.~Fragopoulou, ``Collaborative filtering recommender systems taxonomy,'' \emph{Knowledge and Information Systems}, vol.~64, no.~1, pp. 35--74, 2022.

\bibitem{thorat2015survey}
P.~B. Thorat, R.~M. Goudar, and S.~Barve, ``Survey on collaborative filtering, content-based filtering and hybrid recommendation system,'' \emph{International Journal of Computer Applications}, vol. 110, no.~4, pp. 31--36, 2015.

\bibitem{walek2020hybrid}
B.~Walek and V.~Fojtik, ``A hybrid recommender system for recommending relevant movies using an expert system,'' \emph{Expert Systems with Applications}, vol. 158, p. 113452, 2020.

\bibitem{da2020recommendation}
A.~Da’u and N.~Salim, ``Recommendation system based on deep learning methods: a systematic review and new directions,'' \emph{Artificial Intelligence Review}, vol.~53, no.~4, pp. 2709--2748, 2020.

\bibitem{wu2020sse}
L.~Wu, S.~Li, C.-J. Hsieh, and J.~Sharpnack, ``Sse-pt: Sequential recommendation via personalized transformer,'' in \emph{Proceedings of the 14th ACM conference on recommender systems}, 2020, pp. 328--337.

\bibitem{he2020lightgcn}
X.~He, K.~Deng, X.~Wang, Y.~Li, Y.~Zhang, and M.~Wang, ``Lightgcn: Simplifying and powering graph convolution network for recommendation,'' in \emph{Proceedings of the 43rd International ACM SIGIR conference on research and development in Information Retrieval}, 2020, pp. 639--648.

\bibitem{shuai2022review}
J.~Shuai, K.~Zhang, L.~Wu, P.~Sun, R.~Hong, M.~Wang, and Y.~Li, ``A review-aware graph contrastive learning framework for recommendation,'' in \emph{Proceedings of the 45th international ACM SIGIR conference on research and development in information retrieval}, 2022, pp. 1283--1293.

\bibitem{hussien2021recommendation}
F.~T.~A. Hussien, A.~M.~S. Rahma, and H.~B.~A. Wahab, ``Recommendation systems for e-commerce systems an overview,'' in \emph{Journal of Physics: Conference Series}, vol. 1897, no.~1.\hskip 1em plus 0.5em minus 0.4em\relax IOP Publishing, 2021, p. 012024.

\bibitem{ji2021you}
H.~Ji, J.~Zhu, X.~Wang, C.~Shi, B.~Wang, X.~Tan, Y.~Li, and S.~He, ``Who you would like to share with? a study of share recommendation in social e-commerce,'' in \emph{Proceedings of the AAAI conference on artificial intelligence}, vol.~35, no.~1, 2021, pp. 232--239.

\bibitem{deldjoo2020recommender}
Y.~Deldjoo, M.~Schedl, P.~Cremonesi, and G.~Pasi, ``Recommender systems leveraging multimedia content,'' \emph{ACM Computing Surveys (CSUR)}, vol.~53, no.~5, pp. 1--38, 2020.

\bibitem{wang2020fine}
H.~Wang, F.~Wu, Z.~Liu, and X.~Xie, ``Fine-grained interest matching for neural news recommendation,'' in \emph{Proceedings of the 58th annual meeting of the association for computational linguistics}, 2020, pp. 836--845.

\bibitem{fan2019graph}
W.~Fan, Y.~Ma, Q.~Li, Y.~He, E.~Zhao, J.~Tang, and D.~Yin, ``Graph neural networks for social recommendation,'' in \emph{The world wide web conference}, 2019, pp. 417--426.

\bibitem{yu2021self}
J.~Yu, H.~Yin, J.~Li, Q.~Wang, N.~Q.~V. Hung, and X.~Zhang, ``Self-supervised multi-channel hypergraph convolutional network for social recommendation,'' in \emph{Proceedings of the web conference 2021}, 2021, pp. 413--424.

\bibitem{he2017neural}
X.~He, L.~Liao, H.~Zhang, L.~Nie, X.~Hu, and T.-S. Chua, ``Neural collaborative filtering,'' in \emph{Proceedings of the 26th international conference on world wide web}, 2017, pp. 173--182.

\bibitem{cailightgcl}
X.~Cai, C.~Huang, L.~Xia, and X.~Ren, ``Lightgcl: Simple yet effective graph contrastive learning for recommendation,'' in \emph{The Eleventh International Conference on Learning Representations}.

\bibitem{zhang2021parameterized}
J.~Zhang, S.~Guo, X.~Ma, H.~Wang, W.~Xu, and F.~Wu, ``Parameterized knowledge transfer for personalized federated learning,'' \emph{Advances in Neural Information Processing Systems}, vol.~34, pp. 10\,092--10\,104, 2021.

\bibitem{hu2020personalized}
R.~Hu, Y.~Guo, H.~Li, Q.~Pei, and Y.~Gong, ``Personalized federated learning with differential privacy,'' \emph{IEEE Internet of Things Journal}, vol.~7, no.~10, pp. 9530--9539, 2020.

\bibitem{shamsian2021personalized}
A.~Shamsian, A.~Navon, E.~Fetaya, and G.~Chechik, ``Personalized federated learning using hypernetworks,'' in \emph{International Conference on Machine Learning}.\hskip 1em plus 0.5em minus 0.4em\relax PMLR, 2021, pp. 9489--9502.

\bibitem{pillutla2022federated}
K.~Pillutla, K.~Malik, A.-R. Mohamed, M.~Rabbat, M.~Sanjabi, and L.~Xiao, ``Federated learning with partial model personalization,'' in \emph{International Conference on Machine Learning}.\hskip 1em plus 0.5em minus 0.4em\relax PMLR, 2022, pp. 17\,716--17\,758.

\bibitem{mansour2020three}
Y.~Mansour, M.~Mohri, J.~Ro, and A.~T. Suresh, ``Three approaches for personalization with applications to federated learning,'' \emph{arXiv preprint arXiv:2002.10619}, 2020.

\bibitem{wu2020fedhome}
Q.~Wu, X.~Chen, Z.~Zhou, and J.~Zhang, ``Fedhome: Cloud-edge based personalized federated learning for in-home health monitoring,'' \emph{IEEE Transactions on Mobile Computing}, vol.~21, no.~8, pp. 2818--2832, 2020.

\bibitem{wang2020optimizing}
H.~Wang, Z.~Kaplan, D.~Niu, and B.~Li, ``Optimizing federated learning on non-iid data with reinforcement learning,'' in \emph{IEEE INFOCOM 2020-IEEE conference on computer communications}.\hskip 1em plus 0.5em minus 0.4em\relax IEEE, 2020, pp. 1698--1707.

\bibitem{li2021fedsae}
L.~Li, M.~Duan, D.~Liu, Y.~Zhang, A.~Ren, X.~Chen, Y.~Tan, and C.~Wang, ``Fedsae: A novel self-adaptive federated learning framework in heterogeneous systems,'' in \emph{2021 International Joint Conference on Neural Networks (IJCNN)}.\hskip 1em plus 0.5em minus 0.4em\relax IEEE, 2021, pp. 1--10.

\bibitem{li2021model}
Q.~Li, B.~He, and D.~Song, ``Model-contrastive federated learning,'' in \emph{Proceedings of the IEEE/CVF conference on computer vision and pattern recognition}, 2021, pp. 10\,713--10\,722.

\bibitem{fallah2020personalized}
A.~Fallah, A.~Mokhtari, and A.~Ozdaglar, ``Personalized federated learning with theoretical guarantees: A model-agnostic meta-learning approach,'' \emph{Advances in neural information processing systems}, vol.~33, pp. 3557--3568, 2020.

\bibitem{yang2020fedsteg}
H.~Yang, H.~He, W.~Zhang, and X.~Cao, ``Fedsteg: A federated transfer learning framework for secure image steganalysis,'' \emph{IEEE Transactions on Network Science and Engineering}, vol.~8, no.~2, pp. 1084--1094, 2020.

\bibitem{arivazhagan2019federated}
M.~G. Arivazhagan, V.~Aggarwal, A.~K. Singh, and S.~Choudhary, ``Federated learning with personalization layers,'' \emph{arXiv preprint arXiv:1912.00818}, 2019.

\bibitem{liang2020think}
P.~P. Liang, T.~Liu, L.~Ziyin, N.~B. Allen, R.~P. Auerbach, D.~Brent, R.~Salakhutdinov, and L.-P. Morency, ``Think locally, act globally: Federated learning with local and global representations,'' \emph{arXiv preprint arXiv:2001.01523}, 2020.

\bibitem{zhu2021data}
Z.~Zhu, J.~Hong, and J.~Zhou, ``Data-free knowledge distillation for heterogeneous federated learning,'' in \emph{International conference on machine learning}.\hskip 1em plus 0.5em minus 0.4em\relax PMLR, 2021, pp. 12\,878--12\,889.

\bibitem{smith2017federated}
V.~Smith, C.-K. Chiang, M.~Sanjabi, and A.~S. Talwalkar, ``Federated multi-task learning,'' \emph{Advances in neural information processing systems}, vol.~30, 2017.

\bibitem{sattler2020clustered}
F.~Sattler, K.-R. M{\"u}ller, and W.~Samek, ``Clustered federated learning: Model-agnostic distributed multitask optimization under privacy constraints,'' \emph{IEEE transactions on neural networks and learning systems}, vol.~32, no.~8, pp. 3710--3722, 2020.

\bibitem{wu2021self}
J.~Wu, X.~Wang, F.~Feng, X.~He, L.~Chen, J.~Lian, and X.~Xie, ``Self-supervised graph learning for recommendation,'' in \emph{Proceedings of the 44th international ACM SIGIR conference on research and development in information retrieval}, 2021, pp. 726--735.

\bibitem{xia2022hypergraph}
L.~Xia, C.~Huang, Y.~Xu, J.~Zhao, D.~Yin, and J.~Huang, ``Hypergraph contrastive collaborative filtering,'' in \emph{Proceedings of the 45th International ACM SIGIR conference on research and development in information retrieval}, 2022, pp. 70--79.

\bibitem{ren2023disentangled}
X.~Ren, L.~Xia, J.~Zhao, D.~Yin, and C.~Huang, ``Disentangled contrastive collaborative filtering,'' in \emph{Proceedings of the 46th International ACM SIGIR Conference on Research and Development in Information Retrieval}, 2023, pp. 1137--1146.

\bibitem{roy2022systematic}
D.~Roy and M.~Dutta, ``A systematic review and research perspective on recommender systems,'' \emph{Journal of Big Data}, vol.~9, no.~1, p.~59, 2022.

\bibitem{xia2021survey}
Q.~Xia, W.~Ye, Z.~Tao, J.~Wu, and Q.~Li, ``A survey of federated learning for edge computing: Research problems and solutions,'' \emph{High-Confidence Computing}, vol.~1, no.~1, p. 100008, 2021.

\bibitem{hulora}
E.~J. Hu, P.~Wallis, Z.~Allen-Zhu, Y.~Li, S.~Wang, L.~Wang, W.~Chen \emph{et~al.}, ``Lora: Low-rank adaptation of large language models,'' in \emph{International Conference on Learning Representations}.

\bibitem{lu2020meta}
Y.~Lu, Y.~Fang, and C.~Shi, ``Meta-learning on heterogeneous information networks for cold-start recommendation,'' in \emph{Proceedings of the 26th ACM SIGKDD international conference on knowledge discovery \& data mining}, 2020, pp. 1563--1573.

\bibitem{wei2021contrastive}
Y.~Wei, X.~Wang, Q.~Li, L.~Nie, Y.~Li, X.~Li, and T.-S. Chua, ``Contrastive learning for cold-start recommendation,'' in \emph{Proceedings of the 29th ACM International Conference on Multimedia}, 2021, pp. 5382--5390.

\bibitem{afchar2022explainability}
D.~Afchar, A.~Melchiorre, M.~Schedl, R.~Hennequin, E.~Epure, and M.~Moussallam, ``Explainability in music recommender systems,'' \emph{AI Magazine}, vol.~43, no.~2, pp. 190--208, 2022.

\bibitem{lyu2022knowledge}
Z.~Lyu, Y.~Wu, J.~Lai, M.~Yang, C.~Li, and W.~Zhou, ``Knowledge enhanced graph neural networks for explainable recommendation,'' \emph{IEEE Transactions on Knowledge and Data Engineering}, vol.~35, no.~5, pp. 4954--4968, 2022.

\bibitem{lei2020interactive}
W.~Lei, G.~Zhang, X.~He, Y.~Miao, X.~Wang, L.~Chen, and T.-S. Chua, ``Interactive path reasoning on graph for conversational recommendation,'' in \emph{Proceedings of the 26th ACM SIGKDD international conference on knowledge discovery \& data mining}, 2020, pp. 2073--2083.

\bibitem{chen2023bias}
J.~Chen, H.~Dong, X.~Wang, F.~Feng, M.~Wang, and X.~He, ``Bias and debias in recommender system: A survey and future directions,'' \emph{ACM Transactions on Information Systems}, vol.~41, no.~3, pp. 1--39, 2023.

\bibitem{sun2020multi}
R.~Sun, X.~Cao, Y.~Zhao, J.~Wan, K.~Zhou, F.~Zhang, Z.~Wang, and K.~Zheng, ``Multi-modal knowledge graphs for recommender systems,'' in \emph{Proceedings of the 29th ACM international conference on information \& knowledge management}, 2020, pp. 1405--1414.

\bibitem{liu2024multimodal}
Q.~Liu, J.~Hu, Y.~Xiao, X.~Zhao, J.~Gao, W.~Wang, Q.~Li, and J.~Tang, ``Multimodal recommender systems: A survey,'' \emph{ACM Computing Surveys}, vol.~57, no.~2, pp. 1--17, 2024.

\bibitem{he2024co}
X.~He, S.~Liu, J.~Keung, and J.~He, ``Co-clustering for federated recommender system,'' in \emph{Proceedings of the ACM on Web Conference 2024}, 2024, pp. 3821--3832.

\bibitem{ribero2022federating}
M.~Ribero, J.~Henderson, S.~Williamson, and H.~Vikalo, ``Federating recommendations using differentially private prototypes,'' \emph{Pattern Recognition}, vol. 129, p. 108746, 2022.

\bibitem{MLSYS2020_1f5fe839}
T.~Li, A.~K. Sahu, M.~Zaheer, M.~Sanjabi, A.~Talwalkar, and V.~Smith, ``Federated optimization in heterogeneous networks,'' in \emph{Proceedings of Machine Learning and Systems}, I.~Dhillon, D.~Papailiopoulos, and V.~Sze, Eds., vol.~2, 2020, pp. 429--450.

\bibitem{deng2020adaptive}
Y.~Deng, M.~M. Kamani, and M.~Mahdavi, ``Adaptive personalized federated learning,'' \emph{arXiv preprint arXiv:2003.13461}, 2020.

\bibitem{luo2023gradma}
K.~Luo, X.~Li, Y.~Lan, and M.~Gao, ``Gradma: A gradient-memory-based accelerated federated learning with alleviated catastrophic forgetting,'' in \emph{Proceedings of the IEEE/CVF Conference on Computer Vision and Pattern Recognition}, 2023, pp. 3708--3717.

\bibitem{radford2019language}
A.~Radford, J.~Wu, R.~Child, D.~Luan, D.~Amodei, I.~Sutskever \emph{et~al.}, ``Language models are unsupervised multitask learners,'' \emph{OpenAI blog}, vol.~1, no.~8, p.~9, 2019.

\bibitem{bommasani2021opportunities}
R.~Bommasani, D.~A. Hudson, E.~Adeli, R.~Altman, S.~Arora, S.~von Arx, M.~S. Bernstein, J.~Bohg, A.~Bosselut, E.~Brunskill \emph{et~al.}, ``On the opportunities and risks of foundation models,'' \emph{arXiv preprint arXiv:2108.07258}, 2021.

\bibitem{achiam2023gpt}
J.~Achiam, S.~Adler, S.~Agarwal, L.~Ahmad, I.~Akkaya, F.~L. Aleman, D.~Almeida, J.~Altenschmidt, S.~Altman, S.~Anadkat \emph{et~al.}, ``Gpt-4 technical report,'' \emph{arXiv preprint arXiv:2303.08774}, 2023.

\bibitem{10787102}
K.-H. Huang, H.~P. Chan, Y.~R. Fung, H.~Qiu, M.~Zhou, S.~Joty, S.-F. Chang, and H.~Ji, ``From pixels to insights: A survey on automatic chart understanding in the era of large foundation models,'' \emph{IEEE Transactions on Knowledge and Data Engineering}, pp. 1--20, 2024.

\bibitem{liang2024foundation}
Y.~Liang, H.~Wen, Y.~Nie, Y.~Jiang, M.~Jin, D.~Song, S.~Pan, and Q.~Wen, ``Foundation models for time series analysis: A tutorial and survey,'' in \emph{Proceedings of the 30th ACM SIGKDD conference on knowledge discovery and data mining}, 2024, pp. 6555--6565.

\bibitem{alayrac2022flamingo}
J.-B. Alayrac, J.~Donahue, P.~Luc, A.~Miech, I.~Barr, Y.~Hasson, K.~Lenc, A.~Mensch, K.~Millican, M.~Reynolds \emph{et~al.}, ``Flamingo: a visual language model for few-shot learning,'' \emph{Advances in neural information processing systems}, vol.~35, pp. 23\,716--23\,736, 2022.

\bibitem{yang2023uniaudio}
D.~Yang, J.~Tian, X.~Tan, R.~Huang, S.~Liu, X.~Chang, J.~Shi, S.~Zhao, J.~Bian, X.~Wu \emph{et~al.}, ``Uniaudio: An audio foundation model toward universal audio generation,'' \emph{arXiv preprint arXiv:2310.00704}, 2023.

\bibitem{kojima2022large}
T.~Kojima, S.~S. Gu, M.~Reid, Y.~Matsuo, and Y.~Iwasawa, ``Large language models are zero-shot reasoners,'' \emph{Advances in neural information processing systems}, vol.~35, pp. 22\,199--22\,213, 2022.

\bibitem{zhao2024llm}
J.~Zhao, W.~Wang, C.~Xu, Z.~Ren, S.-K. Ng, and T.-S. Chua, ``Llm-based federated recommendation,'' \emph{arXiv preprint arXiv:2402.09959}, 2024.

\bibitem{zhuang2023foundation}
W.~Zhuang, C.~Chen, and L.~Lyu, ``When foundation model meets federated learning: Motivations, challenges, and future directions,'' \emph{arXiv preprint arXiv:2306.15546}, 2023.

\bibitem{ren2024advances}
C.~Ren, H.~Yu, H.~Peng, X.~Tang, A.~Li, Y.~Gao, A.~Z. Tan, B.~Zhao, X.~Li, Z.~Li \emph{et~al.}, ``Advances and open challenges in federated learning with foundation models,'' \emph{arXiv preprint arXiv:2404.15381}, 2024.

\bibitem{hong2021federated}
J.~Hong, Z.~Zhu, S.~Yu, Z.~Wang, H.~H. Dodge, and J.~Zhou, ``Federated adversarial debiasing for fair and transferable representations,'' in \emph{Proceedings of the 27th ACM SIGKDD Conference on Knowledge Discovery \& Data Mining}, 2021, pp. 617--627.

\bibitem{xu2023bias}
Y.-Y. Xu, C.-S. Lin, and Y.-C.~F. Wang, ``Bias-eliminating augmentation learning for debiased federated learning,'' in \emph{Proceedings of the IEEE/CVF Conference on Computer Vision and Pattern Recognition}, 2023, pp. 20\,442--20\,452.

\bibitem{zhang2020explainable}
Y.~Zhang, X.~Chen \emph{et~al.}, ``Explainable recommendation: A survey and new perspectives,'' \emph{Foundations and Trends{\textregistered} in Information Retrieval}, vol.~14, no.~1, pp. 1--101, 2020.

\bibitem{chen2024large}
J.~Chen, Z.~Liu, X.~Huang, C.~Wu, Q.~Liu, G.~Jiang, Y.~Pu, Y.~Lei, X.~Chen, X.~Wang \emph{et~al.}, ``When large language models meet personalization: Perspectives of challenges and opportunities,'' \emph{World Wide Web}, vol.~27, no.~4, p.~42, 2024.

\bibitem{kunaver2017diversity}
M.~Kunaver and T.~Po{\v{z}}rl, ``Diversity in recommender systems--a survey,'' \emph{Knowledge-based systems}, vol. 123, pp. 154--162, 2017.

\bibitem{castells2021novelty}
P.~Castells, N.~Hurley, and S.~Vargas, ``Novelty and diversity in recommender systems,'' in \emph{Recommender systems handbook}.\hskip 1em plus 0.5em minus 0.4em\relax Springer, 2021, pp. 603--646.

\end{thebibliography}
}

% \bibliographystyle{IEEEtran}

% \bibitem{ref1}
% {\it{Mathematics Into Type}}. American Mathematical Society. [Online]. Available: https://www.ams.org/arc/styleguide/mit-2.pdf

% \end{thebibliography}
\vfill

\end{document}